%% file: root.tex
\documentclass[journal,twoside,web,twocolumn]{ieeecolor} 

\let\labelindent\relax
\usepackage{generic}
\usepackage{amsthm}
\usepackage{booktabs} 
\usepackage{hyperref}       
\usepackage{amssymb}
\usepackage{nicefrac}       
\usepackage{microtype}      
\usepackage{mathtools}
\usepackage{cases}

\usepackage{ifthen}
\usepackage{psfrag}
\usepackage{epstopdf}
\usepackage{listings}
\usepackage{lipsum}
\usepackage{enumitem} 
\usepackage{thm-restate}
\usepackage{comment}
\usepackage{cite}

\usepackage{algorithmic}
\usepackage{algorithm}
\usepackage{graphicx}
\usepackage{textcomp}
\def\BibTeX{{\rm B\kern-.05em{\sc i\kern-.025em b}\kern-.08em
		T\kern-.1667em\lower.7ex\hbox{E}\kern-.125emX}}
\newcommand{\eps}{\varepsilon}

\newcommand{\Talgo}{\relax\ifmmode {T_p} \else $T_p$\fi}
\newcommand{\jum}{\relax\ifmmode {\eta} \else $\eta$\fi}
\newcommand{\linfnorm}[1]{\relax\ifmmode {\|#1\|_\infty} \else $||#1||_\infty$\fi}

\DeclarePairedDelimiter\ceil{\lceil}{\rceil}
\DeclarePairedDelimiter\floor{\lfloor}{\rfloor}

\newcommand{\nnreals}{\mathbb{R}_{\geq 0}}
\newcommand{\dist}{d}
\newcommand{\reals}{\mathbb{R}}

\newcommand{\K}{\mathcal K}

\newcommand {\R}{\mathbb R}

\newcommand{\Z}{\mathcal Z}

\newcommand{\modetopar}{\psi} 
\newcommand{\sw}{\mathit{sw}}

\newcommand{\reach}{\mathit{Reach}}

\newcommand{\inputset}{\mathcal{U}}
\newcommand{\stateset}{\mathcal{X}}
\newcommand{\zeps}{\mathit{z\eps}}
\newcommand{\palpha}{\mathit{p\alpha}}
\newcommand{\infd}{\mathit{d}}

\newcommand{\sayan}[1]{\textcolor{blue}{}} 
\newcommand{\hussein}[1]{{#1}}

\newcommand{\est}{_{\text{est}}}
\newcommand{\sest}{_{\text{sest}}}
\newcommand{\sep}{^{\dagger}}
\newcommand{\Zsep}{\mathcal{Z}\sep}

\newcommand{\estepsz}{\est^0}
\newcommand{\tseq}{tseq}
\newcommand{\estexp}{_{\text{exp}}}

\newtheorem{Assumption}{Assumption}
\newtheorem{definition}{Definition}
\newtheorem{theorem}{Theorem}
\newtheorem{proposition}{Proposition}
\newtheorem{corollary}{Corollary}
\newtheorem{remark}{Remark}
\newtheorem{Lemma}{Lemma}
\newtheorem{example}{Example}

\usepackage{accents}
\newcommand{\ubar}[1]{\underaccent{\bar}{#1}}

\newcommand{\swin}{\psi}
\newcommand{\usig}{u}

\DeclareFontEncoding{FMS}{}{}
\DeclareFontSubstitution{FMS}{futm}{m}{n}
\DeclareFontEncoding{FMX}{}{}
\DeclareFontSubstitution{FMX}{futm}{m}{n}
\DeclareSymbolFont{fouriersymbols}{FMS}{futm}{m}{n}
\DeclareSymbolFont{fourierlargesymbols}{FMX}{futm}{m}{n}
\DeclareMathDelimiter{\VERT}{\mathord}{fouriersymbols}{152}{fourierlargesymbols}{147}

\newcommand{\ts}{\textsuperscript}

\title{State Estimation of Continuous-time Dynamical Systems with Uncertain Inputs with Bounded Variation: Entropy, Bit Rates, and Relation with Switched Systems
	\author{Hussein Sibai and Sayan Mitra, \IEEEmembership{senior member}
		\thanks{Submitted on April 5th, 2021, resubmitted on November 13, 2022, accepted February 11, 2023.
		Hussein Sibai is with the Computer Science and Engineering department at  Washington University in St. Louis and Sayan Mitra is with the Coordinated Science Laboratory at the University of Illinois at Urbana-Champaign. Emails: sibai@wustl.edu and mitras@illinois.edu.}
}}
\begin{document}

\maketitle

\begin{abstract}
\hussein{We extend the notion of {\em estimation entropy} of autonomous dynamical systems proposed by Liberzon and Mitra~\cite{LM:TAC2018} to nonlinear dynamical systems with uncertain inputs with bounded variation. We call this new notion the $\eps$-{\em estimation entropy} of the system and show that it lower bounds the bit rate needed for state estimation.
$\eps$-estimation entropy represents the exponential rate of the increase of the minimal number of functions that  are adequate for $\eps$-approximating any trajectory of the system.
 We show that alternative entropy definitions using spanning or separating trajectories bound ours from both sides. On the other hand, we show that  other commonly used definitions of entropy, for example the ones in \cite{LM:TAC2018}, diverge to infinity.
 Thus, they are potentially not suitable for systems with uncertain inputs.  We derive an upper bound on $\eps$-estimation entropy and estimation bit rates,  and evaluate it for two examples. We present a state estimation algorithm that constructs a function that approximates a given trajectory up to an $\eps$ error, given time-sampled and quantized measurements of state and input. 
We investigate the relation between $\eps$-estimation entropy and a previous notion for switched nonlinear systems and derive a new upper bound for the latter, showing the generality of our results on systems with uncertain inputs. 
}
\sayan{Long version (to be removed): 
Finding the minimal bit rate needed to estimate the evolving state of a dynamical system is an essential problem for monitoring and control. Such a bit rate is related to the rate at which the dynamical system itself generates information, or equivalently, its trajectories become more diverse. The notion of topological entropy for dynamical systems measures such an information generation rate. Various definitions have been shown to lower-bound the minimal bit rates for different estimation and control tasks. 
 In this paper, we extend one of these notions, denoted by {\em estimation entropy} by Liberzon and Mitra~\cite{LM:TAC2018}, to nonlinear dynamical systems with slowly-varying inputs to lower bound the bit rates needed to estimate their states. Our entropy definition, which we denote by $\eps$-estimation entropy, represents the rate of exponential increase of the number of functions needed to approximate the trajectories of the system up to a specified $\eps$ error.
 We show that alternative entropy definitions using spanning or separating trajectories bound ours from both sides.
On the other hand, we show that the existing definitions of entropy that consider supremum over all $\eps$ or require exponential convergence of estimation error are not suitable for systems with inputs.
%
Since the actual value  of entropy is generally hard to compute, we derive an upper bound and evaluate it for two examples. We show that as the bound on the input variation decreases, we recover a previously known bound on estimation entropy for autonomous nonlinear systems. For the sake of computing the bound, we present an algorithm that, given sampled and quantized measurements from a trajectory and its generating input signal, constructs a function that approximates the trajectory up to an $\eps$ error. We show that this algorithm can also be used for state estimation.
We relate the computed bound with a previously known upper bound on the entropy for switched nonlinear systems. We show that a bound on the divergence between the different modes of a switched system is needed to get a meaningful bound on its entropy. Finally, we derive an upper bound  on the entropy of a differential inclusion having solutions that include those of a given switched system using our bound for systems with inputs.}
%

\end{abstract}


\input{intro}
\input{prelims}
\input{Entropydefinition}

\input{Epsilonzero}

\input{def_motivation}
\input{BitRateandEntropy}
\input{UpperBound}
\input{discrepancyFunction}

\input{approxFunction}
\input{noise_systems}

\input{ExpConvergence}

\input{conclusion}

\section*{Acknowledgment}
We thank Daniel Liberzon and the anonymous reviewers for providing detailed and insightful
comments that enhanced the results and presentation of this paper.

\begin{IEEEbiography}[{	\includegraphics[height=1.25in, width=1in]{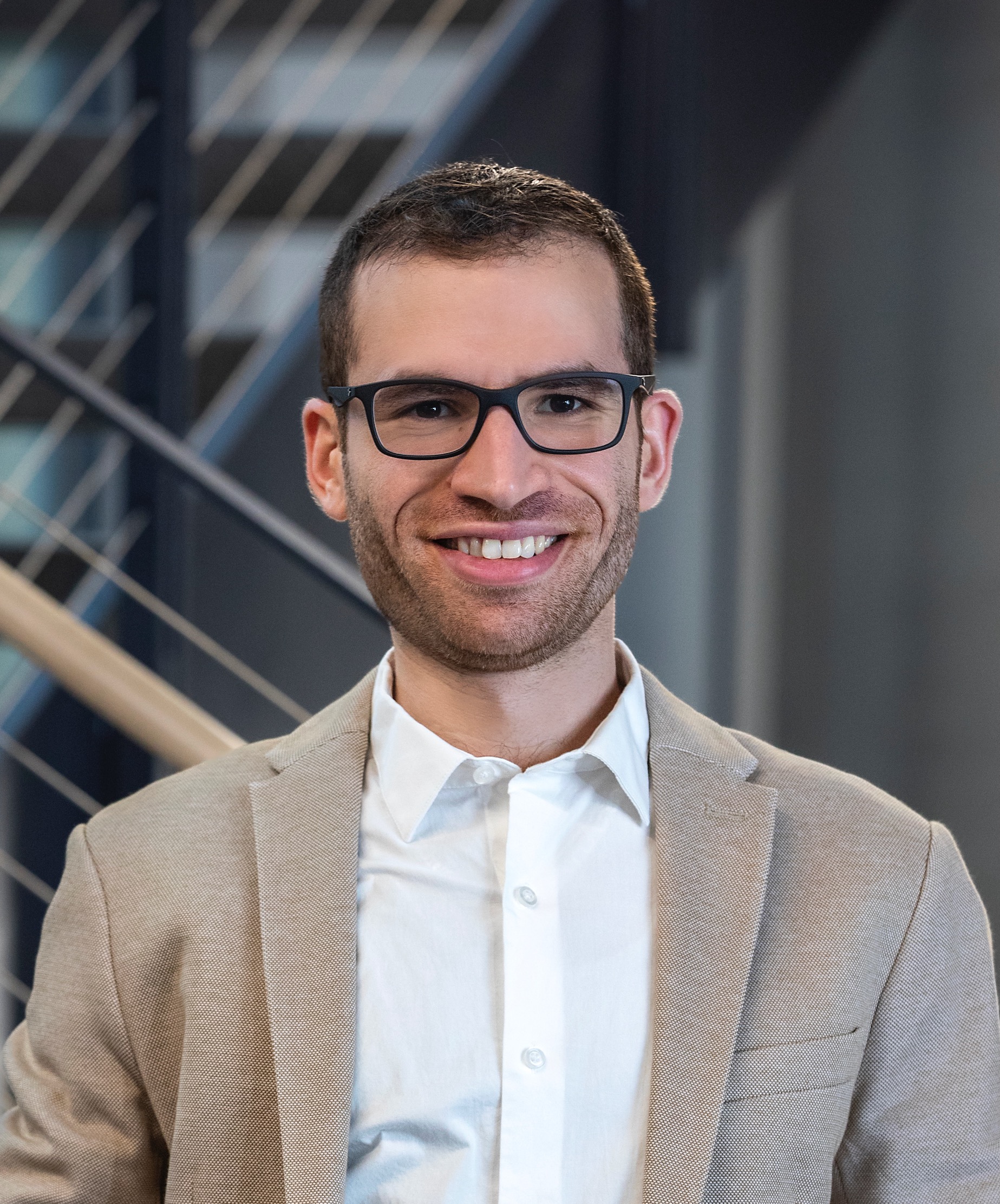}}]{Hussein Sibai}
	is an Assistant Professor in the Computer Science and Engineering department at Washington University in St. Louis. He was a postdoctoral scholar at UC Berkeley. He obtained his Ph.D. in Electrical and Computer Engineering from the University of Illinois Urbana-Champaign (UIUC) in December 2021. He received his bachelor's degree in Computer and Communication Engineering from the American University of Beirut and a master's degree in Electrical and Computer Engineering from UIUC. His research interests are in formal methods, control theory, and machine learning. Hussein won the best poster award in HSCC 2018 and best paper nominations at HSCC 2017 and ATVA 2019. His work has been recognized by the Rambus fellowship, the Ernest A. Reid fellowship, the MAVIS Future Faculty fellowship, and the ACM SIGBED gold medal for the graduate student research competition in CPS Week 21.
\end{IEEEbiography}

\begin{IEEEbiography}[{\includegraphics[height=1.25in, width=1in]{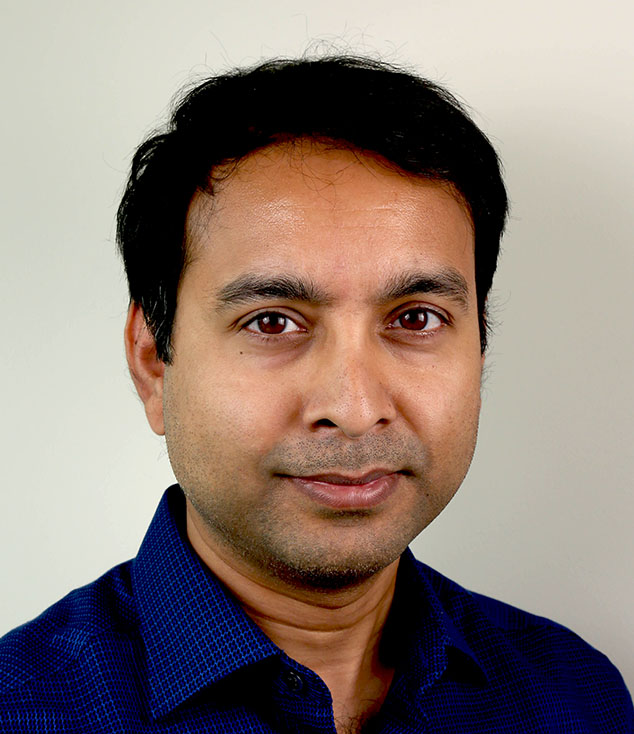}}]{Sayan Mitra}
	(M'01--SM'13)
	received his  undergraduate degree in Electrical Engineering from Jadavpur University, Kolkata, India, a Masters Degree in Computer Science from Indian Institute of Science, Bangalore, and a Ph.D. degree in EECS from MIT in 2007.
	After one year of postdoctoral work at the Center for Mathematics of Information at the California Institute of Technology,  he joined the University of Illinois at Urbana-Champaign, where he is now an  Professor in the Electrical and Computer Engineering Department.
	His research interests include formal methods, hybrid systems, autonomous systems, and verification of cyber-physical systems and their applications.
	His work has been recognized by National Science Foundation's CAREER Award, Air Force Office of Scientific Research Young Investigator Program Award,  IEEE-HKN C. Holmes MacDonald Outstanding Teaching Award and several best paper awards. He has served as the Program co-chair of the \textsc{20th International Conference on Hybrid Systems}.
\end{IEEEbiography}

\bibliographystyle{IEEEtran}
\bibliography{sayan1,hussein}

\appendices

\input{appendix}

\input{LinearSystems}
\input{PendulumExample}

\end{document}

%% file: intro.tex
\section{Introduction}
\label{sec:intro}
State estimation is a fundamental problem for controlling and monitoring dynamical systems. 
In most application scenarios, the estimator has to work with plant state information sent by a sensor over a  channel with finite bit rate. If a certain accuracy is required from the estimator, then a natural question is to ask: {\em what is the minimal bit rate of the channel for the estimator to support this accuracy requirement?} This question has been investigated for both stochastic and non-stochastic system models and channels. In the non-stochastic setting, the point of view of topological entropy has proven to be a fruitful line of investigation. In particular, it has been used for deriving the minimal necessary bit-rates for closed, i.e., autonomous, non-switched, and switched systems (see, for example~\cite{LM015:HSCC,LM:TAC2018,sibai-mitra-2017}). This paper contributes in this line of investigation and proposes answers to the above question for continuous-time dynamical systems with \hussein{uncertain inputs with bounded variation}. 
This class includes a wide variety of systems, for example: those with open-loop control, those with external disturbances, and autonomous systems as a special case. 



State estimation of 
systems with uncertain inputs is a more challenging problem than that of autonomous nonlinear switched and non-switched ones of \cite{LM015:HSCC,sibai-mitra-2017}. That is because even if the uncertainty about the state can be made to decrease over time using sensor measurements, the uncertainty about the input may not decrease. The input can change slowly and the continuous effect of the uncertain input prevents the uncertainty about the state from going to zero.
In contrast, the uncertainty can be made to decrease exponentially over time for autonomous non-switched systems \cite{LM:TAC2018}, as well as for autonomous switched ones between their mode switches \cite{sibai-mitra-2017}.
We contend this challenge of  changing input using a weaker notion of estimation, akin to that in~\cite{Savkin-Petersen-2003}, that only requires the error to be bounded by a constant $\eps>0$.

%
Given two estimation error parameters $\eps > 0$ and $\alpha \geq 0$, we first define a notion of topological entropy $h\est(\eps,\alpha)$ for systems with uncertain inputs that represents the rate of increase over time of the minimal number of functions needed to approximate all system trajectories up to an exponentially decaying error of $\eps e^{-\alpha t}$, for any $t \geq 0$.  We suggest an alternative entropy notion $h\est^*(\eps,\alpha)$, where we restrict the approximating functions to be trajectories of the system. We further consider a third notion $h\est^\dagger(\eps,\alpha)$ that represents the rate of increase over time of the number of trajectories that pairwise violate the approximation error bound $\eps e^{-\alpha t}$, at some $t \geq 0$. We show that all three notions are close, and thus would be viable alternative options. 

Then, we show that requiring exponential decay of estimation error, i.e., an $\alpha$ that is strictly positive, leads to an infinite entropy $h\est(\eps,\alpha)$, even for a simple linear one-dimensional system with piece-wise constant input (Section~\ref{sec:expconvergence}). Moreover, we show that $h\est(\eps,\alpha)$ of the same system grows to infinity as $\eps$ goes to zero (Section~\ref{sec:epsilonzero}). 
On the other hand, we show that $h\est(\eps,0)$ for that simple system is finite for any strictly positive $\eps$ (Corollary~\ref{cor:simple_sys_entropy_finite}).  
These observations imply that the definition of entropy for autonomous systems~(e.g. \cite{Savkin-Petersen-2003}), that takes the limit of $\eps$ as it goes to zero, or that of~\cite{LM015:HSCC, LM:TAC2018} that requires exponential convergence of error, are not suitable for systems with inputs. We thus use $h\est(\eps)$ as a weaker notion of estimation entropy, which we name $\eps$-{\em estimation entropy}, 
from there forward, as well as $h\est^*(\eps)$ and $h\est^\dagger(\eps)$ for the alternative notions, implicitly fixing $\alpha$ to zero and $\eps$ to a strictly-positive real number.
 
	
%
We show
 that there is no state estimation algorithm with a fixed bit rate smaller than $h\est(\eps)$ (Section~\ref{sec:bitrate_entropy}). 
While computing $h\est(\eps)$ exactly is generally hard, we compute an upper bound. To do that, we use local discrepancy functions to upper-bound the sensitivity of a trajectory of a nonlinear system with uncertain inputs to changes in its initial state and in its input signal. Then, we present a procedure (Algorithm~\ref{fig:approxFunct}) that, given sampled and quantized states and inputs, constructs a function that estimates the trajectory up to an $\eps$ error. This procedure is of independent interest, as it can also be used as a state estimation algorithm, provided that the uncertain input signal can be sampled. We count the number of trajectories that can be constructed by this procedure for different initial states and input signals, up to a time bound $T$. The rate of exponential increase of this number as $T$ increases gives an upper bound on entropy, \hussein{ and the bit rate needed for state estimation}. 

The upper bound is presented in terms of  the state and the input dimensions $n$ and $m$, global  bounds on the norms of the Jacobian matrices of the vector field with respect to the state and the input, $M_x$ and $M_u$, and two constants $\mu$ and $\jum$ that bound the variation of the input signal over time (Proposition~\ref{prop:noise_entropy_upperbound_lip}). \hussein{ We consider two approaches in bounding the variation of the input signal (Definition~\ref{def:inputset}). The first bounds the standard {\em total variation} of the signal over any time interval  $[t, t + \tau]$ 
	by $\mu \tau + \eta$. The other is a less-restrictive one that bounds the {\em pointwise variation} 
	of the signal at any two time instants $t$ and $t + \tau$ by $\mu \tau + \eta$.} 
\hussein{ We show that all of our results in the paper hold for both cases.}
\sayan{(removed: Roughly, $\jum$ upper bounds the size of the jumps in the input signal and $\mu$ constrains the number of large jumps in a short amount of time.)} 
We further show that if the variation of the input, \hussein{with either definition}, goes to zero, the upper bound on entropy approaches  $\frac{nL_x}{\ln 2}$, where $L_x$ is the Lipschitz constant of the vector field with respect to the state, exactly matching that computed in \cite{LM015:HSCC} for autonomous systems, with the exponential decay of estimation error parameter $\alpha$ being equal to zero (Corollary~\ref{cor:relationToPrevBound}).
Finally, we rewrite the upper bound on $h\est(\eps)$ in a similar format to that of autonomous switched systems which we presented in \cite{sibai-mitra-2017}.
Such rewriting shows the similarities and differences between the construction of the bounds in the two cases (Section~\ref{sec:si_and_ss_upper_bound_relation}). In addition, in Section~\ref{sec:si_abstract_ss}, we over-approximate the solutions of autonomous switched systems with differential inclusions. Then, we model these inclusions as systems with inputs and obtain a new upper bound on entropy of autonomous switched systems, as an alternative to that we presented in \cite{sibai-mitra-2017}. 

\hussein{We presented some of the results in this paper in \cite{sibai-mitra-2017, sibai-mitra-2018}, particularly, earlier versions of Proposition~\ref{prop:min_bit_rate}, Proposition~\ref{prop:noise_entropy_upperbound_lip},  Lemma~\ref{lm:local_discrep_func}, Lemma~\ref{lm:approxSetBound}, and Theorem~\ref{thm:switched_entropy_upper_bound}. The main novel contributions of this paper are: (1) Comparing the alternative entropy notions using approximating, spanning, and separated sets; (2) showing that the entropy of the simple system $\dot{x} = u$ is infinite as $\eps \rightarrow 0$ and for $\alpha > 0$ and a similar result for the simple switched system $\dot{x} = \sigma x$, where $\sigma$ switches between two positive reals, showing the necessity for bounded divergence between modes for the finiteness of entropy; 
(3) correcting the discrepancy function presented in \cite{sibai-mitra-2018} which was assuming that the largest eigenvalue of the symmetric part of the Jacobian of the dynamics with respect to the state is non-negative as well as assuming 2-norm while being applied as an $\infty$-norm bound;
(4) applying the bound on entropy of systems with uncertain inputs to obtain a bound on entropy of differential inclusions and autonomous switched systems, and thus relating the bounds in \cite{sibai-mitra-2017} and \cite{sibai-mitra-2018}. Finally, we 
 improve the presentation of both papers \cite{sibai-mitra-2017} and \cite{sibai-mitra-2018}. }

\subsection{Related Work}
\label{sec:relatedwork}

Several definitions of topological entropy for control systems have been proposed to bound the data rates necessary for control over limited-bandwidth communication channels \cite{nair-entropy-tac-ncs,nair-...-survey,colonius-kawan-09,colonius-kawan-nair-equivalence,katok-haselblatt} for tasks including stabilization \cite{nair-entropy-tac-ncs,yang2015stabilizing,Yang-Liberzon-2016,yang_liberzon_stabilization_switched_tac2018}, invariance \cite{Rungger-Zamani-2017,Tomar_Zamani_Networkinvarianceentropy_2020,Tomar_Zamani_numerical_2020}, and state estimation~\cite{sibai-mitra-2017,sibai-mitra-2018,LM:TAC2018,LM015:HSCC}, and types of systems including discrete-time~\cite{Rungger-Zamani-2017}, continuous-time \cite{colonius-kawan-09}, switched \cite{sibai-mitra-2017,Yang-Liberzon-2016,MSThesis:Schmidt,Guosong_Schmidt_Liberzon_Hespanha_2020,Yang_Schmidt_Liberzon_CDC2018,Yang_Hespanha_Liberzon_hscc2019,Vicinansa_Liberzon_regular_switched_entropy_2019}, and intercornnected~\cite{Tomar_Zamani_Networkinvarianceentropy_2020,liberzon_interconnected_entropy_2021}.
In \cite{LM015:HSCC, LM:TAC2018}, Liberzon and Mitra introduce the notion of
estimation entropy for autonomous nonlinear systems. They
define it in terms of the number of trajectories needed to
approximate all other trajectories starting from a compact initial
set up to an exponentially decaying error. They establish an
upper bound of $n(Lx+\alpha)$ on estimation entropy, where $n$ is $\ln 2$
the dimension of the system and $L_x$ is the Lipschitz constant of the vector field. In \cite{sibai-mitra-2017}, we extend the notion of estimation entropy to switched nonlinear systems, derive a corresponding upper bound, and construct a state estimation algorithm. In this paper, in Section~\ref{sec:switched_systems}, we relate that entropy definition for autonomous switched systems and its bounds with those of systems with inputs. We first presented the latter upper bound in \cite{sibai-mitra-2018}, and restate it here for completeness. We further show in this paper that an assumption we made to derive the upper bound on entropy of autonomous switched systems in~\cite{sibai-mitra-2017} is {\em necessary} by presenting a simple linear system that violates the assumption, and has an infinite entropy (Section~\ref{sec:finite_d}).

%% file: prelims.tex
\section{Preliminaries}
\label{sec:prelims}
We denote the infinity norm of a real vector $v \in \mathbb{R}^n$ by  $\|v\|$, \hussein{its 2-norm as $\|v\|_2$,} and its transpose by $v^\intercal$. We denote the \hussein{  infinity norm of a real matrix $A \in \mathbb{R}^{n\times n}$ by $\VERT A \VERT$, its  2-norm by $\VERT A \VERT_2$}, and its largest eigenvalue by $\lambda_{max}(A)$. If $A$ is symmetric positive definite, then $\lambda_{max}(A) = \VERT A \VERT_2$.

$B(v,\delta)$ is a $\delta$-ball--closed hypercube of radius $\delta$--centered at $v$. For a hyperrectangle $S \subseteq \mathbb{R}^n$ and $\delta > 0$,  $\mathit{grid}(S,\delta)$, is a collection of $2\delta$-separated points along axis parallel planes such that the $\delta$-balls around these points cover $S$. In that case, we say that the grid is of size $\delta$.  Given $x \in \mathbb{R}^n$ and $C = \mathit{grid}(S,\delta)$, we define $quantize(x,C)$ to be the nearest point to $x$ in $C$.
   
We denote by $[n_1;n_2]$, the set of integers in the interval $[n_1, n_2]$, inclusive, and by $[n_2]$ the set $[1;n_2]$. We denote the cardinality of a finite set $S$ by $|S|$. We denote the diameter of a compact set $S \subset \mathbb{R}^n$ by $\mathit{diam}(S) = \max_{x_1,x_2 \in S}\|x_1 - x_2\|$. For any set $S$ and positive integer $m$, $S^m$ is the $m$-way Cartesian product $S \times S \dots \times S$.  Given two sets $S_1$ and $S_2$, we define $S_1 \oplus S_2 := \{ x_1 + x_2\ |\ x_1 \in S_1, x_2 \in S_2 \}$.
A continuous function $\gamma: \nnreals \rightarrow \nnreals$ belongs to class-$\mathcal{K}$ if it is strictly increasing and $\gamma(0) = 0$. 

For all $t \in \nnreals$, we define the right and left hand limits of a function $u : \nnreals \rightarrow \reals^m$ at $t$ as follows:
\begin{align*}
	\usig(t^+) = \lim_{\tau \rightarrow t^+} \usig(\tau) \mbox{ and } \usig(t^-) = \lim_{\tau \rightarrow t^-} \usig(\tau).
\end{align*}
If $t$ is a point of discontinuity \hussein{and $u$ is piecewise-right-continuous}, we call it a {\em switch} of $u$, and define $\usig(t) = \usig(t^+)$.


In this paper, we will consider dynamical systems with \hussein{uncertain} input signals  \hussein{whose variations are bounded by an affine function of time}. \hussein{We consider two types of variation: pointwise and total.} Such signals are defined as follows.

\begin{definition}[\hussein{Signals with affine-bounded variation} \sayan{(removed: slowly-varying signals)}]
\label{def:inputset}
 Given $\mu \geq 0$, $\jum \geq 0$, and a compact set $U \subset \reals^m$, we define $\mathcal{U}^p(\mu, \jum)$, with the $p$-superscript referring to the word ``{\em pointwise}'', to be the set of all piecewise-\hussein{right-}continuous functions $\usig: \nnreals \rightarrow \reals^m$  with affine-bounded pointwise variation, i.e.,
\begin{align}
\label{variationbound}
u(0) \in U \text{ and } \|\usig(t + \tau) - \usig(t) \| \leq \mu \tau + \jum,
\end{align} 
for all $t$ and $\tau \geq 0$. \hussein{
We define $\mathcal{U}^s(\mu, \jum)$, with the $s$-superscript referring to the word ``{\em slow}'', to be the set of all piecewise-right-continuous functions $\usig: \nnreals \rightarrow \reals^m$ with affine-bounded total variation, i.e.,
	\begin{align}
		\label{def:totalvariationbound}
		u(0) \in U \text{ and } \int_{t}^{t+\tau}\| d\usig \| \leq \mu \tau + \jum,
	\end{align}  
for all $t$ and $\tau \geq 0$, where $ \int_{t}^{t+\tau}\| d\usig \|$ is the total variation of $\usig$ defined as follows:  $ \int_{t}^{t+\tau}\| d\usig \|  := $
\begin{align}
\label{def:totalvariationdistance}
&\int_{t}^{d_1} \|\dot{\usig}(s) \| ds + \int_{d_k}^{t+\tau} \|\dot{\usig}(s) \| ds + \sum_{i=1}^{k-1} \int_{d_i}^{d_{i+1}} \|  \dot{\usig}(s)  \| ds \nonumber \\
  & +\| \usig(t+\tau) - \usig((t+\tau)^{-}) \|  +  \sum_{i=1}^{k}  \| \usig(d_i) - \usig(d_i^-) \|,
\end{align}
where $d_1, d_2, \dots, d_k$ are the points of discontinuity in $u$ between $t$ and $t+\tau$.
}
\end{definition}
\hussein{Note that for any $\mu$ and $\jum \geq 0$, $\mathcal{U}^s(\mu, \jum) \subseteq \mathcal{U}^p(\mu, \jum)$. This can be shown as follows: fix any $\mu$, $\jum$, $t$,  and $\tau \geq 0$, and consider any $\usig \in \mathcal{U}^s(\mu, \jum)$, then  $ \|\usig(t + \tau) - \usig(t) \| = \|\int_{t}^{t+\tau} d \usig \| \leq \int_{t}^{t+\tau} \| d \usig \| \leq  \mu \tau + \jum$. Thus, $\usig \in  \mathcal{U}^p(\mu, \jum)$.
}

\sayan{removed: Definition~\ref{def:inputset} means that for any input signal $u \in \mathcal{U}^p(\mu, \jum)$ and any $t$ and $\tau \geq 0$, $\usig(t + \tau)$ should belong to the truncated $m$-dimensional cone with radius $\jum$ and rate of divergence $\mu$ (removed figure 1 that explains $\mathcal{U}^p$)}. We abbreviate $\mathcal{U}^p(\mu,\jum)$ and $\mathcal{U}^s(\mu,\jum)$  with $\mathcal{U}^p$ and $\mathcal{U}^s$ when $\mu$ and $\jum$ are clear from context.  
\hussein{The set $\mathcal{U}^p(\mu,\jum)$ contains functions that vary arbitrarily fast, while having a finite number of switches in any finite time interval, within an $\eta$ bound. 
}  $\mathcal{U}^p(\mu,0)$ is the class of globally Lipschitz continuous functions with Lipschitz constant $\mu$. In general, knowing that $u$ cannot \hussein{deviate} \sayan{(was: vary)} too much,
\sayan{(removed: i.e., having few points of discontinuity or small gradient,)} 
 can be expressed by setting $\mu$ and $\jum$ to smaller values. 
 Roughly, $\jum$ restricts the maximum norm of a jump \hussein{at any time instant} and $\mu$ restricts \hussein{the distance between the values of a signal at different time instances.} \sayan{(removed: the number of large jumps  in a short time interval. )}
 \hussein{However, it is more common in the control and dynamical systems literature to bound the total variation of functions instead of their deviations.  Accordingly, in addition to $\mathcal{U}^p$, we introduced $\mathcal{U}^s$ which bounds the total variation. Signals in $\mathcal{U}^s$ are upper-bounded in {\em how many} switches or jumps they have within any time interval as well as  the norm of their gradients. Thus, we call them {\em slowly-varying signals}. We present results for both cases in this paper. When deriving lower bounds on entropy, we assume that the input signals are in $\mathcal{U}^s$, and when deriving upper bounds, we assume they are in $\mathcal{U}^p$. Consequently, since $\mathcal{U}^s \subseteq \mathcal{U}^p$, all the results in the paper hold for both cases. }


The slow variation constraint \hussein{in $\mathcal{U}^s$} is \hussein{the same} \sayan{(removed: similar)} to that of \hussein{Theorem 2} \sayan{(was: Assumption 1)} of \cite{Gao-Liberzon-Basar-2015}  which was made on the variation of the system matrix of a time-varying linear dynamical system to relate its stability conditions to those of a switched linear dynamical system with slow switching. Also, it is similar to the slow switching assumption made by Hesphana and Morse in \cite{Hespanha-Morse-1999} to prove the stability of switched systems with stable subsystems. In this paper, we use the bound on the variation of signals in $\mathcal{U}^p$ to derive an upper bound on the entropy of dynamical systems with inputs \hussein{in $\mathcal{U}^p$ and $\mathcal{U}^s$, since $\mathcal{U}^s \subseteq \mathcal{U}^p$}. Further, we use it to relate the upper bound on entropy of autonomous switched systems with slow switching (having minimum dwell-time constraints) with that of dynamical systems with inputs with affine-bounded pointwise variation in Sections~\ref{sec:si_and_ss_upper_bound_relation} and \ref{sec:si_abstract_ss}. 

%% file: Entropydefinition.tex
\section{Entropy for open dynamical systems}
\label{gen_inst}
We consider a dynamical system of the form:
\begin{align} 
\dot{x}(t) = f(x(t), \usig(t)), \label{sys:noise}
\end{align}
where $t \geq 0$ and $f: \mathbb{R}^n \times \mathbb{R}^m \rightarrow \mathbb{R}^n$. The function $f$ is \hussein{locally} Lipschitz, and has piecewise-continuous Jacobian matrices $J_x(x,u) = \frac{\partial f}{\partial x}(x,u)$ and $J_u(x,u) = \frac{\partial f}{\partial u}(x,u)$, with respect to the first and second arguments, respectively. When $f$ is globally Lipschitz, we denote its global Lipschitz constants by $L_{x}$ and $L_{u}$.

For any initial state in 
and measurable input signal 
\hussein{we assume that} the solution of system (\ref{sys:noise}) exists for any $t \geq 0$, is unique, and depends continuously on the initial state~\cite{khalil-book-3ed}. We denote this solution, or {\em trajectory}, by $\xi_{x_0,u} : \nnreals \rightarrow \R^n$. In the rest of this section, we fix a compact set of initial states $K \subset \mathbb{R}^n$ and the set of input functions $ \mathcal{U}^p(\mu, \jum)$. \hussein{ The same results hold if the set of input signals was $\mathcal{U}^s(\mu, \jum)$ instead.}

The reachable set of states of system~(\ref{sys:noise}) starting from $K$, having input signals from $\inputset$, and running for a time horizon $T$, is defined as follows:
\begin{align} 
	\reach(K,\inputset, T) := \{&x \in \reals^n\ |\ \exists x_0 \in K, u \in \mathcal{U}, t\in[0,T], \nonumber\\
	&\xi_{x_0,u}(t) = x\}.
\end{align}

\begin{example}[Dubin's vehicle]
	\label{eg:car_trans}
	\normalfont Consider a car moving at a constant speed $v$. We describe its dynamics as follows:
	\begin{align}
		\label{eq:dubin}
		\dot{x}_1 = v \cos x_3, \dot{x}_2 = v \sin x_3, \dot{x}_3 = u. 
	\end{align}
	where $(x_1,x_2)$ is the position of the car and $x_3$ is its heading angle. $u$ is a control signal defining the steering velocity. The Jacobians of the car dynamics with respect to $x$ and $u$ are as follows:
	\begin{align}
	J_x = 
	\begin{bmatrix}
	0 &0 &-v\sin x_3 \\
	0 &0 &v \cos x_3 \\
	0 &0 &0
	\end{bmatrix} 
	\text{ and }
	J_u = 
\begin{bmatrix}
	0\\
	0 \\
	1
\end{bmatrix}. 
	\end{align}
Therefore, the dynamics are globally Lipschitz with Lipschitz constants:
\begin{align}
L_x = \VERT J_x \VERT \leq v \text{ and } L_u = \VERT J_u \VERT = 1. 
\end{align}
\end{example}

\subsection{Approximating functions and $(\eps, \alpha)$-estimation entropy}

Let us fix the error parameters $\eps > 0$ and $\alpha \geq 0$. Given a $T > 0$, $x_0 \in K$ and $\usig \in \mathcal{U}^p(\mu, \jum)$, we say that a function $z: [0,T] \rightarrow \mathbb{R}^n$ is $(T,\eps,\alpha)$-{\em approximating} for the trajectory $\xi_{x_0,u}$ over the interval $[0,T]$, if
\begin{align}
\label{eq:approx_traj_eps}
\| z(t) -  \xi_{x_0,u}(t) \| \leq \eps e^{-\alpha t},
\end{align} 
for all $t \in [0,T]$.
We say that a set of functions $\mathcal{Z} := \{ z\ |\ z : [0,T] \rightarrow \R^n \}$ is $(T,\eps, \alpha)$-{\em approximating} for system (\ref{sys:noise}), if for every $x_0 \in K$ and $\usig \in \mathcal{U}^p$, there exists a $(T,\eps,\alpha)$-approximating function $z \in \mathcal{Z}$ for the trajectory $\xi_{x_0,u}$ over $[0,T]$. \hussein{In this paper, we mostly follow the same notation used by Liberzon and Mitra in \cite{LM:TAC2018} in their definition of estimation entropy for autonomous dynamical systems.}  We denote the minimal cardinality of such an approximating set by $s\est(T,\eps,\alpha)$. \hussein{The subscript $\mathit{est}$ corresponds to the word {\em estimation} and the letter $s$ stands for the word {\em set}.}

\begin{definition}
\label{def:entropy}
The $(\eps,\alpha)$-{\em estimation entropy} of system (\ref{sys:noise}) is defined as follows:
\begin{align}
\label{eq:entropy_def}
h\est(\eps, \alpha) := \limsup_{T\rightarrow \infty}\frac{1}{T} \log s\est(T,\eps,\alpha).
\end{align}
\end{definition}
The estimation entropy $h\est$ represents the exponential growth rate over time of the number of distinguishable trajectories of the system. Hence, $h\est$ also represents the bit rate need to be sent by the sensor so that the estimator can construct a ``good" estimate of the state. 

\hussein{
\begin{remark}
The value of $(\eps, \alpha)$-estimation entropy has the same value if the $\infty$-norm in inequality~(\ref{eq:approx_traj_eps}) is replaced by any norm. This follows from the equivalence of norms in finite-dimensional spaces. More specifically, since for any two norms $\|\cdot \|_a$ and $\|.\|_b$, for all $x \in \reals^n$, there exists two positive constants $c_1$ and $c_2$ such that $c_1 \|x\|_b \leq \|x\|_a \leq c_2 \|x\|_b$, $s\est$ will only be multiplied by a constant factor under change of norms. That factor will be eliminated when divided by $T$, as $T$ goes to infinity in (\ref{eq:entropy_def}).
\end{remark}
}

\hussein{
	The following remark relates the above notion of approximating sets to the sample complexity of {\em reachable sets\/}. The latter is not a central theme in the current paper but is an important concept in control and verification. 
}
\begin{remark}
	Fix a time horizon $T > 0$. Assume that there is a method to generate a $(T,\eps,\alpha)$-approximating function for any given trajectory of system~(\ref{sys:noise}).
	 Then, $s\est(T,\eps,\alpha)$ is also the minimum number of simulations needed to {\em approximate} the reachset $\reach(K,\inputset, T)$ of system~(\ref{sys:noise}) up to an $\eps e^{-\alpha t}$ over-approximation error, for any $t \in [0,T]$.
Formally, let $\Z$ be the $(T,\eps,\alpha)$-approximating set with cardinality $s\est(T,\eps,\alpha)$ and define an over-approximation of the reachable set
	\begin{align*}
		&\reach_{\eps,\alpha}(K,\inputset, T) := \{x + c\ |\ \exists x_0 \in K, u \in \mathcal{U}, t\in[0,T], \nonumber\\
		&\hspace{1.3in}  c \in B_{\eps e^{-\alpha t}}, \xi_{x_0,u}(t) = x\} 
	\end{align*}
and another over-approximation using $\Z$
\begin{align*}
&R_{\eps,\alpha} := \{z(t) + c\ |\ z \in \Z, t \in [0,T],  c \in B_{\eps e^{-\alpha t}}\},
\end{align*}
	where $B_{\eps e^{-\alpha t}} := B(0, \eps e^{-\alpha t} )$. 
	Then, \[\reach(K,\inputset, T) \subseteq R_{\eps,\alpha} \subseteq \reach_{2\eps,\alpha} (K,\inputset, T).\] 
\end{remark}

\section{Alternative entropy notions}
\label{sec:feasible_entropy_notions}

In this section, we show that two alternative definitions of entropy are comparable in the sense that a bound on one of them results in bounds on the others\footnote{ \hussein{This section is a novel contribution of this paper.}}.

\subsection{Approximating trajectories instead of functions}
\label{sec:spanning_set}
In this section, we modify the definition of entropy $h\est$ by restricting the approximating functions to be trajectories of system~(\ref{sys:noise}). Following previous notation (e.g. \cite{LM015:HSCC, LM:TAC2018}), we call the resulting restricted approximating sets {\em spanning} sets. Formally, we say that a set $\Z^* \subseteq K \times \inputset$ is $(T,\eps, \alpha)$-{\em spanning}, if for every $x_1 \in K$ and $\usig_1 \in \inputset$, there exists a pair $(x_2,u_2) \in \Z^*$ such that the trajectory $\xi_{x_2,\usig_2}$ is a $(T,\eps,\alpha)$-approximating function for the trajectory $\xi_{x_1,u_1}$ over $[0,T]$. The minimum cardinality of a spanning set is denoted by $s\est^*(T,\eps,\alpha)$. If we propagate this restriction to Definition~\ref{def:entropy}, the resulting definition of entropy would be as follows:
\begin{definition}
	\label{def:traj_entropy}
	The spanning sets-based $(\eps,\alpha)$-{\em estimation entropy} of system (\ref{sys:noise}) is defined as follows:
	\begin{align}
	h\est^{*}(\eps,\alpha) := \limsup_{T\rightarrow \infty}\frac{1}{T} \log s\est^*(T,\eps,\alpha).
	\end{align}
\end{definition}

The following inequality is an application of the fact that any $(T,\eps,\alpha)$-spanning set is a $(T,\eps,\alpha)$-approximating set: 
\begin{align}
\label{eq:app_less_spanning}
	s\est(T,\eps,\alpha) \leq s\est^*(T,\eps,\alpha).
\end{align}



\subsection{Separated sets instead of approximating ones}

In this section, we provide an entropy definition based on the concepts of $(T,\eps,\alpha)$-{\em separated} trajectories and $(T, \eps,\alpha)$-{\em separated} sets.

Fix any error parameters $\eps > 0$ and $\alpha \geq 0$. Given $T \geq 0$, two initial states $x_1$ and $x_2$ and two input signals $\usig_1$ and $\usig_2$, the two trajectories $\xi_{x_1,\usig_1}$ and $\xi_{x_2,\usig_2}$ are $(T,\eps,\alpha)$-{\em separated} iff there exists $t \in [0,T]$ such that
\begin{equation}
\label{eq:separated_traj}
\| \xi_{x_1,\usig_1}(t) - \xi_{x_2,\usig_2}(t) \| > \eps e^{-\alpha t}.
\end{equation}

A set of pairs of initial states and input signals  $\Zsep \subseteq K \times \mathcal{U}^p$ is called a $(T,\eps, \alpha)$-{\em separated} set if the trajectories corresponding to any two pairs in $\Zsep$ are $(T,\eps,\alpha)$-separated.
We denote the largest cardinality of a $(T,\eps,\alpha)$-separated set by $s\sep\est(T,\eps,\alpha)$.

\begin{definition}
\label{def:separated_entropy}
The separated sets-based $(\eps,\alpha)$-{\em estimation entropy} of system (\ref{sys:noise}) is defined as follows:
\begin{align}
h\est\sep(\eps,\alpha) = \limsup_{T\rightarrow \infty}\frac{1}{T} \log s\est\sep(T,\eps,\alpha).
\end{align}
\end{definition}

The following two lemmas are analogous to Lemmas 1 and 2 in \cite{LM015:HSCC}. They draw the relation between approximating, spanning, and separated sets.

\begin{Lemma}
	\label{lm:s_eps_less_n_eps}
	For all $K$, $\mathcal{U}^p$,  $\eps$, $\alpha$, and $T$,
	\begin{align}
	s\est^*(T,\eps,\alpha) \leq s\est\sep(T,\eps,\alpha).
	\end{align}
\end{Lemma}
\begin{proof}
	The lemma follows from the observation that every maximal $(T,\eps,\alpha)$-separated set $\Zsep$ is also $(T,\eps,\alpha)$-\hussein{spanning} \sayan{(was: approximating)} set. In fact, if there is a pair of initial state $x_1 \in K$ and input signal $u_1 \in \mathcal{U}^p$ such that there is no pair $(x_2,u_2) \in \Zsep$ where $\xi_{x_2,u_2}$ is $(T,\eps,\alpha)$-approximating  for  $\xi_{x_1,u_1}$, then $(x_1,u_1)$ and any pair in $\Zsep$ violate (\ref{eq:approx_traj_eps}) for some $t \in [0,T]$. Therefore, $\xi_{x_1,u_1}$ can be added to $\Zsep$, which contradicts its maximality.
\end{proof}

\begin{Lemma}
	\label{lm:n_less_s}
	For all $K$, $\mathcal{U}^p$,  $\eps$, $\alpha$, and $T$,
	\begin{align}
	s\sep\est(T,2\eps,\alpha) \leq s\est(T,\eps,\alpha).
	\end{align}	
\end{Lemma}
\begin{proof}
	Consider an arbitrary $(T,\eps,\alpha)$-approximating set $\Z$ and an arbitrary $(T,2\eps,\alpha)$-separated set $\Zsep$. Assume, for the sake of contradiction, the cardinality of $\Zsep$ is larger than the cardinality of $\Z$. Then, since $\Z$ is a $(T,\eps,\alpha)$-approximating set, there exist two pairs $(x_1,u_1)$ and $(x_2,u_2) \in \Zsep$ such that (\ref{eq:approx_traj_eps}) is satisfied with the same $z \in \Z$. Hence, 
	\begin{align*}
	&\|\xi_{x_1,u_1}(t) - \xi_{x_2,u_2}(t) \| \\
	&\leq \|z(t) - \xi_{x_1,u_1}(t)\| + \|z(t) - \xi_{x_2,u_2}(t)\| \\
	&\hspace{0.5in} \mbox{[by triangular inequality]}\\
	&\leq \eps + \eps = 2\eps.
	\end{align*}
	Thus, the two trajectories are not $(T,2\eps,\alpha)$-separated, contradicting the assumption that $\Zsep$ is a $(T,2\eps,\alpha)$-separated set.
\end{proof}

The following theorem shows the relation between all of the introduced entropy definitions so far.

\begin{theorem}
	\label{thm:hest_bounded_hsep}
For all $K$, $\mathcal{U}^p$,  $\eps$, $\alpha$, and $T$, 
\begin{align}
h\est\sep(2\eps,\alpha) \leq h\est(\eps,\alpha) \leq h\est^*(\eps,\alpha) \leq h\est\sep(\eps,\alpha).
\end{align}
\end{theorem}
\begin{proof}
The first inequality follows from Lemma~\ref{lm:n_less_s}, the second one follows from (\ref{eq:app_less_spanning}), and the third one from Lemma~\ref{lm:s_eps_less_n_eps}.
\end{proof}

%% file: EpsilonZero.tex
\section{Infeasible alternative Entropy notions}
\label{sec:entro_motivation}
In this section, we show that the $(\eps,\alpha)$-estimation entropy $h\est(\eps,\alpha)$ of a simple dynamical system is infinite if $\eps$ approaches zero or if $\alpha > 0$. 
On the other hand, we show in Corollary~\ref{cor:simple_sys_entropy_finite} that  $h\est(\eps, 0)$ of that system is finite for any fixed $\eps > 0$. 
We conclude that it is not meaningful to take the supremum over $\eps > 0$ or choose an $\alpha > 0$, in contrast with the definition of estimation entropy of autonomous systems \cite{LM015:HSCC}\footnote{\hussein{The results presented in this section are novel contributions of this paper.}}.

Consider the simple dynamical system:
\begin{align}
\label{sys:simple}
\dot{x}(t) = u(t),
\end{align}
where $t \geq 0$, the initial state $x(0)$ is fixed to any real number, and the input signal $u$ will be chosen from sets that we will construct for each variant of entropy definition.

It is worth noting that system~(\ref{sys:simple}) has been shown earlier to not being exponentially stabilized by any finite number of control signals, and thus having an infinite stabilization entropy \cite{colonius-siam-2012}. A modification of the stabilization property to include a constant stabilization error, similar to our constant estimation error, is made in \cite{colonius-siam-2012} to ensure finiteness of entropy.
%

The sets of input signals that we use in this section share the same structure described in the following definition.

\begin{figure}[h]
	\centering
	\includegraphics[scale=0.3]{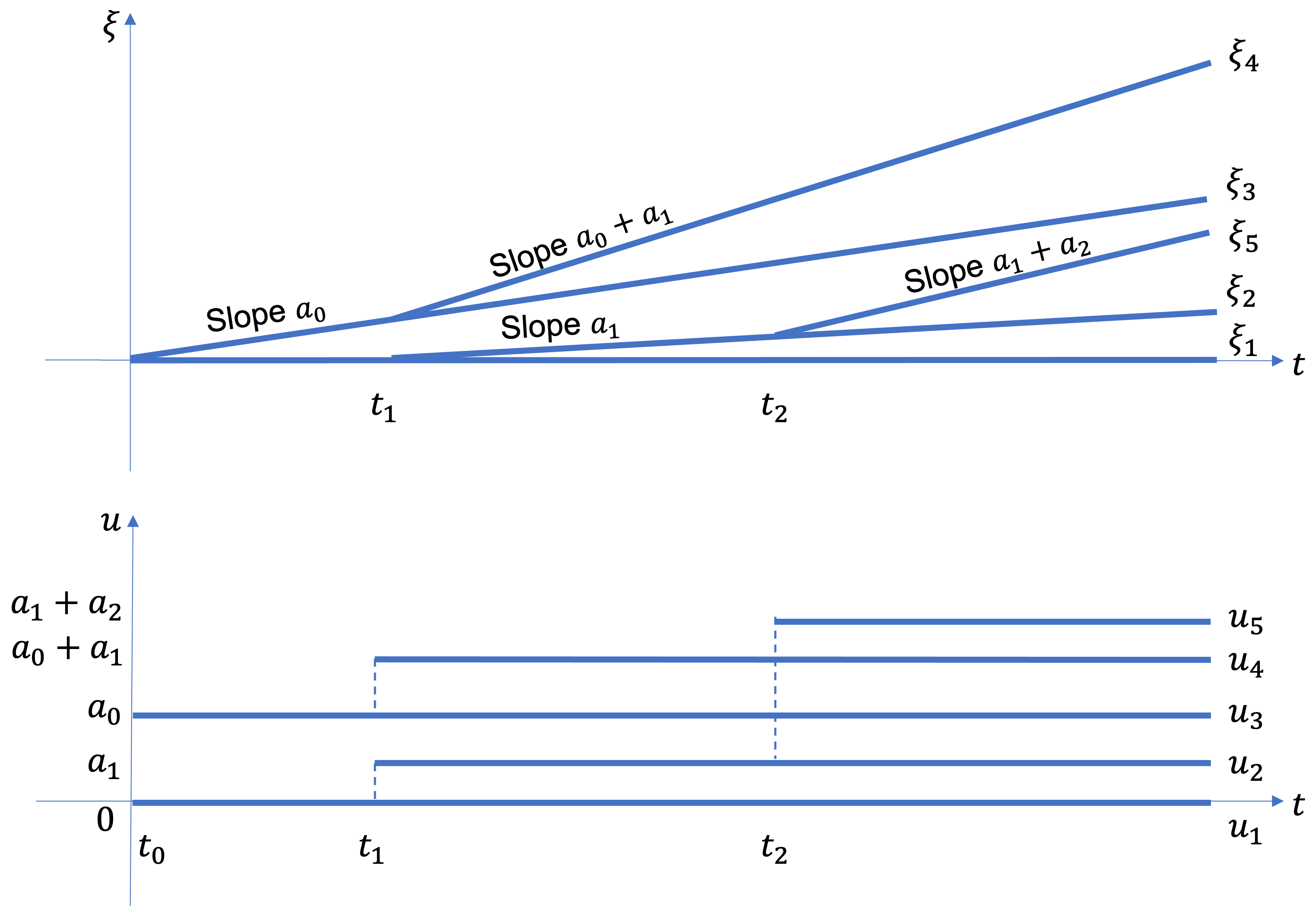}
	\caption{
	\sayan{(was Construction of the set of input signals in Definition~\ref{def:pwconstantab}. Only four such signals are shown: $u_1$ that is equal to $a$ over $[t_0,t_5]$, $u_2$ that differs from $u_1$ by being equal to $b$ over $[t_3, t_4]$, $u_3$ that differs from $u_1$ by being equal to $b$ over $[t_1,t_2]$, and $u_4$ that differs from $u_3$ by being equal to $b$ over  $[t_3,t_4]$  (bottom). Corresponding trajectories of system~(\ref{sys:simple}) (top).)}
	\hussein{ Construction of an example set of input signals following Definition~\ref{def:pwconstantab}. Only five such signals are shown: $u_1$ that is equal to $0$ at all times, $u_2$ that differs from $u_1$ by being equal to $a_1$ starting from $t_1$, $u_3$ that is equal to $a_0$ at all times,  $u_4$  that differs from $u_3$ by being equal to $a_0 + a_1$ starting from $t_1$, and $u_5$ that differs from $u_2$ by being equal to $a_1 + a_2$ starting from  $[t_2]$  (bottom). Corresponding trajectories of system~(\ref{sys:simple}) (top).} \label{fig:separationset_constraint} \vspace{-0.25in} }
\end{figure}

\begin{definition}
	\label{def:pwconstantab}
	 \sayan{(removed: Fix any two constants $a$ and $b \in \reals$ such that $a > b$ and let $W = \{a,b\}$.)} Given \hussein{a sequence of $l$ real numbers $A = a_0, \dots, a_{l-1}$ and } an increasing sequence of $l+1$ time points $\tseq = t_0,\ldots,t_l$, \sayan{(removed: and a string $se \in A^l$,)} 
	\hussein{we define the set $\mathit{SE}$ of sequences (SE for SEquences) of length $l$ that are of the form: $\mathit{se}[0] = 0$ or $\mathit{se}[0] = a_0$ and for any $i >0$, $\mathit{se}[i] = \mathit{se}[i-1]$ or $\mathit{se}[i] = \mathit{se}[i-1] + a_i$. Following such a construction, we define}  a piecewise-constant signal $u : [t_0,t_l) \rightarrow \reals$ \hussein{generated by a sequence $\mathit{se}$} as follows: 
	for all $i <l$ and $t \in [t_i,t_{i+1})$,
	\begin{align}
	u_{\tseq,se}(t) = \mathit{se}[i].
	\end{align}  
	We denote the set of all such piecewise-right-constant signals, with $\tseq$ and $A$ fixed, by \hussein{$\mathcal{U}_{\tseq}^{A} = \underset{se \in \mathit{SE}}{\bigcup} u_{\tseq,se}$.} \sayan{(was  $\mathcal{U}_{\tseq} = \underset{se \in W^l}{\bigcup} u_{\tseq,se}$).}
	%
\end{definition}
Observe that the cardinality of \hussein{$\mathcal{U}_{\tseq}^{A}$} is $2^l$. Moreover, 
 \hussein{each of $\mathcal{U}^p(0,  \sum_{i = 0}^{l-1} a_i)$ and $\mathcal{U}^s(0,\sum_{i = 0}^{l-1} a_i)$} \sayan{(was $\mathcal{U}^p(0, a - b)$)} contains \hussein{$\mathcal{U}_{\tseq}^{A}$}. Example signals from $\inputset_{\tseq}^{A}$ with the corresponding trajectories of system~(\ref{sys:simple}) are shown in Figure~\ref{fig:separationset_constraint}.


%


	\vspace{-0.15in}
\subsection{Adding $\lim_{\eps \rightarrow 0}$ to entropy definition}
\label{sec:epsilonzero}

In this section, we will discuss the first variant of entropy definition: taking the limit as $\eps$ goes to zero. Such a variant is considered in \cite{LM:TAC2018} and represents the case where the allowed estimation error $\eps$ is arbitrarily small. We will show that the resulting notion of entropy is infinite for system~(\ref{sys:simple}) with inputs constructed using Definition~\ref{def:pwconstantab}.

Using Theorem~\ref{thm:hest_bounded_hsep}, we get the following inequality:
\begin{align}
\label{eq:h_less_n}
\lim_{\eps \rightarrow 0} h\est(\eps,\alpha)  &\geq  \lim_{\eps \rightarrow 0} h\est\sep(2\eps,\alpha) \nonumber \\
&= \lim_{\eps \rightarrow 0} \limsup_{T\rightarrow \infty}\frac{1}{T} \log s\sep\est(T,2\eps,\alpha).
\end{align}

Hence, it is sufficient to show that the RHS of (\ref{eq:h_less_n}) is infinite. 
 We will construct a $(T,2\eps,\alpha)$-separated set of size \hussein{$O(2^{\nicefrac{T}{3\sqrt{\eps}}})$} \sayan{(was  $O(2^{\nicefrac{T}{3\eps}})$)} which when substituted in (\ref{eq:h_less_n}) would evaluate to infinity. 

To construct the separated set, \hussein{for any fixed $\eps$ and $T$,} we use the family of input signals \hussein{$\mathcal{U}_{\tseq}^{A}$, where $A = \sqrt{\eps}, \dots, \sqrt{\eps}$, } 
$\tseq_{\zeps} = \{0, \tau, 2\tau, \dots \floor{T/\tau}\tau \}$, and $\tau$
\hussein{$=3 \sqrt{\eps}$}.
\hussein{By such choices of $A$ and $\tau$, the total variation of any signal in $\mathcal{U}_{\tseq}^{A}$ is upper-bounded by $\sqrt{\eps} \floor{T/\tau}$ which is equal to $\sqrt{\eps} \floor{T/(3\sqrt{\eps})} = O(T)$.} \hussein{Therefore, there exist some constants $\mu$ and $\eta$, such that for any $\eps > 0$ and $T > 0$, $\mathcal{U}_{\tseq}^{A}$ is a subset of $\mathcal{U}^s(\mu, \eta)$, and consequently,  a subset of $\mathcal{U}^p(\mu, \eta)$ as well.}
	 


\begin{Lemma}
\label{lm:sep_construct_eps_0}
Fix $T$ and $\eps > 0$. Consider any initial state $x_0 \in \mathbb{R}$ and the set of input signals $\mathcal{U}_{\tseq}^A$.
The resulting trajectories of system~(\ref{sys:simple}) form a $(T,2\eps,0)$-separated set.
\end{Lemma}

\begin{proof}
Consider any two signals $u_1, u_2 \in \mathcal{U}_{\tseq}$ with corresponding sequences $\mathit{se}_1$ and $\mathit{se}_2$ with the same prefix up till the $j^{\mathit{th}}$ value for some $j \in [\floor{T/\tau}-1]$, and with different values in the $(j+1)^{\mathit{th}}$ index. That is $se_1[j+1]\neq se_2[j+1]$. The resulting trajectories $\xi_{x_0,u_1}$ and $\xi_{x_0,u_2}$ from these input signals and the common initial state $x_0$ are $(T,2\eps,0)$-separated.
To see this, first note that the solution of the system is: $\xi_{x,u}(t) = \hussein{\xi_{x,u}(j\tau)} + u(j\tau)(t-j\tau)$, for $t \in [j\tau, (j+1) \tau)$, since $u(t)$ is constant over that interval. Then,
  the distance 
\hussein{$\|\xi_{\xi_{x,u_1}(j\tau),se_1[j+1]}(t - j\tau) - \xi_{\xi_{x,u_2}(j\tau),se_2[j+1]}(t - j\tau)\|$}
between the two trajectories that start from $\xi_{x,u_1}(j\tau)$, or equivalently, $\xi_{x,u_2}(j\tau)$, and have $se_1[j+1]$ and $se_2[j+1]$, respectively, as input for $\tau$ time units is \hussein{$a_{j+1}\tau$, which is equal to $\sqrt{\eps}\tau$}. By our choice of $\tau$ \hussein{ to be equal to $3 \sqrt{\eps}$}, the distance is $3\eps$. Hence, the distance $\|\xi_{x_0,u_1}((j+1)\tau) - \xi_{x_0,u_2}((j+1)\tau)\|$ between $\xi_{x_0,u_1}$ and $\xi_{x_0,u_2}$ at $t = (j+1)\tau$ is equal to $3\eps$, strictly larger than $2\eps$. 
Since it is enough for two trajectories to have a distance larger than $2\eps$ at a single point in time in the interval $[0,T]$ to be considered $(T,2\eps,0)$-separated, $\xi_1$ and $\xi_2$ are $(T,2\eps,0)$-separated. Hence, all the trajectories resulting from the initial state $x_0$ and $\mathcal{U}_{\tseq}$ are $(T,2\eps,0)$-separated. 
\end{proof}

\begin{theorem}
	\label{thm:infiniteentropy_zeroepsilon}
$\lim_{\eps \rightarrow 0} h\est(\eps,\alpha)$ of system (\ref{sys:simple}) with the input set $\inputset_{\tseq}^A$ is infinite. 
\end{theorem}
\begin{proof}
 
 The number of trajectories in the separated set constructed in Lemma~\ref{lm:sep_construct_eps_0} is equal to $|\mathcal{U}_{\tseq}^A|$ which is $2^{\floor{\nicefrac{T}{\tau}}}$. 
 Moreover, observe that any $(T,2\eps,0)$-separated set is also a $(T,2\eps,\alpha)$-separated set, for any $\alpha >0$. Hence, for any $\alpha > 0$, $s\sep\est(T,2\eps,\alpha) \geq s\sep\est(T,2\eps,0)$. 
 Therefore, $\lim_{\eps \rightarrow 0} h\est(\eps,\alpha)$ 
  is infinite by a simple substitution of $s_{\est}\sep(T,2\eps,\alpha)$ with $2^{|\tseq_\zeps|} =$ \hussein{$2^{\floor{\nicefrac{T}{3\sqrt{\eps}}}}$} \sayan{(was $2^{\floor{\nicefrac{T(a-b)}{3\eps}}}$)}, the size of the $(T,2\eps,0)$-separated set we constructed in the previous lemma, in inequality (\ref{eq:h_less_n}).
 
  
\end{proof}

It follows that taking the supremum over $\eps > 0$ would likely result in an infinite entropy for most systems with inputs, while it might be finite for any fixed $\eps > 0$. Accordingly, we parameterize $h\est$ with $\eps$ as in Definition \ref{def:entropy}. \hussein{ It is worth noting that Matveev and Savkin in Theorem 2.3.17 in~\cite{Matveev2009}  presented a sufficient condition for the topological entropy, with $\sup_{\eps> 0}$ in its definition, of a discrete-time system with uncertain inputs to be infinite. Their sufficient condition is a function of the trajectories of the system.
	In fact, their result can be generalized to our setting of continuous-time systems of the form of system~(\ref{sys:noise}) in the following lemma. Before stating it, it is worth mentioning that there is no bound on the variation of the input signals in this lemma. }
\begin{Lemma}
\label{lm:generalization_matveev_2009}
\hussein{Fix a set of input signals $\mathcal{U}$
	and any trajectory $\xi_{x, u}$ of system~(\ref{sys:noise}) with an arbitrary $x \in \reals^n$ and $u \in \mathcal{U}$. Assume that  $\exists \tau > 0$ and a vector $a_0 \neq 0$ in $\reals^n$, such that for any $c_1, c_2,  \dots$ in the interval $[0,1]$, there exists a trajectory $\xi_{x',u'}$, that for any positive integer $j$, is equal to $\xi_{x,u}(j \tau) + c_j a_0$. Then, $\lim_{\eps \rightarrow 0} h\est(\eps,\alpha)$ of system (\ref{sys:noise}) is infinite. }
\end{Lemma}
\begin{proof}
	\hussein{The proof is almost the same proof as that of Theorem 2.3.17 in \cite{Matveev2009}. Given the assumptions in the lemma, we can construct a $(l\tau, \frac{\|a_0\|}{M}, 0)$-separated set, for any $M$ and $l \geq 1$, as follows: First, choose $c_i = \frac{i-1}{M}$, for all $i \in [M]$. Then, consider the trajectories of system~(\ref{sys:noise}) which satisfy $\xi_{x,u'}(j\tau) = \xi_{x,u}(j \tau) + c_i a_0$, for $i \in [M]$ and $j \in [l]$. Such trajectories exist by the second assumption in the lemma. Each trajectory in this set is $(l\tau, \frac{\|a_0\|}{M}, 0)$-separated from the other trajectories in the set. The cardinality of the set is equal to $M^l$. Taking the limit of $\eps$ going to zero is equivalent to taking the limit of $M$ going to infinity. By substituting $s_{\est}\sep(T,2\eps,\alpha)$ by $M^l$ and $T$ by $l\tau$ in the RHS of inequality~(\ref{eq:h_less_n}) results in it being infinite. }
\end{proof}
\hussein{The assumption in Lemma~\ref{lm:generalization_matveev_2009} would be satisfied if the system is {\em locally reachable} along a trajectory that is {\em separated from the origin}, as shown in Theorem 2.3.17 in \cite{Matveev2009}. Extending Definition 2.3.13 in \cite{Matveev2009} to continuous-time systems, system~(\ref{sys:noise}) is said to be {\em locally reachable} along a trajectory $\xi_{x,u}$ if two positive constants $\delta$ and $\tau$ exist such that for any positive $t$  and any $a$ and $b \in \reals^n$ such that 
$\|\xi_{x,u}(t) - a\| \leq \delta \|\xi_{x,u}(t)\|$ and $\|\xi_{x,u}(t + \tau) - b\| \leq \delta \|\xi_{x,u}(t + \tau)\|$,
another trajectory $\xi_{x',u'}$ of system~(\ref{sys:noise}) exists and satisfies 
$\xi_{x',u'}(t) = a$ and  $\xi_{x',u'}(t + \tau) = b$. Extending Definition 2.3.14 in \cite{Matveev2009} to continuous time systems, a trajectory $\xi_{x,u}$ of system~\ref{sys:noise} is said to be {\em separated from the origin} if there exists a positive constant $\delta_0$ such that $\|\xi_{x,u}(t)\| \geq \delta_0$, for all $t \geq 0$.
}
\subsection{Requiring exponentially convergent estimation error }
\label{sec:expconvergence}
In this section, we prove that $h\est(\eps,\alpha)$ of system~(\ref{sys:simple}) with a particular input set is infinite for any strictly positive $\alpha$.  \hussein{We present two alternative proofs: the first is a corollary of Theorem~\ref{thm:infiniteentropy_zeroepsilon} and the second is done by constructing a new separated set.} \hussein{We start with the first and easier result, the corollary of Theorem~\ref{thm:infiniteentropy_zeroepsilon}. This corollary is suggested by one of the anonymous reviewers of the paper. }
\begin{corollary}
\hussein{For any $\eps$ and $\alpha > 0$, $h\est(\eps,\alpha)$ of system (\ref{sys:simple}) with the input set $\cup_{\eps > 0} \inputset_{\tseq, \eps}^A$ is infinite, where $ \inputset_{\tseq, \eps}^A$ is the input set constructed for a specific $\epsilon$ in Lemma~\ref{lm:sep_construct_eps_0}. }
\end{corollary}
\begin{proof}
\hussein{Recall from (\ref{eq:entropy_def}) that $h\est(\eps,\alpha; K)  = \limsup_{T\rightarrow \infty}\frac{1}{T} \log s\est(T,\eps,\alpha; K)$. Thus, for any finite time bound $\tau$, $ h\est(\eps,\alpha; K)$ is greater than or equal to:
	\begin{align}
		\label{cor_proof:decomposing_approximating_sets}
		& \limsup_{T\rightarrow \infty}\frac{1}{T} \log \big(s\est(\tau,\eps,\alpha; K)  \nonumber\\ 
		&\hspace{0.3in} s\est(T - \tau,\eps e^{-\alpha \tau},\alpha; \reach(K, \inputset, [\tau,\tau]))\big) \nonumber 
		\end{align}
	\begin{align}
		& =  \limsup_{T\rightarrow \infty}\frac{1}{T} \log s\est(T - \tau,\eps e^{-\alpha \tau},\alpha; \reach(K, \inputset, [\tau,\tau])) \nonumber \\
		& = h\est(\eps e^{-\alpha \tau},\alpha; \reach(K, \inputset, [\tau,\tau])),
	\end{align} 
where $\reach(K, \inputset, [\tau,\tau])$ represents the reachable states {\em exactly} at time $\tau$.  The inequality follows from the fact that to construct a $(T,\eps,\alpha; K)$-approximating set, one can construct a $(\tau,\eps,\alpha; K)$-approximating set and a $(T - \tau,\eps e^{-\alpha \tau},\alpha; \reach(K, \inputset, [\tau,\tau]))$-approximating one. Then, for each function in the former set, concatenate it with a function in the latter set.	}

\hussein{Since $\lim_{\eps \rightarrow 0} h\est(\eps, \alpha; K)$ of system (\ref{sys:simple}) is infinity for any $K$, $\lim_{\tau \rightarrow \infty} h\est(\eps e^{-\alpha \tau},\alpha; \reach(K, \inputset, [\tau,\tau]))$ is infinity as well. Therefore, $h\est(\eps, \alpha; K)$ is infinite. }
\end{proof}
Our \hussein{second} proof relies on designing a separated set, as in the previous section. Since we require the error to exponentially decrease with rate $\alpha > 0$ in this section, we can decrease the time intervals  between switches in the input signals of Definition~\ref{def:pwconstantab} as well as their values while still ensuring that the resulting set of trajectories is $(T,2\eps,\alpha)$-separated. \hussein{That way, we can pack more separated trajectories in the same time interval while satisfying the slow-variation constraint.}

\hussein{Specifically, we use the set of input signals $\mathcal{U}_{\palpha}$ that results from Definition~\ref{def:pwconstantab} with the sequence of time instants $\tseq_\palpha$ being of the form $t_0=0$ and $t_i = \sum_{j=1}^{i}v_j$, where $i \leq l$, $l$ is the largest index where $v_l$ is less than $T$, $v_1 = \sqrt{2\eps}$, and $v_{i+1} = v_i e^{-\alpha v_i/2}$ (or equivalently $v_{i+1} = v_1 e^{- \frac{\alpha}{2} \sum_{j=1}^{i}v_j}$). Observe that for all $i$, $v_{i+1} < v_{i}$, and $v_i$ is the time interval between $t_{i-1}$ and $t_{i}$. We choose the sequence $A$ in the definition of $\mathcal{U}_{\palpha}$  to be of the form $v_1, \dots, v_l$, which are the same values used to define the switches above.}

\sayan{was (Specifically, we use the set of input signals $\mathcal{U}_{\palpha}$ that results from Definition~\ref{def:pwconstantab} with the sequence of time instants $\tseq_\palpha$ being of the form $t_0=0$ and $t_i = \sum_{j=1}^{i}v_j$, where $i \leq l$, $l$ is the largest index where $v_l$ is less than $T$, $v_1 = \nicefrac{2\eps}{(a - b)}$, and $v_{i+1} = v_i e^{-\alpha v_i}$ (or equivalently $v_{i+1} = v_1 e^{- \alpha \sum_{j=1}^{i}v_j}$). Observe that for all $i$, $v_{i+1} < v_{i}$, and $v_i$ is the time interval between $t_{i-1}$ and $t_{i}$.)}

\vspace{-0.15in}
\begin{Lemma}
	\label{lm:udefinition}
	Fix $T$, $\eps$, and $\alpha > 0$. Consider any initial state $x_0 \in \mathbb{R}$ and the set of input signals $\mathcal{U}_{\palpha}$.
	The resulting trajectories of system~(\ref{sys:simple}) form a ($T,2\eps,\alpha$)-separated set.
\end{Lemma}
\begin{proof}
	First, consider any two trajectories having initial state $x_0$ and input signals that are equal to \hussein{ $v_1$ and $0$} \sayan{(removed: $a$ and $b$)} for the first $v_1$ time units, respectively. Then, \hussein{$|\xi_{x_0,v_1}(v_1) - \xi_{x_0,0}(v_1)|$} \sayan{(was $|\xi_{x_0,a}(v_1) - \xi_{x_0,b}(v_1)|$)} $=  (v_1 - 0) v_1 = 2\eps > 2 \eps e^{-\alpha v_1}$, where the last inequality follows from the fact that \hussein{$v_1 > 0$} \sayan{(was: $v_1 \geq 0$)}. Now, consider any two signals $u_1, u_2 \in \mathcal{U}_{\palpha}$ with corresponding sequences $\mathit{se}_1$ and $\mathit{se}_2$ with the same prefix up till the $j^{\mathit{th}}$ value for some $j \in [l]$, and with different ones in the $(j+1)^{\mathit{th}}$ index, i.e., $se_1[j+1]\neq se_2[j+1]$. The resulting trajectories $\xi_{x_0,u_1}$ and $\xi_{x_0,u_2}$ from these input signals and the common initial state $x_0$ are $(T,2\eps,\alpha)$-separated. 
	To see this, observe that for any $i \geq 1$,  \hussein{ and any $x_0 \in \reals$, $|\xi_{x_0,u_1}(t_i) - \xi_{x_0,u_2}(t_i)|  = |\xi_{x,a_i + u_1(t_{i-1})}(v_i) - \xi_{x,u_1(t_{i-1})}(v_i)| =  |\xi_{x,v_i + u_1(t_{i-1})}(v_i) - \xi_{x,u_1(t_{i-1})}(v_i)|  =  |x + (v_i +u_1(t_{i-1})) v_i - x - u_1(t_{i-1}) v_i| = v_i^2$} \sayan{(was ($|\xi_{x_0,u_1}(t_j) - \xi_{x_0,u_2}(t_j)|  = |\xi_{x,a}(v_j) - \xi_{x,b}(v_j)| =  |x + a v_j - x - b v_j|$))} $= 2 \eps e^{-\alpha \sum_{j=1}^{i-1} v_j} > 2 \eps e^{-\alpha \sum_{j=1}^{i} v_j}$, where $x = \xi_{x_0,u_1}(t_{i-1})$. 
	Hence, any two trajectories resulting from two input signals corresponding to two different strings $se_1$ and $se_2 \in \mathit{SE}$ are $(T,2\eps,\alpha)$-separated. 
	Therefore, this set of trajectories is $(T,2\eps,\alpha)$-separated set.
\end{proof}

Now that we have proven that the set of trajectories with input signals from $\inputset_{\palpha}$ is a $(T,2\eps,\alpha)$-separated set, we compute a lower bound on $|\tseq_\palpha|$.
 This will determine a lower bound on the size of the separated set, and consequently provide a lower bound on entropy of (\ref{sys:simple}).
We start by bounding $t_i$ with a function of $i$, the index of the switch.

\begin{Lemma}
	\label{lm:countingws}
	The $i^{\mathit{th}}$ time instant in $\tseq_\palpha$ is upper-bounded as follows: $t_i = \sum_{j=1}^{i} v_j \leq$ \hussein{$\frac{2}{\alpha}\ln (\frac{2\alpha \sqrt{2\eps} i}{2 - \alpha \sqrt{2\eps}} + 1)$} \sayan{(was $ \frac{1}{\alpha} \ln (\frac{2 \alpha \eps i}{a - b - \alpha \eps} + 1))$}. Moreover, the cardinality of $\tseq_\palpha$ over the time horizon $[0,T]$  is lower-bounded as follows: $|\tseq_\palpha| \geq$ \hussein{$\floor{\frac{\sqrt{2\eps} -\alpha \eps}{2\alpha \eps}(e^{\alpha T} - 1)}$} \sayan{(was $\floor{\frac{a - b -\alpha \eps}{2\alpha \eps}(e^{\alpha T} - 1)}$)}.
\end{Lemma}
\begin{proof}
	To upper bound the sum of $v_i$'s, we upper bound the sequence $\{v_i\}$ with another sequence $\{w_i\}$, whose sum is easier to compute. 
	Recall that $v_1 = $ \hussein{$\sqrt{2\eps}$} \sayan{(was $\frac{2\eps}{(a - b)}$)} and $v_{i+1} =$ \hussein{$v_i e^{-\alpha v_i /2 }$} \sayan{(was $v_i e^{-\alpha v_i}$)} for $i \geq 1$. Since  $\alpha v_i > 0$, $v_{i+1} <$ \hussein{$\frac{v_i}{1 + \alpha v_i / 2 }$} \sayan{(was $\frac{v_i}{1 + \alpha v_i}$)}. Now, let us define $w_1 = v_1$ and $w_{i+1} =$ \hussein{$\frac{w_i}{1 + \alpha w_i/2}$} \sayan{(was $\frac{w_i}{1 + \alpha w_i}$)}, for all $i > 1$. \hussein{(this change is propagated in the proof below without red markings.)}
	
	{\em Claim:} $v_i \leq w_i$, for all $i\geq 1$.
	
	{\em Proof of claim:}
	 We will proceed by induction. The base case is when $i=1$ and we know that $v_1 \leq w_1$. Let $g(y) = \frac{y}{1 + \alpha y/2}$ for $y \in \mathbb{R}$. Then, $\frac{d}{d y} g(y) =  \frac{1}{(1 + \alpha y/2)^2} > 0$ and thus is an increasing function. Hence, assuming that $v_{i} \leq w_i$, we get that $v_{i+1} < \frac{w_i}{1 + \alpha w_i/2} = w_{i+1}$, since we know that $v_{i+1} < \frac{v_i}{1 + \alpha v_i /2}$.  Hence, by induction, $v_i \leq w_i$ for all $i \geq 1$.
	
	{\em Claim:} $w_{i} = \frac{2}{\alpha( i + c)}$, where $c = \frac{2 - \alpha w_1}{\alpha w_1}$, for all $i\geq 1$. 
	
	{\em Proof of claim:} We again proceed by induction: for $i = 1$, $\frac{2}{\alpha(1 + c)} = w_1$. Assume that $w_i = \frac{2}{\alpha(i + c)}$, then $w_{i+1} = \frac{\frac{2}{\alpha( i + c)}}{\frac{2(\alpha /2)}{\alpha(i + c)} + 1} = \frac{2}{\alpha((i+1) + c)}$. Hence, $\sum_{j=1}^{i} w_j = \frac{2}{\alpha} \sum_{j= 1}^{i} \frac{1}{j + c}$. 
	
	Now we proceed to prove the main lemma. First, note that $c > -1$ since $w_1$ and $\alpha$ are positive.
	By convexity of $\frac{1}{x}$, $\frac{1}{x} < \int_{x-\nicefrac{1}{2}}^{x+\nicefrac{1}{2}} \frac{ds}{s}$ for $x \geq \nicefrac{1}{2}$. Thus,
	\[
	\frac{1}{j + c} < \int_{j - \nicefrac{1}{2}}^{j + \nicefrac{1}{2}} \frac{dy}{y + c}, \mbox{for\ } j \geq 1/2 - c.
	\]
	Hence, the time of the $i^{\mathit{th}}$ switch can be bounded as follows:
	\begin{align*}
	\sum_{j=1}^{i}\frac{1}{j + c} 
	&<  \sum_{j=1 }^{i} \int_{j-\nicefrac{1}{2}}^{j+\nicefrac{1}{2}} \frac{dy}{y + c} 
	= \int_{\nicefrac{1}{2}}^{i+\nicefrac{1}{2}}\frac{dy}{y + c}\\
	&= \ln \frac{i + c + \nicefrac{1}{2}}{c + \nicefrac{1}{2}} 
	=  \ln (\frac{i}{c + \nicefrac{1}{2}} + 1)\\
	&=  \ln (\frac{2\alpha w_1 i}{2 - \alpha w_1/2} + 1)
	=\hussein{\ln (\frac{4\alpha \sqrt{2\eps} i}{4 - \alpha \sqrt{2\eps}} + 1).}\  \sayan{(\text{was } \ln (\frac{2\alpha \frac{2\eps}{a-b} i}{2 - \alpha \frac{2\eps}{a-b}} + 1)}\\
	\end{align*}
	Therefore, $\sum_{j=1}^{i} v_j \leq \sum_{j=1}^{i} w_j \leq$ \hussein{$\frac{2}{\alpha}\ln (\frac{4\alpha \sqrt{2\eps} i}{4 - \alpha \sqrt{2\eps}} + 1)$.} \sayan{(was: $\frac{1}{\alpha} \ln (\frac{2 \alpha \eps i}{a - b - \alpha \eps} + 1)$)}.
	%

	
Now that we have the upper bound of $t_i$ as a function of $i$, we can lower bound the number of switches that an input signal is allowed before a specified time bound $T$ by \hussein{$\floor{\frac{4 -\alpha \sqrt{2\eps}}{4\alpha \sqrt{2\eps}}(e^{\alpha T/2} - 1)}$}. \sayan{(was $\floor{\frac{a - b -\alpha \eps}{2\alpha \eps}(e^{\alpha T} - 1)}$).}
\end{proof}

\begin{Lemma}
\label{lm:nonzeno-proof}
\hussein{
The sequence $\{t_i\}_i$ diverges to infinity, i.e., $\lim_{i\rightarrow \infty} t_i = \infty$, and the set of inputs in $\mathcal{U}_{\palpha}$ are not Zeno.}
\end{Lemma}

\begin{proof}
	\hussein{
The proof is by contradiction. By definition, $t_{i+1} = t_i + v_{i+1} = t_i + v_1 e^{-\alpha t_{i}/2}$.  Assume that $\lim_{i\rightarrow \infty} t_i = c$, for some constant $c \in \nnreals$. Then, $c = c + v_1 e^{-\alpha c/2}$, a contradiction.}
\end{proof}

	\begin{corollary}
	\hussein{	The total variation  of any signal in $\mathcal{U}_{\palpha}$ with number of switches that is $O(e^{\alpha T})$ is $O(T)$.}
	\end{corollary}
\begin{proof}
	\hussein{The total variation of any signal in $\mathcal{U}_{\palpha}$ is upper bounded by  $\sum_{i=0}^{N-1} a_i = \sum_{i=1}^{N} v_i$, 
 where $N$ is the number of switches in the signal. }
\hussein{ Thus, we get that the total variation is upper bounded by the same bound on $t_i$ in Lemma~\ref{lm:countingws}, i.e., $\frac{2}{\alpha}\ln (\frac{2\alpha \sqrt{2\eps} N}{2 - \alpha \sqrt{2\eps}} + 1)$.}
 \hussein{Thus, if we only consider signals in $\mathcal{U}_{\palpha}$ with number of switches $N$ that are $O(e^{\alpha T})$, the total variation would be $O(T)$. Therefore, there exist constants $\mu$ and $\eta$ such that those signals belong to $\mathcal{U}^s(\mu, \eta)$ and $\mathcal{U}^p(\mu, \eta)$.}
\end{proof}

\begin{theorem}
	\label{thm:exp_conv_entropy_infinite}
	For any $\eps$ and $\alpha > 0$, $h\est(\eps,\alpha)$ of system (\ref{sys:simple}) with the input set $\inputset_{\palpha}$, even when restricted to signals with $O(e^{\alpha T})$ switches and thus having linearly-bounded total variation, is infinite.
\end{theorem}
\begin{proof}
	 First, there are $2^{|\tseq_\palpha|}$ different sequences with values in $\{a,b\}$. Following the construction of Lemma~\ref{lm:udefinition}, we can conclude that there is the same number of input signals and ($T,2\eps,\alpha$)-separated trajectories over the interval $[0,T]$.
	
	By substituting \hussein{$O(e^{\alpha T})$} 
	\sayan{(was: the bound on $|\tseq_\palpha|$ in Lemma~\ref{lm:countingws})}
	 in the analysis above, we get that $s\est\sep(T,2\eps,\alpha) \geq 2^{O(e^{\alpha T})}$ and hence by Lemma~\ref{lm:n_less_s}, $s\est(T,\eps,\alpha) \geq 2^{O(e^{\alpha T})}$. Substituting this lower bound in (\ref{eq:entropy_def}) results in $h\est(\eps,\alpha) = \infty$.
	
	
\end{proof}

We conclude from this section that requiring  $\alpha > 0$ or $\eps$ approaching zero is not suitable for systems with inputs. Therefore, we assume that $\alpha = 0$ and write $h\est(\eps,0)$ to be $h\est(\eps)$ and denote the resulting notion by $\eps$-estimation entropy, for conciseness. We drop $\alpha$ as well from all definitions of approximating, spanning, and separated sets, and $h\est\sep$ and $h\est^*$.

%% file: BitRateandEntropy.tex
\section{Relation between $\eps$-estimation entropy and the bit rate of estimation algorithms}
\label{sec:bitrate_entropy}
\begin{figure}[t!]
	\centering
	\includegraphics[height=0.5in, width=3.5in]{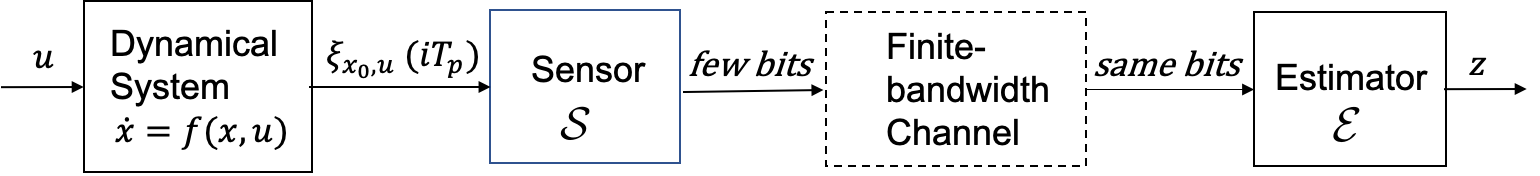}
	%
	
	
	\caption{\small The flow of information from the system to the sensor to the estimator. For any $t \geq 0$, $\| z(t) - \xi_{x_0,u}(t)\| \leq \eps$.}
	\label{fig:setup}
\end{figure}

In this section, we show that for any $\eps > 0$, there is no state estimation algorithm for system (\ref{sys:noise}) that can guarantee a maximum of $\eps$ estimation error when receiving state estimates at a fixed bit rate smaller than the $\eps$-estimation entropy $h\est(\eps)$ of the system.
First, let us define state estimation algorithms.
\begin{definition}

	A state estimation algorithm for system (\ref{sys:noise}) with a fixed bit rate is a pair of functions $(\mathcal{S},\mathcal{E})$, where $\mathcal{S} : \mathbb{R}^n\ \times\ Q_s \rightarrow \Gamma\ \times\ Q_s$ represents the sensor, $\mathcal{E} : \Gamma\ \times\ Q_e \rightarrow ([0,\Talgo] \rightarrow \mathbb{R}^n)\ \times\ Q_e$ represents the estimator,  $\Talgo$ is the sampling time, $\Gamma$ is an alphabet, and $Q_s$ and $Q_e$ are the sets of internal states of the sensor $\mathcal{S}$ and estimator $\mathcal{E}$, respectively. 
\end{definition} 
	Given $\eps$ and $\Talgo > 0$, sensor $\mathcal{S}$ samples the state of the system each $\Talgo$ time units, starting from $t = 0$. For any $i \geq 0$, at $t = i\Talgo$, $\mathcal{S}$ sends to  $\mathcal{E}$ a symbol from $\Gamma$ representing an approximation of the state $\xi_{x_0,u}(i\Talgo)$. Finally, $\mathcal{E}$ maps the received symbol to a $(\Talgo,\eps)$-approximating function of the trajectory for the time interval $[i\Talgo, (i+1)\Talgo]$ (see Figure~\ref{fig:setup}).

Now, let us define the bit rate of an estimation algorithm:
 \[b_r(\eps) := \limsup_{T \rightarrow \infty}\frac{1}{T} \sum_{i=0}^{\floor{\nicefrac{T}{\Talgo}}}\log |\Gamma| = \frac{\log |\Gamma|}{\Talgo}. \]

Given any time bound $T$, the average bit rate of the state estimation algorithm is the total number of bits it sends 
 divided by $T$. It sends $\log |\Gamma|$ bits every $\Talgo$ time units.
\begin{proposition}
	\label{prop:min_bit_rate}
	For any $\eps > 0$, there is no fixed bit rate state estimation algorithm for system (\ref{sys:noise})  with a bit rate smaller than $h\est(\eps)$ that for any $T>0$, generates a $(T,\eps)$-approximating function for any input trajectory.
\end{proposition}
\begin{proof}
	The proof is similar to the proof of Proposition 2 in \cite{sibai-mitra-2017}. It is based on showing that if any estimation algorithm has a bit rate $b_r(\eps)$ less than $h\est(\eps)$, then for any time bound $T$, the set of functions $z$ (see Figure~\ref{fig:setup}) that can be constructed by that algorithm is a $(T, \eps)$-approximation set with a smaller cardinality than $s\est(T,\eps)$, the supposed minimal cardinality, leading to a contradiction.
	 
	Specifically, for the sake of contradiction, assume that there exists an estimation algorithm with a bit rate smaller than $h\est(\eps )$ and constructs $(T, \eps)$-approximating functions for input trajectories.
	 Recall that $h\est(\eps ) =\limsup_{T\to\infty}\nicefrac{1}{T} \log s\est(T,\eps)$. Then, 
	\begin{align*}
		b_r(\eps) &< h\est(\eps), \\
		\limsup_{T \rightarrow \infty}\frac{1}{T} \sum_{i=0}^{\floor{\nicefrac{T}{\Talgo}}}\log |\Gamma| &< \limsup_{T \rightarrow \infty}\frac{1}{T} \log s\est(T, \eps ), \text{ and} \\
		\limsup_{T \rightarrow \infty} \frac{(\floor{\nicefrac{T}{\Talgo}} + 1)\log |\Gamma|}{T} &<  \limsup_{T \rightarrow \infty} \frac{1}{T}\log s\est(T, \eps ).
	\end{align*}
	Hence, for a sufficiently large $T'$, we should have  $\frac{(l + 1)\log |\Gamma|}{T'} <  \frac{1}{T'}\log s\est(T', \eps )$, where $l = \floor{\nicefrac{T'}{\Talgo}}$. Consequently, we get the inequality $|\Gamma|^{l + 1} < s\est(T',\eps)$. However, $|\Gamma|^{l+1}$ is the number of possible sequences of length $l+1$ that can be sent by the sensor over $l+1$ iterations. There are $l+1$ instead of $l$ iterations over the interval $[0,T']$ since the sensor starts sending the codewords at $t = 0$. Hence, the number of functions that can be constructed by the estimator is upper bounded by $|\Gamma|^{l+1}$. 
	
	For any given trajectory, the output of the estimator is a corresponding $(T',\eps)$-approximating function over the interval $[0,T']$. This is true since the estimator should be able to construct a $(T',\eps)$-approximating function for the corresponding trajectory of the system over the interval $[0, (l+1)\Talgo)$ given the symbols sent by the sensor in the first $l+1$ iterations. Hence, the set of functions that can be constructed by the estimator defines a $(T',\eps)$-approximating set. However, $s\est(T',\eps)$ is the minimal cardinality of such a set. Therefore, the set of functions that can be constructed by the algorithm is a $(T',\eps)$-approximating set with a cardinality smaller than $s\est$, the supposed minimal one, which is a contradiction.
\end{proof}

%% file: UpperBound.tex
\section{Entropy Upper Bound and Algorithm}
\label{sec:main_upperbound}

In this section, we derive an upper bound on the entropy $h\est(\eps)$ of system (\ref{sys:noise}) in terms of its sensitivity to its initial state and input and the required bound on the estimation error.

%% file: discrepancyFunction.tex
\subsection{Local input-to-state discrepancy function construction}
\label{sec:discrepancyfunction}
We use a modified version of the definition of local input-to-state discrepancy as introduced in~\cite{Huang-Fan-Mitra-16} in order to upper bound the distance between any two trajectories. 
We relax the original definition to include piece-wise continuous input signals and piece-wise continuous Jacobian matrices.

\begin{definition}(Local IS Discrepancy). \hussein{ Fix an arbitrary compact subset  $\mathcal{X}\subset \reals^n$, time interval $[t_0,t_1] \subset \nnreals$, and a set of piecewise-right-continuous functions $\mathcal{U}$ mapping $[t_0, t_1]$ to $\reals^m$}. A function $V: \mathcal{X}^2 \rightarrow \nnreals$  is a {\em local input-to-state (IS) discrepancy function\/}  for system (\ref{sys:noise}) over $\mathcal{X}$ and $[t_0,t_1]$ if:
	\begin{enumerate}[label=(\roman*)]
	\item  there exist class-$\K$  functions $\bar{\alpha},\ubar{\alpha}$ such that for any  $x,x' \in \mathcal{X}$, $\ubar{\alpha}(\|x - x'\|) \leq V(x,x') \leq \bar{\alpha}(\|x - x'\|)$, and
	\item there exists a class-$\K$ function in the first argument and continuous in the second argument $\beta:\nnreals \times \nnreals \rightarrow \nnreals$ and a  class-$\K$ function $\gamma: \nnreals\rightarrow \nnreals$ such that for any $x_0, x_0' \in \mathcal{X}$ and $u, u' \in \mathcal{U}$, if $\xi_{x_0,u}(t)$ and $\xi_{x_0',u'}(t) \in \mathcal{X}$ for all  $t \in [t_0,t_1]$, then  for any such $t$,
	\begin{align} 
	&V(\xi_{x_0',u'}(t), \xi_{x_0, u}(t)) \leq \nonumber \\
	&\hspace{0.05in}\beta(\|x_0 - x_0'\|, t - t_0) + \int_{t_0}^{t}\gamma(\|u(s) - u'(s)\|)ds.
	\label{eq:isdisc}
	\end{align}
	\end{enumerate}
\end{definition}

The local discrepancy function $V$ together with $\beta$ and $\gamma$ give the sensitivity of the trajectories of the system to changes in the initial state and the input. The functions $\bar{\alpha},\ubar{\alpha},\beta,\gamma$ are sometimes called witnesses of the local IS discrepancy $V$.
Techniques for computing local discrepancy functions have been presented~\cite{Huang-Fan-Mitra-16, Fan_Mitra_2016, Fan_Mitra_2017}.

The following is a straight-forward generalization of Lemma 15 of~\cite{Huang-Fan-Mitra-16} to handle systems with piece-wise continuous inputs, instead of just continuous ones. \hussein{ In contrast with \cite{Huang-Fan-Mitra-16}, we consider the distance between the trajectories squared as the discrepancy function instead of the distance itself. This is a result we presented earlier in \cite{sibai-mitra-2018}, but had two issues: first, it was false when $G_x$ is negative, and second, it was used as a bound on the squared $\infty$-norm while it was a bound on the squared 2-norm of the difference between the two trajectories. In this paper, we correct both errors by adding a $\sup_{\tau \in [t_0,t_1]}$ in the definition of $\gamma$ to correct the first error and multiply the definition of $G_x$ by $n$ and that of $G_u$ by $\sqrt{m}$ to correct the second one. We present the proof  in Appendix~\ref{sec:ISdiscrepancy}. } 
\begin{restatable}{Lemma}{discrepancy}
	\label{lm:local_discrep_func}
	The function $V(x,x') := \| x - x'\|^2$ is a local IS discrepancy for system (\ref{sys:noise}) over any compact set of states $\mathcal{X} \subset \mathbb{R}^n$, time interval  $[t_0 ,t_1]\subseteq \nnreals$, \hussein{and piecewise-right-continuous inputs $\mathcal{U}$}  with
	\begin{align*}
		&\beta(y,t -t_0) := e^{2 G_x (t - t_0)}y^2 \text{ and } \\
		&\hspace{0.5in} \gamma(y) := G_u^2\hussein{(\sup_{\tau \in [t_0,t_1]}}e^{2 G_x (t_1 - \tau)}\hussein{)} y^2, \text{where } t \in [t_0,t_1],
	\end{align*}
	\begin{align}
	\label{def:lm:local_discrep}
	G_x &:= \hussein{\nicefrac{n}{2}}\bigg(\underset{u \in \mathcal{U},x \in \mathcal{X}}{{\underset{t \in [t_0,t_1]}{\sup}}} \hussein{2}\lambda_{max} \big( \frac{J_x(x,u(t)) + J_x(x,u(t))^T}{2}\big) + \hussein{1}\bigg), \nonumber \\
	G_u &:= \hussein{\sqrt{m}}\underset{u \in \mathcal{U}, x \in \mathcal{X}}{{\underset{t \in [t_0,t_1]}{\sup}}}  \VERT J_u(x,u(t))\VERT. \nonumber 
	\\
	\end{align}
\end{restatable}
If $f$ is globally Lipschitz continuous in both arguments, one can infer that $G_x$ and $G_u$ are finite for any compact set of states $\mathcal{X}$ and time interval $[t_0,t_1]$. In that case, we will denote any global upper bounds on $G_x$ and $G_u$ of any $\mathcal{X}$ and time interval $[t_0, t_1]$, by $M_x$ and $M_u$, respectively. An example of such bounds is presented in the following proposition, with the proof in Appendix~\ref{sec:ISdiscrepancy}. 
\begin{restatable}{proposition}{boundsMx}
	\label{prop:computingMxMu}
If $f$ is globally Lipschitz continuous in both arguments, then for any time interval $[t_0,t_1] \subset \nnreals$ and compact set $\mathcal{X} \subset \mathbb{R}^n$, $ G_x \leq \hussein{n(}nL_x + \frac{1}{2}\hussein{)}$ and $G_u \leq m\sqrt{m} L_u$, 
where $L_x$ and $L_u$ are the global Lipschitz constants of $f$. 
\end{restatable}


Further, it is shown in \cite{Fan_Mitra_2016} that if $f$ has a continuous Jacobian, one can get tighter local bounds on $G_x$ and $G_u$ that depend on the set of input functions $\mathcal{U}(\mu, \jum)$, the compact set $\mathcal{X}$, and the interval $[t_0,t_1]$. 

Therefore, for any $\tau > 0$, $t \in [0,\tau]$, $x_0, x_0' \in \mathbb{R}^n$, and $u, u' \in \mathcal{U}(\mu, \jum)$, the squared distance $\| \xi_{x_0,u}(t) - \xi_{x_0',u'}(t)\|^2$, is upper bounded by:
\begin{align}
\label{eq:discrepFunction}
e^{2 M_x t}\|x_0 - x_0'\|^2 + M_u^2 e^{2 \hussein{M_x^+} t} \int_{0}^{t} \| u(s) - u'(s) \|^2 ds, 
\end{align}
where $\hussein{M_x^+ = \max\{M_x, 0\}}$.

\begin{example}[Dubin's vehicle IS Discrepancy]
\label{eg:car_discrepancy}
\normalfont
Consider the vehicle of example~\ref{eg:car_trans}. Then, its
	\begin{align}
	\frac{J_x + J_x^T}{2}  = 
	\begin{bmatrix}
		0 &0 &-\frac{v}{2} \sin x_3 \\
		0 &0 &\frac{v}{2} \cos x_3 \\
		-\frac{v}{2} \sin x_3 &\frac{v}{2} \cos x_3 &0
	\end{bmatrix}, 
\end{align}
and has eigenvalues $0$ and $\pm \frac{v}{2}$.
Hence, $G_x =\hussein{\nicefrac{n}{2} (\nicefrac{2v}{2} + 1) = \frac{3(v+1)}{2}}$ \sayan{(was $(v +  1)/2$)}. Moreover, $G_u =\hussein{ \sqrt{m} }\VERT J_u \VERT = 1$.
If we use Proposition~\ref{prop:computingMxMu} to upper bound $G_x$ and $G_u$, we get $M_x = \hussein{n(n L_x + \nicefrac{1}{2}) = 3(3 v + \nicefrac{1}{2})}$  and  $M_u = m \sqrt{m} L_u = 1$. Here $G_x$ and $G_u$ do not depend on $\stateset$ or on the time interval and thus they can replace $M_x$ and $M_u$ in inequality~(\ref{eq:discrepFunction}).
\end{example}


%% file: approxFunction.tex
\subsection{Approximating set construction}
\label{sec:approxFunction}
 Let us fix $\eps > 0$ throughout this section. We will describe a procedure (Algorithm \ref{fig:approxFunct}) that, given a time bound $T>0$, an initial state $x_0 \in K$ and an input signal $u \in \mathcal{U}^p(\mu, \jum)$, constructs a $(T,\eps)$-approximating function for the trajectory $\xi_{x,u}$ over the time interval $[0,T]$. It follows that the set of functions that can possibly be constructed by Algorithm~\ref{fig:approxFunct} for different $x_0\in K$ and $u \in \mathcal{U}^p(\mu, \jum)$ is a $(T,\eps)$-approximating set for system~(\ref{sys:noise}). We will show in the next section how to obtain an upper bound on entropy by deriving an upper bound on the cardinality of this set. \hussein{That upper bound on entropy will be an upper bound on the minimal bit rate of state estimation as well, as Algorithm~\ref{fig:approxFunct} is a state estimation algorithm that requires a fixed bit rate equal to that upper bound.}

 \begin{algorithm}
 	\begin{algorithmic}[1]
 		\renewcommand{\algorithmicrequire}{\textbf{Input:}}
 		\STATE {\textbf{input:} $T$, $\Talgo$, $\delta_x$, $\delta_u$}
 		\STATE {$S_{x,0} \gets K$; $S_{u,0} \gets U$;} \label{Szero}
 		\STATE {$C_{x,0} \gets\ \mathit{grid}(S_{x,0},\delta_x)$;} \label{Cx_definition}
 		\STATE {$C_{u,0} \gets\ \mathit{grid}(S_{u,0},\delta_u)$;} \label{Cu_definition}
 		\STATE {$i \gets 0$;}
 		\WHILE{$i \leq \floor{\frac{T}{\Talgo}}$} 
 		\STATE {$x_i \gets \xi_{x_0,u}(i\Talgo)$; $u_i \gets  u(i\Talgo)$;} 
 		\STATE $q_{x,i} \gets quantize(x_i, C_{x,i})$; \label{quant}
 		\STATE $q_{u,i} \gets quantize(v_i, C_{u,i})$; \label{quantU}
 		\STATE {$z_{i} \gets \xi_{q_{x,i},q_{u,i}}$;} \label{z_construction}
 		\STATE {$i \gets i + 1$;} \COMMENT{parameters for next iteration}
 		\STATE $S_{x,i} \gets B(z_{i-1}(\Talgo^-),\eps)$; \label{Sx_construction}
 		\STATE $S_{u,i} \gets B(q_{u,i-1},\jum + \mu \Talgo + \delta_u)$; \label{Su_construction}
 		\STATE $C_{x,i} \gets \mathit{grid}(S_{x,i},\delta_{x})$; \label{Cx_construction}
 		\STATE $C_{u,i} \gets \mathit{grid}(S_{u,i},\delta_{u})$; \label{Cu_construction}
 		\STATE wait($\Talgo$);
 		\ENDWHILE
 		\renewcommand{\algorithmicrequire}{\textbf{Output:}}
 		\STATE {\textbf{output:} $\{z_i([0,\Talgo)) : i\in [0;\floor{\frac{T}{\Talgo}}-1]\} \cup \{z_{\floor{\frac{T}{\Talgo}}}([0,\Talgo])\}$}
 		
 	\end{algorithmic}
 	\caption{Construction of a $(T,\eps)$-approximating function.\label{fig:approxFunct}}
 \end{algorithm} 
	
	 The procedure (Algorithm~\ref{fig:approxFunct}) is parameterized by a time horizon $T>0$, a sampling period $\Talgo>0$, and two constants $\delta_x$ and $\delta_u > 0$. 
	 The procedure also uses the set of initial states  $K$, the set of initial inputs $U$, a particular initial state $x_0 \in K$, and an input signal $u \in \mathcal{U}^p(\mu, \jum)$ for system~(\ref{sys:noise}). 
	 The output is a piece-wise continuous function $z:[0,T]\rightarrow \reals^n$ that is constructed iteratively over each $[i\Talgo, (i+1)\Talgo)$ interval for $i \in [0;\floor{\frac{T}{\Talgo}}]$. 
	 Later we will infer several constraints on the parameters such that the output  $z$ is indeed a $(T,\eps)$-approximating function for the given trajectory $\xi_{x_0,u}$.
	 
	 Initially, $S_{x,0}$ is assigned to the initial set $K$. Similarly, $S_{u,0}$ is assigned to the initial input set $U$.
	 $C_{x,0}$ is a grid of size $\delta_x$ over $K$ and 
	 $C_{u,0}$ is a grid of size $\delta_u$ over $U$. 
	 At the $i^{\mathit{th}}$ iteration, $i \in [0;\floor{\frac{T}{T_p}}]$,
	$x_i$ and $u_i$ store the values $\xi_{x_0,u}(i\Talgo)$ and $u(i\Talgo)$, the state and input at the time $i\Talgo$, respectively.
	 Then, $q_{x,i}$ and $q_{u,i}$ are set to be the quantization of $x_i$ and $u_i$ with respect to $C_{x,i}$ and $C_{u,i}$, respectively.
	 With slight abuse of notation, we will also denote the function of time that maps the interval $[0,\Talgo)$ to $q_{u,i}$ by $q_{u,i}$, as in line \ref{z_construction}, for example. 
	 The variable $z_i$ stores the trajectory that results from running system~(\ref{sys:noise}) starting from initial state $q_{x,i}$, with input signal $q_{u,i}$, and running for $\Talgo$ time units. After that, $i$ is incremented by $1$ and the next iteration variables  $S_{x,i}$, $S_{u,i}$, $C_{x,i}$, and $C_{u,i}$ are initialized. 
	 Finally, the procedure outputs the concatenation of the $z_i$'s, for all $i \in [0;\floor{\frac{T}{\Talgo}}]$ that is denoted by the function $z:[0,T] \rightarrow \mathbb{R}^n$. Note that for $i=\floor{\frac{T}{\Talgo}}$, we consider $z_i$ to have a domain that is closed from the right as well.
	 
	 In the following lemma, we show that if the parameters of the procedure $\Talgo$, $\delta_x$ and $\delta_u$, are small enough, then the output is a $(T,\eps)$-approximating function for $\xi_{x_0,u}$.  
	

%
Before stating the lemma, let us define the following function
\begin{align}
g_c(\delta_x, \delta_u, \Talgo) :=&g_{c,x}(\delta_x,\Talgo) + g_{c,u}(\delta_u,\Talgo),	\label{eq:feasible_set}
\end{align}
where $g_{c,x}(\delta_x,\Talgo) := \delta_x^2 e^{2M_x\Talgo}$ and 
$g_{c,u}( \delta_u, \Talgo) := M_u^2 e^{2\hussein{M_x^+}\Talgo} \big( \frac{1}{3}\mu^2\Talgo^3 + (\delta_u + \jum)\mu \Talgo^2  +(\delta_u + \jum)^2\Talgo \big)$. The subscript $c$ is added since the functions would be used to specify the {\em constraints} that the parameters should satisfy for the output of Algorithm~\ref{fig:approxFunct} to be a $(T,\eps)$-approximating function.


 Since $g_c$  \hussein{is continuous in $\Talgo$, $\delta_x$, and $\delta_u$, non-negative, }
\sayan{(was:  strictly increases when $\Talgo$, $\delta_x$, or $\delta_u$ increase)},
 and is equal to zero when \hussein{$\delta_x = $} $\Talgo = 0$, then for any $\eps > 0$, there exist $\Talgo$, $\delta_x$, and $\delta_u >0$ such that $g_c(\delta_x, \delta_u, \Talgo) \leq \eps^2$.
	 \begin{Lemma}
	 	\label{lm:proveApproxFunc}
	 Given $\eps > 0$, fix $\delta_x$, $\delta_u$, and $\Talgo$ such that $g_c(\delta_x,\delta_u,\Talgo) \leq \eps^2$.
	 For any $x_0 \in K$ and $u \in \mathcal{U}^p(\mu, \jum)$,
	  for all $i \in [0;\floor{\frac{T}{\Talgo}}]$, and for all $t \in [i\Talgo,(i+1)\Talgo]$,
	 \begin{enumerate}[label=(\roman*)]
	 \item $u_i \in S_{u,i}$,
	 \item $\|u(t) - q_{u,i}\|\leq \mu (t - i\Talgo) + \jum + \delta_u$,
	 \item $x_i \in S_{x,i}$, and
	 \item $\| z_i(t - i\Talgo) - \xi_{x_i, u_i}(t - i\Talgo)\| \leq \eps$,
	 \end{enumerate}
  	where $u_i(t) := u(i\Talgo + t)$, the $i^{\mathit{th}}$ piece of $u$ of size $\Talgo$.
	 \end{Lemma}
	\begin{proof}
	First, $\|u_0 - q_{u,0} \|\leq \delta_u$ since the $u(0) = u_0 \in U$ and $q_{u,0}$ is its quantization of $u_0$ with respect to a grid of resolution $\delta_u$. Moreover, $\| u(t) - u(0)\| \leq \mu t + \jum$ by Definition~\ref{def:inputset}. But, by triangle inequality, $\| u(t) - q_{u,0}\|  \leq  \|u(t) - u_0\| + \|u_0 - q_{u,0} \| \leq \mu t + \jum + \delta_u$. Hence, $u_1 \in S_{u,1}$. Fix $i \geq 1$ and assume that $u_i \in S_{u,i}$. Then, $\| u_i - q_{u,i}\|\leq \delta_u$. Repeating the same analysis for the $i=0$ case, results in $\| u_i(t-i\Talgo) - q_{u,i}\| \leq \mu (t-i\Talgo)  + \jum + \delta_u$. Thus, $u_i \in S_{u,i}$ for all $i$. 
	Second, \hussein{$x_0 \in S_{x,0}$ since $x_0 \in K$. Also, $\| z_0(0) - \xi_{x_0,u_0}(0)\| = \| q_{x,0} - x_0 \| \leq \delta_x \leq \eps$, where the last inequality follows from the assumption that $g_c(\delta_x,\delta_u, \Talgo) \leq \eps^2$. Now,} fix $t \in [0,T]$ and let $i = \floor{\frac{t}{\Talgo}}$.
	Then,
	\begin{align}
	\label{eq:approxFunction_i}
	&\|\xi_{x_i,u_i}(t -i\Talgo) -  \xi_{q_{x,i},q_{u,i}}(t - i\Talgo)  \|^2 \nonumber\\
	&\leq  \| x_i - q_{x,i} \|^2 e^{2 M_x(t - i\Talgo)} \nonumber\\
	&\hspace{0.3in} + M_u^2 e^{2\hussein{M_x^+}(t - i\Talgo)} \int_{i\Talgo}^{t} \|u(s) - q_{u,i} \|^2 ds\nonumber \\
	&\hspace{1in}\mbox{[by (\ref{eq:discrepFunction})]} \nonumber \\
	&\leq \delta_x^2 e^{2M_x(t-i\Talgo)}  \nonumber\\
	&\hspace{0.3in} +  M_u^2 e^{2\hussein{M_x^+}(t - i\Talgo)}\int_{i\Talgo}^{t}  \big( \mu (s-i\Talgo) + \jum + \delta_u)^2 ds, \nonumber \\
	&\leq \delta_x^2 e^{2M_x \Talgo} + M_u^2 e^{2\hussein{M_x^+}\Talgo} \big( \frac{1}{3}\mu^2\Talgo^3 + (\delta_u + \jum)\mu \Talgo^2 \nonumber\\
	&\hspace{0.3in}+ (\delta_u + \jum)^2\Talgo \big) = g_c(\delta_x, \delta_u,\Talgo) \leq \eps^2,
	\end{align}
	where the last inequality follows from the assumption in the lemma statement.
	Therefore, for all $i \in [0;\floor{\frac{T}{\Talgo}}]$ and $t\in [0,T]$, $x_i \in B(z_{i-1}(\Talgo),\eps) = S_{x,i}$.
\end{proof}

\begin{corollary}
\label{cor:z_approx_func}
For any $\delta_x$, $\delta_u$, and $\Talgo$ such that $g_c(\delta_x,\delta_u,\Talgo) \leq \eps^2$, for all $t\in [0,T]$, the output function $z$ of Algorithm~\ref{fig:approxFunct} is a $(T,\eps)$-approximating function to the input trajectory $\xi_{x_0,u}$, i.e.
\begin{align}
\|z(t) - \xi_{x_0,u}(t) \| \leq \eps.
\end{align}
\end{corollary}
	
One can conclude from Corollary~\ref{cor:z_approx_func} that the set of all functions that can be constructed by Algorithm~\ref{fig:approxFunct} for any input trajectory $\xi_{x_0,u}$, where $x_0 \in K$ and $u \in \mathcal{U}^p(\mu, \jum)$, is a $(T,\eps)$-approximating set. In the following lemma, we will compute an upper bound on the number of these functions to obtain an upper bound on $s\est(T,\eps)$.
	
	\begin{Lemma}
		\label{lm:approxSetBound}
	For any $T \geq 0$ and $\delta_x$, $\delta_u$ and $\Talgo$ such that $g_c(\delta_x,\delta_u,\Talgo) \leq \eps^2$, the number of functions that can be constructed by Algorithm \ref{fig:approxFunct} for any $x_0 \in K$ and $u \in \mathcal{U}^p(\mu, \jum)$, is upper-bounded by: 
		\begin{align*}
		\ceil{\frac{diam(K)}{2 \delta_x}}^n  \ceil{\frac{diam(U)}{2 \delta_u}}^m\bigg(\ceil{\frac{\eps}{\delta_x}}^n\ceil{\frac{ \hussein{\mu \Talgo + \eta} }{\delta_u} + 1}^m\bigg)^{\floor{\frac{T}{ T_p }}}.
		\end{align*}
	\end{Lemma}
	\begin{proof}
		To construct a $(T,\eps)$-approximating function for a given trajectory $\xi_{x,u}$, at an iteration $i \in [0; \floor{\frac{t}{\Talgo}}]$, Algorithm \ref{fig:approxFunct} picks one point in $C_{x,i}$ and one in $C_{u,i}$. Hence, the number of different outputs that it can produce is upper bounded by: 
		$\prod_{i=0}^{\floor{\nicefrac{T}{ T_p }}}|C_{x,i}| |C_{u,i}|.$
		
		To construct the grids $C_{x,0}$ and $C_{u,0}$, in each of the $n$  (or $m$) dimensions of the state (or input) space, we partition a segment of length $\mathit{diam}(K)$ (or $\mathit{diam}(U)$) to smaller segments of size $2 \delta_x$ (or $2 \delta_u$).
		Then,
		$|C_{x,0}| \leq \ceil{\frac{\mathit{diam}(K)}{2 \delta_x}}^n$ and $|C_{u,0}| \leq \ceil{\frac{\mathit{diam}(U)}{2 \delta_u}}^m$. Similarly, for all $i > 0$, $S_{x,i} = B(z_{i-1}(\Talgo^-), \eps)$ and $S_{u,i} = B(q_{u,i-1}, \mu \Talgo + \jum + \delta_u)$. Hence,
		$|C_{x,i}| \leq \ceil{\frac{2 \eps}{2 \delta_x}}^n = \ceil{\frac{\eps}{\delta_x}}^n$, since $\mathit{diam}(S_{x,i}) = 2\eps$ and $|C_{u,i}| \leq \ceil{\frac{2(\mu \Talgo + \jum + \delta_u)}{2 \delta_u}}^m = \ceil{\frac{\mu \Talgo + \jum}{\delta_u} + 1}^m$, since $\mathit{diam}(S_{u,i}) = 2(\mu \Talgo + \jum + \delta_u)$. 
		Substituting these values in the bound above  
		leads to the upper bound in the lemma.
	\end{proof}

%% file: noise_systems.tex
\subsection{Entropy upper bound}
\label{sec:upperbound}
 The following proposition gives an upper bound on the entropy of system (\ref{sys:noise}) in terms of $\Talgo$, $\delta_x$ and $\delta_u$. 
  It shows the effect of our choice of the parameters of Algorithm~\ref{fig:approxFunct}. It will also help us recover the bound on estimation entropy $h\est(\eps)$ of autonomous systems in \cite{LM:TAC2018} in Corollary~\ref{cor:relationToPrevBound}.
\begin{proposition}
\label{prop:noise_entropy_upperbound_lip}
For a fixed $\Talgo$, $\delta_x$ and $\delta_u$ where $g_c(\delta_x,\delta_u,\Talgo) \leq \eps^2$,
	the entropy $h\est(\eps )$ of system (\ref{sys:noise}) is upper bounded by $g_o(\delta_x, \delta_u, \Talgo)$ defined to be equal to: 
	\begin{align}
	\label{eq:parameterized_upper_bound}
	\frac{1}{\Talgo} \big( n \log \ceil{\frac{\eps}{\delta_x}} + m \log \ceil{\frac{\mu \Talgo + \jum}{\delta_u} + 1} \big).
	\end{align}
\end{proposition}
\begin{proof}
	We substitute the upper bound on the cardinality of the minimal approximating set obtained in the previous section in definition of $h\est$ in Equation (\ref{def:entropy}) to get: $h\est(\eps) = $
	\begin{align*}
	&\limsup_{T\rightarrow \infty} \frac{1}{T} \log s\est(T,\eps) \\
	&  \leq \limsup_{T\rightarrow \infty} \frac{1}{T} \log |C_{x,0}||C_{u,0}|(|C_{x,1}| |C_{u,1}|)^{\floor{\frac{T}{ T_p }}} \\
	&  \leq \limsup_{T\rightarrow \infty} \frac{1}{T} \log (|C_{x,1}| |C_{u,1}|)^{\floor{\frac{T}{ T_p }}} \\
	&\leq \limsup_{T\rightarrow \infty} \frac{1}{T} \log (\ceil{\frac{\eps}{\delta_x}}^n \ceil{\frac{ \mu \Talgo + \jum}{\delta_u} + 1}^m)^{\floor{\frac{T}{ \Talgo }}} \\
	&\leq \limsup_{T\rightarrow \infty} \frac{\nicefrac{ T }{\Talgo}}{ T }n \log  \ceil{\frac{\eps}{\delta_x}}\\
	&\hspace{0.5in}+ \limsup_{T\rightarrow \infty} \frac{\nicefrac{T}{ \Talgo }}{ T }m \log \ceil{\frac{ \mu \Talgo + \jum}{\delta_u} + 1} \\
	&= \frac{1}{\Talgo} \big( n \log \ceil{\frac{\eps}{\delta_x}} + m \log \ceil{\frac{ \mu \Talgo + \jum}{\delta_u} + 1} \big).
	\end{align*}
\vspace{-0.1in}
\end{proof}
	We show that if the bound on the input norm is negligible, we recover the bound  $\frac{nL_x}{\ln2}$ on entropy  derived in \cite{LM015:HSCC}.
	\begin{corollary}
		\label{cor:relationToPrevBound}
	Given any $\eps > 0$, $\underset{\mu, \jum \rightarrow 0}{\lim} h\est(\eps) \leq \frac{nL_x}{\ln 2}$.
	\end{corollary}
	\begin{proof}
	Fix any $\eps > 0$. Fix a $\jum_0$ and $\mu_0$ such that $\jum_0 + \mu_0 < 1$. Then, fix $\delta_u$ to be equal to $\delta_{u,0} = \jum_0 + \mu_0$. After that, while taking the limits of $\mu$ and $\jum$ going to zero, keep $\delta_u$ fixed to $\delta_{u,0}$, choose $\delta_x = \sqrt{1-(\mu + \jum)}\eps e^{-L_x\Talgo}$, and $\Talgo$ to be such that $ g_{c}(\delta_x,\delta_{u,0},\Talgo)\leq \eps^2$. 
	Hence, by Proposition \ref{prop:noise_entropy_upperbound_lip}, $h\est(\eps ) \leq \frac{1}{\Talgo}\big( n \log \ceil{\frac{\eps}{\delta_x}} + m \log \ceil{\frac{\jum + \mu \Talgo}{\delta_u} + 1}\big) \leq nL_x \log \ceil{\frac{e}{\sqrt{1 - (\mu + \jum)}}} + m \log \ceil{\frac{\jum + \mu \Talgo}{\delta_{u,0}} + 1}$. Moreover, as $\mu$ and $\jum$ decrease to zero, the argument of the second $\log$ approaches one and the upper bound approaches $\frac{nL_x}{\ln2}$.
	\end{proof}

\begin{corollary}
	\label{cor:simple_sys_entropy_finite}
For any $\eps> 0$, $h\est(\eps)$ of system~(\ref{sys:simple}),  \hussein{ with any set of inputs $\mathcal{U}^p(\mu, \eta)$ 
with finite $\mu$ and $\eta$}, is finite.
\end{corollary}
\begin{proof}

 The Jacobian $J_x$ of system~(\ref{sys:simple}) is equal to zero since the state $x$ does not appear on the RHS of the differential equation. Hence, $M_x = 1/2$ using Equation~(\ref{def:lm:local_discrep}). Moreover, $J_u = 1$ and thus $M_u = 1$. Then, there exist $\delta_x$, $\delta_u$ and $\Talgo >0$ that satisfy $g_c(\delta_x,\delta_u,\Talgo) \leq \eps^2$, and $h\est(\eps)$ of system~(\ref{sys:simple}) is upper bounded by $g_o(\delta_x,\delta_u,\Talgo)$, by Proposition~\ref{prop:noise_entropy_upperbound_lip}.
\end{proof}

Corollary~\ref{cor:simple_sys_entropy_finite} is the last piece of the argument that other variants of entropy definitions mentioned in Section~\ref{sec:entro_motivation} are not adequate for \hussein{systems with uncertain inputs.} \sayan{(was: open systems)}

\begin{example}[Dubin's car entropy upper bound]
	\label{eg:car_upper_bound}
	\normalfont
	Consider the car of example~\ref{eg:car_trans} and its IS discrepancy function of example~\ref{eg:car_discrepancy}. In this example, we will compute its upper bound per Proposition~\ref{prop:noise_entropy_upperbound_lip}.
	Suppose $\mu = \pi/4$ and $\eta = \pi/4$ and the needed estimation accuracy $\eps = 0.1$. 
	Let us fix $v = 10$ m/s. Then, $L_x = \nicefrac{(10 + 1)}{2} = 5.5$, \hussein{$G_x = \nicefrac{3}{2} (10 + 1) = 16.5$}, $M_x = \hussein{3 (}3 (10) + \nicefrac{1}{2} \hussein{)} = \hussein{91.5}$ \sayan{(was $30.5$)} and $ G_u = 1$.

	First, we use $G_x$ and $G_u$ for the discrepancy function as a replacement of $M_x$ and $M_u$ in the bound of Proposition~\ref{prop:noise_entropy_upperbound_lip} and in the definition of $g_c$ in Equation~(\ref{eq:feasible_set}).
	Let us fix $\delta_u = \eta = \pi/4$, $T_p = 1.9 \times 10^{-3}$, and $\delta_x = \frac{\eps}{\sqrt{2}} e^{-G_x\Talgo}$. Such assignment satisfy $g_c(\delta_x,\delta_u,\Talgo) = 0.0099 \leq \eps^2 = 0.01$ and results in the bound $g_o(\delta_x, \delta_u,\Talgo) = \hussein{727}$. \sayan{(was $2158$, it was wrong even with the previous value of $M_x$ following the 2-norm.)}
	
	Second, we derive the resulting bound from using the upper bounds on $G_x$ and $G_u$, $M_x$ and $M_u$ of Proposition~\ref{prop:computingMxMu}, instead of $G_x$ and $G_u$. The previous choices of $\delta_x$, $\delta_u$, and $\Talgo$ would result in $g_c(\delta_x,\delta_u,\Talgo) = 0.011 > \eps^2 = 0.01$, which violates the condition in Proposition~\ref{prop:noise_entropy_upperbound_lip}. Instead, we use $\delta_x = \frac{\eps}{\sqrt{2}} e^{-M_x\Talgo}$ and $\Talgo = \hussein{1.5 \times 10^{-3}}$ \sayan{(was $1.7 \times 10^{-3}$)} while keeping $\delta_u = \pi/4$.
	Such an assignment satisfies $g_c(\delta_x,\delta_u,\Talgo) = 0.0098 \leq \eps^2 = 0.01$ and results in the bound $g_o(\delta_x, \delta_u,\Talgo) = \hussein{920}$ \sayan{(was $2411$)}, larger than the one we obtained using $G_x$ and $G_u$.
\end{example}

\begin{example}[Harrier jet]
	\label{eg:harrier_jet}
	\normalfont
	 We study the  Harrier ``jump jet'' model from~\cite{Astrom-2008}. The dynamics of the system is given by:
	\begin{align*}
	&\dot{x}_1 = x_4; \dot{x}_4 = -g \sin x_3 - \frac{c x_4}{m'}  + \frac{u_1}{m'} \cos x_3 - \frac{u_2}{m'} \sin x_3;\\
	&\dot{x}_2 = x_5;  \dot{x}_5 = g(\cos x_3 - 1) - \frac{c x_5}{m'}  + \frac{u_1}{m'} \sin x_3 + \frac{u_2}{m'} \cos x_3;\\
	&\dot{x}_3 = x_6;  \dot{x}_6 = \frac{r}{J} u_1;
	\end{align*}
	where $(x_1,x_2)$ is the position of the center of mass and $x_3$ is the orientation of the aircraft in the vertical plane, and $(x_4,x_5,x_6)$ are their time derivatives, respectively.
	The mass of the aircraft is $m'$, the moment of inertia is $J$, the gravitational constant is $g$, 
	and the damping coefficient is $c$.
	%
	 The inputs $u_1$ and $u_2$ are the force vectors generated by the main downward thruster and the
	maneuvering thrusters. 
	We computed $G_x$ and $G_u$ in Appendix~\ref{sec:harrier_jacobian}.
	 Their values are $G_x =\hussein{n (} \frac{-c}{m'} + \frac{1}{2} \hussein{)}$ and  $G_u \hussein{\leq}  \sqrt{\hussein{m (}\frac{r^2}{J^2}+ \frac{1}{m'^2}\hussein{)}}$.
Fixing $m' = 100$, $g = 9.81$, $r = 5$, $c = 100$, and $J = 50$, we get $G_x = -\frac{\hussein{3}}{2}$ and $G_u = \hussein{0.14}$ \sayan{(was $0.1$)}. Now, suppose that $\mu = 10$ and $\jum = 20$ and the needed estimation accuracy $\eps = 0.5$.
 We choose  $\delta_u = \jum = 20$, $\Talgo = \hussein{3.95 \times 10^{-3}}$ \sayan{(was $7.5 \times 10^{-3})$}, and $\delta_x = \frac{1}{\sqrt{2}}\eps e^{-G_x\Talgo}$.
 Then, using Proposition~\ref{prop:noise_entropy_upperbound_lip}, $g_c(\delta_x,\delta_u,\Talgo) = 0.249 \leq \eps^2 = 0.25$, and thus $h\est(0.5) \leq g_o(\delta_x,\delta_u,\Talgo) = \hussein{699}$ \sayan{(was $693$)}. 
   However, if $\mu = 0.1$ and $\jum = 45$, and choosing $\delta_u = \jum$, $\Talgo = \hussein{0.75 \times 10^{-3}}$ \sayan{(was $1.5 \times 10^{-3})$}, and $\delta_x = \frac{1}{\sqrt{2}}\eps e^{-G_x\Talgo}$. Then, $g_c(\delta_x,\delta_u,\Talgo) = 0.244 \leq \eps^2 = 0.25$ and hence $h\est(0.5) \leq \hussein{3680}$ \sayan{(was $3465$)}. 
   Hence, the bound increased significantly as the bound on the variation of the input signal was relaxed. Suppose now that we take the other extreme, where we restrict the allowed size of jumps in the input signal by decreasing $\jum$ to $0.1$ while allowing large continuous variations by increasing $\mu$ to $20$. In that case, the input signals are almost continuous. 
   If we fix $\delta_u = \jum$, $\Talgo = \hussein{0.35}$\sayan{(was $0.6$)}, and $\delta_x = \frac{1}{\sqrt{2}}\eps e^{-G_x\Talgo}$. Then, $g_c(\delta_x,\delta_u,\Talgo) = 0.247 \leq \eps^2 = 0.25$ and $h\est(0.5) \leq \hussein{11}$\sayan{(was $22$)}, which is significantly smaller than the two previous bounds.
\end{example}

%% file: ExpConvergence.tex
\section{Switched systems: entropy, upper bounds, and relation with systems with uncertain inputs}
\label{sec:switched_systems}
In this section, we define the estimation entropy of general autonomous switched systems and present a corresponding upper bound (Theorem~\ref{thm:switched_entropy_upper_bound})\hussein{, which is a restatement of Theorem 1 in \cite{sibai-mitra-2017}}. In Section~\ref{sec:finite_d}, through an example, we \hussein{present a novel result of this paper showing} \sayan{(was: show)} that an assumption \hussein{we made in  \cite{sibai-mitra-2017}} on the boundedness of the divergence between different modes in the derivation of the upper bound on entropy is reasonable. Then, in Section~\ref{sec:si_and_ss_upper_bound_relation}, we rewrite the upper bound of Proposition~\ref{prop:noise_entropy_upperbound_lip} which we derived for systems with inputs in a similar format to that of Theorem~\ref{thm:switched_entropy_upper_bound} which we derived for switched systems to show the similarities and differences of the two bounds. Finally, we show how switched systems can be modeled by systems with inputs in Section~\ref{sec:si_abstract_ss}. This implies that the upper bound on entropy of systems with inputs can be used for autonomous switched systems. \hussein{All of these results are a novel contribution of this paper.}

We denote a switched system with $N$ modes by:
\begin{align}
\label{sys:switched_system}
\dot{x}(t) = f_\sw(x(t),\sigma(t)),
\end{align}
where $\sigma : \nnreals \rightarrow [N]$, is a piece-wise-right-constant signal called a {\em switching signal}, $f_\sw : \reals^n \times [N] \rightarrow \reals^n$, is \hussein{globally} Lipschitz continuous in the first argument, and $x(t) \in \reals^n$. 
We call $f_\sw$ with a fixed second argument $p \in [N]$ a {\em mode} of the system.  
We define $L_x = \max_{p \in [N]} L_p$, where $L_p$ is the \hussein{global} Lipschitz constant with respect to the state of mode $p \in [N]$. Observe that we do not yet have any constraint on the behavior of $f_\sw$ in different modes. In the case of globally-Lipschitz non-switched systems with inputs, we had the global Lipschitz constant $L_u$, which governs how the system reacts to changes in its input. Here, in Equation~(\ref{def:doft}), we will use a function $d(t)$ that plays a similar role in bounding the difference of behavior resulting from running $f_\sw$ in different modes.

The points of discontinuities in $\sigma$ are called {\em switches}. The switching signal
$\sigma$ has a {\em minimum dwell time\/} $T_d >0$ if at least $T_d$ time units elapse between any two consecutive switches. We fix $T_d$ throughout the section and denote the corresponding set of switching signals as $\Sigma(T_d)$.  The initial state $x(0) \in K$, where $K$ is a compact initial set of states in $\mathbb{R}^n$.  The state trajectory of system~(\ref{sys:switched_system}) with $\sigma \in \Sigma(T_d)$ is denoted by $\xi_{x_0,\sigma}: \nnreals \rightarrow \mathbb{R}^n$, which \hussein{exists} 
\sayan{(was: is assumed to exist)}
 globally for all $t \geq 0$, unique, and continuously depends on its initial state.
 
 Moreover, we denote the set of all reachable states of system~(\ref{sys:switched_system}) under the family of switching signals $\Sigma(T_d)$ as: $\reach_\sw(K, T_d, T):= \{x\ | \ \exists x_0 \in K, \sigma \in \Sigma(T_d), t \in [0,T], \xi_{x_0,\sigma}(t) = x \}$.

Based on our setup in \cite{sibai-mitra-2017}, the entropy of the autonomous switched system~(\ref{sys:switched_system}) is defined as follows:
 First, three positive error parameters $\eps, \alpha$, and $\tau$, and the time interval $T > 0$ are fixed. Second, we redefine the concept of approximating functions. A function $z : \nnreals \rightarrow \reals^n$ is called a {\em $(T,\eps, \alpha, \tau)$-approximating} function of $\xi_{x_0,\sigma}$ if for all $j \geq 0$ and for all $t \in [s_j, s_{j+1})$, 
\begin{align}
\label{e-estimation}
\| z(t) - \xi_{x_0, \sigma}(t) \| \leq 
\begin{cases}
\varepsilon &t \in [s_{j}, s_{j} + \tau), \\
\varepsilon e^{-\alpha (t - (s_j + \tau))} &otherwise,
\end{cases}
\end{align}
where $s_0 = 0, s_1, \dots$ are the switches of $\sigma$. The norm in inequality (\ref{e-estimation}) can be arbitrary. Third, we define $s{\sest}(T,\eps,\alpha,\tau)$ to be the minimal cardinality of an approximating set of all trajectories starting from $K$ and having a switching signal from $\Sigma(T_d)$ over the time interval $[0,T]$.
\begin{definition}
\label{def:s_entropy}
The estimation entropy of the switched system~(\ref{sys:switched_system}) is:
\[h{\sest}(\eps, \alpha, \tau):= \limsup_{T\to\infty}\dfrac 1T \log s{\sest}(T,\eps, \alpha, \tau).\]
\end{definition}


The following function $d(t)$, which we denote as {\em mode divergence}, bounds the distance between any two trajectories of system~(\ref{sys:switched_system}) evolving according to two {\em different\/} modes while starting from any reachable state.
\begin{align}
\label{def:doft}
\dist(t) := &\max_{p_1,p_2 \in [N]} \sup_{x \in \reach_\sw(K, T_d, \infty)} \nonumber \\ 
&\int_{0}^{t} \|f_{\sw}(\xi_{x,p_1}(s),p_1) - f_{\sw}(\xi_{x,p_2}(s),p_2) \| ds.
\end{align}

\begin{Assumption}
	\label{ass:finite_d}
	The mode divergence $d(T_d)$ is finite. 
\end{Assumption}

This assumption can be checked, for example, if the reachset $\mathit{\reach_\sw(K,T_d,\infty)}$ is compact and $\|f_{\sw}(x,p)\|$ is finite for any $p$ and any finite $x$. The reason is that $\|f_{\sw}(x_1,p_1) - f_{\sw}(x_2,p_2)\|$ will be finite for any $x \in \mathit{\reach_\sw(K, T_d, \infty)}$. That can be seen by invoking the triangular inequality to get $\|f_{\sw}(x_1,p_1) - f_{\sw}(x_2,p_2)\| \leq \|f_{\sw}(x_1,p_1)\| + \|f_{\sw}(x_2,p_2)\|$, which is finite for finite $x_1$ and $x_2$.

\hussein{Now we restate Theorem 1 in \cite{sibai-mitra-2017} that presents an upper bound on the estimation entropy of Definition~\ref{def:s_entropy}.}

\begin{theorem}[Theorem 1 in \cite{sibai-mitra-2017}]
	\label{thm:switched_entropy_upper_bound}
The entropy of the switched system~(\ref{sys:switched_system}) has the following upper bound:
\begin{align}
\label{eq:sw_bound}
h\sest(\eps, \alpha, \tau) \leq \frac{(L_x +\alpha) n}{\ln 2} + \frac{\log N}{T_e},
\end{align}
where $T_e$ is the largest real number in $(0,\tau]$ such that 
\begin{align}
\label{eq:d_condition}
d(T_e) \leq \eps(1 - e^{-\alpha(T_d - T_e)}).
\end{align}
\end{theorem} 

\begin{remark}
	\hussein{Following our previous argument from Section~\ref{sec:expconvergence} for systems with inputs and setting } $\alpha=0$, results in the condition \hussein{(\ref{eq:d_condition})} on $d(T_e)$ in Theorem~\ref{thm:switched_entropy_upper_bound} becoming $d(T_e) = 0$. In that case, there is no non-zero $T_e$ that satisfies the condition in Equation (\ref{eq:d_condition}) unless all the modes are identical. Hence, if $\alpha = 0$, then $T_e = 0$ and the entropy upper bound in Equation~(\ref{eq:sw_bound}) becomes infinite, and thus useless. However, with a closer look at the proof of Theorem 1 in \cite{sibai-mitra-2017}, \hussein{this infinite bound is an artifact of the assumption in (\ref{eq:d_condition}). In fact,} the choices of $\alpha$ and $T_e$ in (\ref{eq:d_condition}) and the bound (\ref{eq:sw_bound}) can be replaced by $\hat{\alpha}$ and $\hat{T}_e$ such that $\hat{\alpha}\geq \alpha$ and $d(\hat{T}_e) \leq \eps(1 - e^{-\hat{\alpha}(T_d - \hat{T}_e)})$. Such a bound may be finite even in the case when $\alpha = 0$.
	\end{remark}


\subsection{Finiteness of mode divergence $d(t)$}
\label{sec:finite_d}
In this section, similar to Sections~\ref{sec:epsilonzero} and \ref{sec:expconvergence}, we present an example system with an infinite estimation entropy.  The system we consider is a scalar switched linear system with an unbounded $d(t)$, giving an intuition as to why Assumption~\ref{ass:finite_d} may be necessary. A similar result was derived by Berger and Jungers \cite{berger_jungers_cdc20_infinite_entropy} to show the necessity of dwell time constraints for the finitness of stabilization entropy of switched linear systems.

We first define separated sets for switched systems. 
Two trajectories $\xi_{x_1,\sigma_1}$ and $\xi_{x_2,\sigma_2}$ of switched system~(\ref{sys:switched_system}) are $(T,\eps,\alpha,\tau)$-separated if there exists a $t \in [0,T]$ such that the inequality~(\ref{e-estimation}) is violated when \hussein{both: $z$ in the LHS is replaced by $\xi_{x_1,\sigma_1}$ and $\xi_{x_0,\sigma}$ by  $\xi_{x_2,\sigma_2}$, respectively,
	 {\em and} when $z$ in the LHS is replaced by $\xi_{x_2,\sigma_2}$ and $\xi_{x_0,\sigma}$ by $\xi_{x_1,\sigma_1}$, respectively}. 
\sayan{(was: the LHS is $\| \xi_{x_1,\sigma_1}(t) - \xi_{x_2,\sigma_2}(t)\|$ instead of $\| z(t) - \xi_{x_0,\sigma}\|$)}. A set of trajectories $\Zsep$ is $(T,\eps,\alpha,\tau)$-separated if all pairs of trajectories in $\Zsep$ are $(T,\eps,\alpha,\tau)$-separated. The maximal cardinality of such a set is denoted by $s\sest\sep(T,\eps,\alpha,\tau)$.
\begin{Lemma}
	\label{lm:s-n_less_s_exp}
	For all $K$, $T_d$,  $\eps$, $\alpha$, $\tau_1$, $\tau_2$, and $T> 0$,
	\begin{align}
	\hussein{s\sest\sep(T,2\eps,0,\tau_1)  \leq s\sest(T,\eps,0,\tau_2) \leq s\sest(T,\eps,\alpha,\tau_2)} \sayan{(was: s\sest\sep(T,2\eps,\alpha,\tau) \leq s\sest(T,\eps,\alpha,\tau_2))}.
	\end{align}
\end{Lemma}
\begin{proof}
	\hussein{The first inequality} follows from the same proof as Lemma~\ref{lm:n_less_s} but by replacing the previous definitions of approximating functions and separated trajectories with those of this section. \hussein{The second inequality follows from the fact that for any $\alpha \geq 0$, any $(T,\eps, \alpha, \tau)$-approximating set is also  a $(T,\eps, 0, \tau)$-approximating one.}
\end{proof}

Fix arbitrary two constants $a> b >0$, and consider the functions $f_\sw^l(x,1) = a x$ and $f_\sw^l(x,2) = bx$, and the resulting switched system~(\ref{sys:switched_system}) having them as its modes:
\begin{equation}
\label{eq:exp_switched_system}
\dot{x}(t) = f_\sw^l(x(t),\sigma(t)),
\end{equation} 
where $\sigma: \nnreals \rightarrow \{1,2\}$ is a switching signal with some minimum dwell time $T_d > 0$.

\begin{Lemma}
	\label{lm:infinite-d}
$\forall t > 0, d(t)$ of system~(\ref{eq:exp_switched_system}) is unbounded.
\end{Lemma}
\begin{proof}
Observe that the state of the system $x(t)$ at any time $t$ is lower bounded by $|x_0|e^{bt}$ and upper bounded by $|x_0|e^{at}$.
Since $x_0 \neq 0$, $\reach_\sw(K,T_d,\infty)$ is unbounded. 
Moreover, the difference between the two modes is as follows: $|\xi_{x,a}(t) - \xi_{x,b}(t)| = |e^{at} -e^{ bt}||x|$. 
Therefore, the integral in Equation~(\ref{def:doft}) is $\int_{0}^{t}|f_\sw^l(\xi_{x,a}(s),a) - f_\sw^l(\xi_{x,b}(s),b)| ds = |ae^{at} -be^{bt}||x| = |be^{bt}x||\frac{a}{b}e^{(a-b)t} - 1|$, which diverges to $\infty$ with diverging $x$. Thus, when the $\sup$  over the reachset is taken, the argument of the integral  is infinite and $d(t)$ is unbounded, for any $t>0$.
\end{proof}


We first ignore the minimum dwell time constraint and construct a family of switching signals in a way similar to how we constructed the sets of input signals in Definition~\ref{def:pwconstantab}.  \hussein{However, in contrast with Definition~\ref{def:pwconstantab}, instead of having a sequence $A$ to define the allowed increments in the input signals, the switching signals we consider here only switch between two fixed values $a$ and $b$ at the time instants in $\tseq$.}
\hussein{In a similar construction to that of the input signals in Section~\ref{sec:expconvergence}, } we consider the sequence $\tseq_\infd$ of time instants of the form $t_i = \sum_{j=1}^{i} v_j$ that are less than $T$, where $v_1 = 2\eps/(|x_0|(a-b))$ and $v_{i+1} = v_i e^{-bv_i}$ for $i\geq 1$. We denote the resulting set of switching signals by $\Sigma_{\infd}$.  \hussein{By a similar argument to Lemma~\ref{lm:nonzeno-proof}, $t_i \rightarrow \infty$ as $i \rightarrow \infty$, and thus the input signals do not have Zeno behavior.} We add the dwell time constraint back in Lemma~\ref{lm:dwelltimeconstraint_infinite_entropy}.

The time sequence used here $\tseq_\infd$ is similar to the time sequence $\tseq_\palpha$ we considered in Section~\ref{sec:expconvergence} to prove that the entropy of system~(\ref{sys:simple}) is infinite if $\alpha > 0$. The first time instance in $\tseq_\palpha$ is equal to \hussein{$\sqrt{2\eps}$} \sayan{(was $2\eps/(a-b)$)} while that of $\tseq_\infd$ is $2\eps/((a-b)|x_0|)$. Moreover, the time between time instants of $\tseq_\palpha$ is $v_{i+1} = v_i e^{-\alpha v_i/2}$ while that of $\tseq_\infd$ is $v_{i+1} = v_i e^{-b v_i}$. Consequently, we can follow the same proof of Lemma~\ref{lm:countingws}, to conclude that $|\tseq_\infd| \geq$
\begin{align}
	\label{eq:tseqd_lower_bound}
 \hussein{\floor{\frac{2 - b v_1}{4 b v_1}(e^{b T} - 1)} = \floor{\frac{|x_0|(a-b) - b \eps }{4 b \eps}(e^{b T} - 1)}}.
\end{align}
\begin{Lemma}
	\label{lm:nonlipschitz}
	Fix the constants $T$, $\eps$, $\tau$, and $\alpha > 0$. Consider any nonzero initial state $x_0 \in \mathbb{R}$ and the set of switching signals $\Sigma_{\infd}$.
	The resulting set of trajectories of system~(\ref{eq:exp_switched_system}) form a \hussein{($T,2\eps,0,\tau$)}-separated set \sayan{(was: ($T,2\eps,\alpha,\tau$))}.
\end{Lemma}
\begin{proof}
	First, using a similar proof to that of Lemma~\ref{lm:infinite-d}, $|\xi_{x_0,a}(v_1 ) - \xi_{x_0,b}(v_1)| = |e^{av_1} -e^{ bv_1}||x_0| > (a -  b) v_1|x_0| = 2\eps$. For any switching signal $\sigma$, the state at time $t$ is $\xi_{x_0,\sigma}(t) \geq x_0 e^{bt}$. 
	Consider any two switching signals $\sigma_1, \sigma_2 \in \Sigma_{d}$ with corresponding sequences $\mathit{se}_1$ and $\mathit{se}_2$ with the same prefix up till the $i^{\mathit{th}}$ entry and differ in the $(i+1)^{\mathit{th}}$ one.
	Finally, let $x_i = \xi_{x_0,\sigma_2}(t_i)$. Then,
	\begin{align*}
	&|\xi_{x_0,\sigma_1}(t_{i+1}) - \xi_{x_0,\sigma_2}(t_{i+1})| \\
	&= |x_i| |e^{a v_{i+1}} - e^{b v_{i+1}}|  > |x_i|(a -  b)v_{i+1} \\
	&\geq |x_0| e^{b t_i} (a -  b)v_{i+1}  = |x_0| e^{b t_i} (a -  b) v_1 e^{-b t_i} = 2\eps.
	\end{align*}
	Therefore, any two trajectories starting from $x_0$ that have switching signals corresponding to two different sequences are separated by more than $2\eps$ for at least one time instant. Hence, such set of trajectories is \hussein{$(T,2\eps,0,\tau)$}-separated. 
\end{proof}

\begin{theorem}
	\label{thm:switched_unbounded_entropy}
For any $\eps$, $\alpha$, and $\tau > 0$, $h\sest(\eps, \alpha, \tau)$ of switched system~(\ref{eq:exp_switched_system}) is infinite. 
\end{theorem}
\begin{proof}
This theorem follows from the same proof of Theorem~\ref{thm:exp_conv_entropy_infinite} while replacing $\alpha$ with $b$: the cardinality of the separated set constructed in Lemma~\ref{lm:nonlipschitz} is equal to $2^{|\tseq_\infd|}$. Substituting the lower bound on $|\tseq_\infd|$ in  Equation~(\ref{eq:tseqd_lower_bound}), we get that the separated set cardinality is lower bounded by $2^{O(e^{bT})}$. Hence, the cardinality of the maximum separated set \hussein{$s\sest\sep(T,2\eps,0,\tau)$} is lower bounded by $2^{O(e^{bT})}$. From Lemma~\ref{lm:s-n_less_s_exp}, we get that the minimum cardinality of an approximating set $s\sest(T,\eps,\alpha,\tau)$ is lower bounded by $2^{O(e^{bT})}$. If we substitute this lower bound in Definition~\ref{def:s_entropy}, we get $h{\sest}(\eps, \alpha, \tau) = \infty$.
\end{proof}

Now, we show that even if the minimum dwell time constraint is assumed, the entropy of system~(\ref{eq:exp_switched_system}) is still  infinite.

\begin{Lemma}
	\label{lm:dwelltimeconstraint_infinite_entropy}
	Even if the set of possible of switching signals of system~(\ref{eq:exp_switched_system}) is restricted to have a minimum dwell time of $T_d$, its entropy is infinite. 
\end{Lemma}
\begin{proof}
	We will need to modify $\tseq_\infd$ to satisfy the minimum dwell time constraint.
	 We decompose the interval $[0,T]$ to intervals of the form $[jT_d,(j+1)T_d]$ for all integers $j < \floor{\frac{T}{T_d}}$. Then, we define $\tseq_\infd'$ to be the same sequence of time instants of $\tseq_\infd$ in the intervals with odd $j$s, i.e., $[T_d,2T_d]$, $[3T_d,4T_d]$, etc., and no time instances in the intervals with even $j$s, i.e., $[0, T_d]$, $[2T_d,3T_d]$, etc.
	Then, we consider the trajectories that result from switching signals that switch at \hussein{ most at one} time instant \hussein{ in $\tseq_\infd'$ in each $T_d$ interval with odd $j$s, i.e., at most one instant in $[T_d, 2T_d]$, at most one instant in $[3T_d, 4T_d]$, etc}. 
	Then, we can observe that such switching signals 
	 satisfy the minimum dwell time constraint of $T_d$. 
	
	Moreover, the trajectories of system~(\ref{eq:exp_switched_system}) resulting from such switching signals 
	are $(T,2\eps,\alpha,\tau)$-separated, by the same proof of Lemma~\ref{lm:nonlipschitz}. Hence, a lower bound on \hussein{the number of such signals }\sayan{(was $|\tseq_\infd'|$)}would result in a lower bound on the maximum cardinality $s\sest\sep(T,2\eps,\alpha,\tau)$   of $(T,2\eps,\alpha,\tau)$-separated set, and consequently a lower bound on $h\sest(\eps,\alpha,\tau)$ of system~(\ref{eq:exp_switched_system}).
	 
	 \hussein{Now, we will lower bound the number of time instants in $\tseq_\infd'$ in each $T_d$ interval in $[0,T]$ with odd index. The number of switching signals that satisfy the minimum dwell time constraint we defined above is lower bounded by the product of these numbers. To that end, fix an interval $[jT_d, (j+1)T_d]$, with $j$ being a positive odd integer. Let $t_k$ be the first time instant in $\tseq_\infd'$ in that interval. Recall that by definition, $t_{k+1} - t_k = v_{k+1} = v_1 e^{-bt_k} \leq v_1 e^{-j bT_d}$. Using the same analogy as in Equation~(\ref{eq:tseqd_lower_bound}), the number of time instants in $\tseq_\infd'$ in $[jT_d, (j+1)T_d]$ is lower bounded by $\floor{\frac{2 - b v_{k+1}}{4 b v_{k+1}}(e^{b T_d} - 1)}$, which is lower bounded by $\floor{\frac{2 - b  v_1 e^{-j bT_d}}{4 b  v_1 e^{-j bT_d}}(e^{b T_d} - 1)}$.  If we take the product of such lower bounds for each odd $j$, the result is $O(e^{bT_d} + e^{3bT_d} + \dots  e^{b\floor{T/T_d}} )  = O(e^{b \floor{T/T_d}^2})$ . 
	 	Therefore, $s\sest(T,\eps,\alpha,\tau)$ is still lower bounded by $O(e^{b \floor{T/T_d}^2})$ and the entropy $h\sest(\eps,\alpha,\tau)$ is still infinite. } 
	\sayan{was: Since the time interval between the $i^{\mathit{th}}$ and $(i+1)^{\mathit{st}}$ switch $v_i$ of $\tseq_\infd$ decreases as $i$ increases, the number of switches in $\tseq_\infd$ in an interval $[jT_d,(j+1)T_d]$, is smaller than that of the following interval $[(j+1)T_d,(j+2)T_d]$ for all $j$. 
	Hence, the total number of switches in odd time intervals of size $T_d$ in $\tseq_\infd$ is at least half that in the interval $[0,T]$. Thus, $|\tseq_\infd'| \geq |\tseq_\infd|/2$. Therefore, $s\sest(T,\eps,\alpha,\tau)$ is still lower bounded by $2^{O(e^{bT})}$ and the entropy $h\sest(\eps,\alpha,\tau)$ is still infinite. }
\end{proof}

\vspace{-0.3in}
\subsection{Relating the upper bounds on entropy of autonomous switched systems and non-switched systems with inputs}
\label{sec:si_and_ss_upper_bound_relation}



In the following corollary of Proposition~\ref{prop:noise_entropy_upperbound_lip}, 
we present a \hussein{novel} alternative format of the upper bound on the entropy of non-switched systems with inputs that is similar to the entropy upper bound of autonomous switched systems in Theorem~\ref{thm:switched_entropy_upper_bound}\hussein{, the latter being a previous result from \cite{sibai-mitra-2017}}.  

\begin{corollary}
	\label{cor:specificUpperBound}
	Fix any $\eps$ and $\rho > 0$, and let $\delta_x = \eps^2 e^{-(M_x + \rho)\Talgo}$ and $\delta_u$ and $\Talgo$ be such that  
	\begin{align}
	\label{eq:gcu_condition_sw_bound}
	g_{c,u}(\delta_u, \Talgo) \leq \eps^2 (1 - e^{-\rho \Talgo}),
	\end{align}
	where $g_{c,u}$ is as defined in Equation~(\ref{eq:feasible_set}).
	Then, $h\est(\eps)$ of system (\ref{sys:noise}) is upper bounded by:
	\[
\frac{(M_x + \rho)n}{\ln 2} + \frac{\log P}{\Talgo},
	\]
	where $P = \ceil{\frac{\jum +\mu\Talgo}{\delta_u} + 1}^m$ represents the number of possible quantized inputs at each sampling time.
\end{corollary}
\begin{proof}
	By substituting our choice of $\delta_x$ in $g_{c,x}$, we obtain $g_{c,x}(\delta_x,\Talgo) = \eps^2 e^{-\rho \Talgo}$. Then, $g_c(\delta_x,\delta_u,\Talgo) = g_{c,x}(\delta_x,\Talgo) + g_{c,u}(\delta_u,\Talgo) = \eps^2$. Hence, the condition on $g_c$ in Proposition~\ref{prop:noise_entropy_upperbound_lip} is satisfied. Now, by substituting our choice of $\delta_x$ in the bound in $g_o(\delta_x,\delta_u,\Talgo)$, along with replacing $\ceil{\frac{\jum +\mu\Talgo}{\delta_u} + 1}^m$ with $P$, we get the upper bound.
\end{proof}

Note that $P$ in Corollary~\ref{cor:specificUpperBound} represents the number of cells in the grid in the input space $\mathbb{R}^m$ that Algorithm~\ref{fig:approxFunct} chooses from at each sampling time in the construction of approximating functions. The center of the chosen cell would be the input signal for a sampling period of time. In that sense, the quantized input values act as modes of a switched system whose trajectories are the approximation functions constructed by Algorithm~\ref{fig:approxFunct}. Hence, $P$ affects the bound on $h\est$ of non-switched systems with inputs similar to how $N$, the number of modes, affects the bound on $h\sest$ of  switched systems.


In Theorem~\ref{thm:switched_entropy_upper_bound}, $\alpha$ represents the exponential decay in the uncertainty between mode switches. Since the switching signal has a minimum dwell time of $T_d$, the uncertainty decays enough that even after adding the uncertainty caused by the mode switch, the total uncertainty does not exceed $\eps$. The parameter $\rho$ in Corollary~\ref{cor:specificUpperBound} has a similar meaning. 

\begin{Lemma}
	\label{lm:si_exp_conv_approx_function}
	If $\mu = \jum = 0$, line~\ref{Sx_construction} in Algorithm~\ref{fig:approxFunct} is changed to $S_{x,i} \gets B(z_{i-1}(\Talgo^-),\eps e^{-i\rho\Talgo})$, $\delta_{x,0} = \eps$, and for all iterations $i \geq 1$, $\delta_{x,i} = \delta_{x, i-1} e^{-\rho \Talgo}$, then for all $t \geq 0$, 
	\begin{align}
	\|z(t) - \xi_{x_0,u}(t) \| \leq \eps e^{-\rho t}.
	\end{align}
\end{Lemma}
\begin{proof}
	If the input variation represented by $\jum$ and $\mu$ is zero, then $\delta_u$ can be chosen to be zero and $S_{u,i}$ to be equal to $u(0)$ for any $i\geq 0$. Such a choice of $\delta_{x,i}$ and $\delta_{u,i}$ would still satisfy Lemma~\ref{lm:proveApproxFunc}. 
	In addition, it would result in $g_{c,u}(\delta_u,\Talgo) =0$ implying zero contribution of the input to the uncertainty in the state in Equation~(\ref{eq:approxFunction_i}) and $g_{c,x}(\delta_x,\Talgo) = \eps^2 e^{-2\rho i\Talgo}$, at each iteration $i \geq 0$ of Algorithm~\ref{fig:approxFunct}. 
\end{proof}	

Lemma~\ref{lm:si_exp_conv_approx_function} shows that Algorithm~\ref{fig:approxFunct} is able to generate exponentially converging approximating functions in the absence of input variation. This is similar to the exponential decay of the error between switches in switched systems. 

\subsection{Autonomous switched systems modeled as non-switched systems with inputs}
\label{sec:si_abstract_ss}

In this section, we model autonomous switched systems as non-switched systems with inputs. This modeling allows applying the upper bound on entropy $h\est(\eps)$ of Proposition~\ref{prop:noise_entropy_upperbound_lip} to get \hussein{a novel} upper bound on entropy $h\sest(\eps,\alpha,\tau)$, with $\alpha = 0$, of the switched system~(\ref{sys:switched_system}). We compare the resulting bound with the bound of Theorem~\ref{thm:switched_entropy_upper_bound}.


In our modeling, we embed the $N$ modes of the switched system in a continuous input space and define a new dynamics function $\hat{f}_\sw$ that takes a piece-wise continuous input signal instead of a piece-wise constant one in its second argument. We first define the input space $U$ to be the standard simplex of dimension $N-1$, i.e., $U := \{u \in \reals^{N-1} | u[p-1] \geq 0, \sum_{p\in [2;N]} u[p-1] \leq 1\}$ and $\inputset_\sw$ be the set of all measurable functions mapping time to $U$. Hence, $\hat{f}_\sw: \reals^n \times U \rightarrow \reals^n$. Such a construction is standard in the literature analyzing switched systems (e.g., \cite{MARGALIOT20068}). Observe that bounding the input set to be the standard simplex instead of the unbounded real space of dimension $N$, would not affect the fact that $g_o$ in Proposition~\ref{prop:noise_entropy_upperbound_lip} is an upper bound on $h\est$. 


	
%
%
Now, let us define the RHS of the ODE modeling the dynamics of the new system with input: $\forall x\in \reals^n, \forall u \in \inputset_\sw$,
\begin{align}
\label{sys:abstract}
 \hat{f}_\sw(x(t), u(t)) := f_\sw(x(t),1) + \sum_{p \in [2;N]} g(x(t),p)u(t)[p-1],
\end{align}
where $g(x,p) := f_\sw(x,p) - f_\sw(x,1)$.
The solutions of system~(\ref{sys:abstract}) coincide with the solutions of the differential inclusion $\dot{x} = \mathit{co}(f_\sw(x(t),1), \dots, f_\sw(x(t),N))$, where $\mathit{co}$ denotes the convex hull, by the Filippov's Selection Lemma (e.g., Lemma 1 in \cite{MARGALIOT20068}).   
This 
means that the trajectories of the constructed system (\ref{sys:abstract}) include those of the switched system~(\ref{sys:switched_system}). Thus, applying the bound in Proposition~\ref{prop:noise_entropy_upperbound_lip} on (\ref{sys:abstract}) may lead to a courser upper bound on entropy than that of the switched system~(\ref{sys:switched_system}) derived in Theorem~\ref{thm:switched_entropy_upper_bound}. Yet, it is useful for showing the generality of our results on bounding the entropy of non-switched systems with inputs.

Observe that the Jacobian of $\hat{f}_\sw$ with respect to the state and input at a certain $x(t)$ and $u(t)$ can be written as follows: 
\begin{align}
J_u &= 
\begin{bmatrix}
 g(x(t), 2), \dots,
 g(x(t), N)
\end{bmatrix} , \\
J_x& = J_{x,1}(x(t))\nonumber\\
 &\hspace{0.1in} + \sum_{p \in [2;N]} u(t)[p-1] (J_{x,p}(x(t))-J_{x,1}(x(t))) \nonumber 
\end{align}
where $J_{x,p}$ is the 
Jacobian of $f_\sw(x(t),p)$ with respect to $x$.

Given these $J_x$ and $J_u$, we can compute $G_x$ and $G_u$ in Lemma~\ref{lm:local_discrep_func}. 
Using a similar argument to Proposition~\ref{prop:computingMxMu}, $G_x \hussein{\ \leq n(n L_x + \nicefrac{1}{2})  \leq n(n \max_{p \in [N]} L_p + \nicefrac{1}{2})}$ \sayan{(was $n \sum_{p\in[N]}L_p + N/2$)}. Moreover,   $G_u= \sqrt{m} \VERT J_u \VERT= \sqrt{N-1} \VERT J_u \VERT$  and $\VERT J_u \VERT = \max_{j \in [n]} \sum_{p \in [2;N]} |(J_u)_{j,p}| \leq  \sum_{p \in [2;N]} \|g(x(t),p)\| = \sum_{p \in [2;N]} \|f_\sw(x(t),p) - f_\sw(x(t),1) \|$. Let us define 
$d_{\mathit{co}} := \sqrt{N-1} \sup_{x \in \reach(K, \mathcal{U}_\sw, \infty)} \sum_{p \in [2;N]} \|f_\sw(x,p) - f_\sw(x,1)\|$.
$d_{\mathit{co}}$ is an upper bound on $\VERT J_u\VERT$ for all $x \in \reach(K, \mathcal{U}_\sw, \infty)$.
To use the bound on entropy for non-switched systems with inputs in Proposition~\ref{prop:noise_entropy_upperbound_lip}, we have to represent our input set $\inputset_\sw$ in the format of Definition~\ref{def:inputset}. We do that by over-approximating $\inputset_\sw$ by the set of input signals $\hat{\inputset}_\sw$ \hussein{with affine-bounded pointwise variation}  constructed according to Definition~\ref{def:inputset} with parameters $\mu = 0$, $\eta = 1$, \hussein{i.e., $\inputset^p(0, 1)$, } and initial set of inputs $U$ being the unit box $[0,1]^{N-1}$.
The following theorem is a direct application of Corollary~\ref{cor:specificUpperBound} to system~(\ref{sys:abstract}) after setting $\delta_u$ to be equal to $\eta$.

\begin{theorem}
\label{thm:inclusion_bound}
The entropy $h\est(\eps)$ of system (\ref{sys:abstract}) is upper bounded by:
\[
\hussein{\frac{( n(n \max_{p\in[N]}L_p + \nicefrac{1}{2}) + \rho)n}{\ln 2} + \frac{N-1}{\Talgo},}
\]
where $\rho$ and $\Talgo$ satisfy $(2M_ue^{M_x\Talgo} )^2 \Talgo \leq \eps^2(1 - e^{-\rho\Talgo})$, i.e.,
\hussein{$(2d_{\mathit{co}}e^{n(n \max_{p\in[N]}L_p + \nicefrac{1}{2})\Talgo})^2\Talgo \leq \eps^2 (1 - e^{-\rho\Talgo}).$}
\sayan{(removed: $(2d_{\mathit{co}}e^{(n \sum_{p\in[N]}L_p + N/2)\Talgo})^2\Talgo \leq \eps^2 (1 - e^{-\rho\Talgo}))$}.
\end{theorem}
Note that there is always $\rho$ and $\Talgo$ that satisfy the inequality in Theorem~\ref{thm:inclusion_bound} since although both sides increase with increasing $\Talgo$ and the LHS is zero for $\Talgo = 0$, the LHS is independent of $\rho$ and the RHS strictly increases with $\rho$. The bound is on the entropy of the differential inclusion which contains the trajectories of the switched system. It grows as \hussein{$O(n^3)$, $O(\max_{p\in[N]}L_p)$, and $O(N)$} \sayan{(removed: $O(n^2)$, $O(\sum_{p\in[N]}L_p)$, and $O(N)$)} compared to the $O(n)$, $O(\max_{p\in[N]}L_p)$, and $O(\log N)$ of the bound in Theorem~\ref{thm:switched_entropy_upper_bound}. For $\Talgo$ when compared to $T_e$ in Theorem~\ref{thm:switched_entropy_upper_bound}, they have a comparable type of bound as we establish that $d(t) \leq t d_{co} e^{\max_{p \in [N]} L_p t}$  in Appendix~\ref{sec:dt_vs_du}. 
This shows that our results  for non-switched systems with inputs generalize to wide set of applications, including differential inclusions and switched systems.

%% file: conclusion.tex
\section{Conclusion} 
\label{sec:conclusion}
We presented a notion of topological entropy as a lower bound on the bit rate needed to estimate the state of nonlinear \sayan{(removed: open)}dynamical systems with \sayan{(removed: slowly)}\hussein{uncertain inputs with variation upper-bounded by an affine function of time}. 
We considered several alternative definitions of entropy and showed that two of the candidates are not meaningful as they diverge to infinity.
We computed an upper bound on entropy.
We provided another form of the upper bound
 that matches the form of that of switched systems we presented in Theorem 1 in \cite{sibai-mitra-2017}. Moreover, we modeled switched systems as dynamical systems with \sayan{(removed: slowly-varying)}\hussein{uncertain inputs}. Consequently, we were able to use the upper bound we computed to get an upper bound on entropy of switched systems.
Finally, we showed that an assumption made in Theorem 1 in \cite{sibai-mitra-2017} on the difference between the modes of the switched system is needed for a meaningful entropy  definition.


%% file: appendix.tex
\section{Appendix: Proofs of Lemmas}
\label{sec:ISdiscrepancy}

This generalization of the mean-value theorem is  used in the construction of the local IS discrepancy functions in~\cite{Huang-Fan-Mitra-16} restricted to time-invariant systems rather than general time variant ones.
\begin{proposition}
\label{prop:mean_value_thm}
For any  differentiable $f: \mathbb{R}^n \times \mathbb{R}^m \rightarrow \mathbb{R}^n$, for any $x, x' \in \mathbb{R}^n$, any $u , u' \in \mathbb{R}^m$: $f(x', u') - f(x, u) =$
\begin{align*}
& \bigg( \int_{0}^{1} J_x (x + (x' - x)s, u') ds \bigg) (x' - x) \\
&+ \bigg( \int_{0}^{1} J_u (x, u + (u' - u)\tau) d\tau \bigg) (u' - u).
\end{align*}
\end{proposition}

\discrepancy*
\begin{proof} 
	Let $x$ and $x' \in K$, and $u$ and $u' \in \mathcal{U}$. Define $y(t) = \xi_{x',u'}(t) - \xi_{x,u}(t)$ and $v(t) = u'(t) - u(t)$. For a $t \in \nnreals$, using proposition (\ref{prop:mean_value_thm}), we have
	\begin{align}
	\label{def:proof:lm:ydot}
	\dot{y}(t) &= f(\xi_{x',u'}(t),u'(t)) - f(\xi_{x,u}(t),u(t)), \nonumber\\
	&= \bigg( \int_{0}^{1} J_x (\xi_{x,u}(t) + sy(t), u'(t)) ds \bigg) y(t)\nonumber \\
	&\hspace{0.5in}+ \bigg( \int_{0}^{1} J_u (\xi_{x,u}(t), u(t) + v(t)\tau) d\tau \bigg) v(t).
	\end{align}
	We write $ J_x (\xi_{x,u}(t) + sy(t), u'(t))$ as $J_x(t,s)$ or simply $J_x$ when the dependence on $t$ and $s$ is clear from context. Similarly, $J_u (\xi_{x,u}(t), u(t) + v(t)\tau)$ is written as $J_u(t, \tau)$ or $J_u$. Then, differentiating $\|y(t)\|_2^2$ with respect to $t$ leads to: $\frac{d}{d t}\|y(t)\|_2^2 =$
	\begin{align}
	\label{def:proof:lm:local_discrep}
	& \frac{d}{d t} (y(t)^T y(t)) = \dot{y}(t)^Ty(t) + y(t)^T\dot{y}(t) \nonumber \\
	&= y(t)^T \big( \int_{0}^{1} (J_x^T + J_x)  ds \big) y(t) + v(t)^T\big(\int_{0}^{1} J_u^T d\tau\big) y(t) \nonumber\\
	&\hspace{0.5in} + y(t)^T\big(\int_{0}^{1} J_u d\tau\big) v(t) \nonumber\\
	&\hspace{1in} \mbox{[substituting $\dot{y}(t)$ with (\ref{def:proof:lm:ydot})]} \nonumber \\
	&\leq y(t)^T \big( \int_{0}^{1} (J_x^T + J_x)  ds \big) y(t) + y(t)^Ty(t) \nonumber \\
	&\hspace{0.5in} + \big( \big(\int_{0}^{1} J_u d\tau\big) v(t) \big)^T\big( \big(\int_{0}^{1} J_u d\tau\big) v(t) \big),
	\end{align}
	where the inequality follows from the fact that for all $w,z \in \mathbb{R}^n$, $w^Tz + z^Tw \leq w^Tw + z^Tz$, since $0 \leq (z - w)^T(z - w) = z^Tz - w^Tz - z^Tw + w^Tw$.
	Let $\lambda_{J} (\mathcal{X}) = \sup_{x \in \mathcal{X}} \lambda_{max}(\frac{J_x + J_x^T}{2})$ be the upper bound of the eigenvalues of the symmetric part of $J_x$ over $\mathcal{X}$, so $J_x + J_x^T \preceq 2 \lambda_{J}(\mathcal{X}) I$. 
	 Thus, (\ref{def:proof:lm:local_discrep}) becomes:
	\begin{align*}
	\frac{d}{d t}\|y(t)\|_2^2 &\leq (2\lambda_{J}(\mathcal{X}) + 1)\|y(t)\|_2^2 + \|\big(\int_{0}^{1} J_u d\tau\big) v(t)\|_2^2 \\
	&\sayan{removed: \leq 2G_x \|y(t)\|_2^2 + (G_u\|v(t)\|_2)^2,}
	\end{align*}
	for $t \in [t_0,t_1]$. 
	\hussein{Since for any $x \in \reals^n$, $\| x \|_\infty \leq \|x\|_2 \leq \sqrt{n} \|x\|_\infty$, 
	\begin{align*}
		\frac{d}{d t}\|y(t)\|^2 &\leq n (2\lambda_{J}(\mathcal{X}) + 1) \|y(t)\|^2 + m \|\big(\int_{0}^{1} J_u d\tau\big) v(t)\|^2 \\
		&\leq 2G_x \|y(t)\|^2 + (G_u\|v(t)\|)^2,
	\end{align*}
}
	Finally, by integrating both sides of the above inequality from $t_0$ to $t$, we get: 
	\hussein{
		\begin{align}
		\|y(t)\|^2 & \leq e^{2G_x (t-t_0)}\|y(t_0)\|^2 + \int_{t_0}^{t} (G_u e^{G_x (t-\tau)} \|v(\tau)\|)^2d\tau \nonumber \\
		&\leq e^{2G_x (t-t_0)}\|y(t_0)\|^2 \nonumber \\ 
		&\hspace{0.5in} + \int_{t_0}^{t} (G_u (\sup_{s \in [t_0,t_1]} e^{G_x (t-s)}) \|v(\tau)\|)^2d\tau.
		\end{align}}
	\sayan{(was:
	$
	\|y(t)\|^2 \leq e^{2G_x (t-t_0)}\big( \|y(t_0)\|^2 + \int_{t_0}^{t} (G_u \|v(\tau)\|)^2d\tau \big).
	$)}
	
\end{proof}

\boundsMx*

\begin{proof}
First, $J_u$ and $J_x$ exist since $f$ is differentiable in both arguments. Second, 
\sayan{(removed: note that $\VERT J_u \VERT_2 \leq \sqrt{m} \VERT J_u \VERT$, where)}
 $\VERT J_u \VERT = \max_{i \in [n]} \sum_{j = 1}^{m} |(J_u)_{i,j}|$, and $(J_u)_{i,j}$ is the entry in the $i^{th}$ row and $j^{th}$ column of $J_u$. Moreover, since for all $i \in [n],j \in [m]$, $|(J_u)_{i,j}| \leq L_u$, by Lipschitz continuity of $f$ with respect to $u$, then $\VERT J_u \VERT \leq m L_u$.
  \sayan{(removed: Hence, $\VERT J_u \VERT_2 \leq m \sqrt{m} L_u$.)}
  Similarly, one can prove that $\VERT J_x \VERT_\infty \leq n L_x$, since the number of columns is $n$ instead of $m$.
Therefore,
\begin{align*}
&G_x \leq \hussein{\nicefrac{n}{2} ( 2 }\VERT\frac{J_x + J_x^T}{2}\VERT + \hussein{ 1 )} \leq \hussein{\nicefrac{n}{2} (2} \frac{\VERT J_x\VERT + \VERT J_x^T\VERT}{2} + \hussein{ 1 )}\\
&\hspace{0.5in} \leq \hussein{n}(nL_x + \frac{1}{2}),\\
&\mbox{and } G_u \leq \hussein{\sqrt{m} \VERT J_u \VERT \leq }\ m\sqrt{m}L_u.
\end{align*}
\end{proof}

%% file: LinearSystems.tex
	\section{systems with Linear Inputs}
	\label{sec:linear}
 In this section, we provide tighter bounds on entropy $h\est$ than that of Proposition~\ref{prop:noise_entropy_upperbound_lip} for systems where the input affects the dynamics linearly. Formally, we consider dynamical systems of the form:
	\begin{align}
	\label{sys:linearInput}
	\dot{x}(t) = f(x(t)) + u(t),
	\end{align}
	where the initial state $x_0 \in K$ and $u \in \mathcal{U}(\mu, \jum)$, as before. 
	
	We will establish a new IS-discrepancy function designed to utilize the linear relation between the input and the state dynamics of the system. Then, we will use Algorithm~\ref{fig:approxFunct} to construct $(T,\eps)$-approximating functions for the trajectories of system~(\ref{sys:linearInput}) using the new IS-discrepancy function. After that, we will show that the number of approximating functions that can be constructed by the modified algorithm is the same as that of Lemma~\ref{lm:approxSetBound} in terms of its parameters $\delta_x$, $\delta_u$ and $\Talgo$. However, larger values of these parameters would suffice to get $(T,\eps)$-approximating function.
	  
	\subsection{Input-to-state discrepancy function construction for systems with linear inputs}
	In this section, we will show that we can get a tighter upper bound on the distance between two different trajectories than that of (\ref{eq:discrepFunction}). 
	For any two initial states $x_0, x_0'\in K$, two input signals $u,u' \in \mathcal{U}(\mu, \jum )$, and for all $t \in \nnreals$, the distance between the two resulting trajectories can be written as:
	\begin{align}
	\label{eq:linearDiscrepFunc}
	&\| \xi_{x_0,u}(t) - \xi_{x_0',u'}(t) \| \nonumber\\
	&= \| x_0 + \int_{0}^{t} \big(f(\xi_{x_0,u}(s)) + u(s)\big) ds\nonumber\\
	&\hspace{0.5in}- x_0' - \int_{0}^{t} \big(f(\xi_{x_0',u'}(s)) + u'(s)\big) ds \| \nonumber\\
	&\leq \| x_0 - x_0' \| + \int_{0}^{t} \| f(\xi_{x_0,u}(s)) - f(\xi_{x_0',u'}(s)) \| ds \nonumber\\
	&\hspace{0.5in}+ \int_{0}^{t}\| u(s) - u'(s) \|ds \nonumber\\
	&\hspace{1in}\mbox{[by triangular inequality]} \nonumber\\
	&\leq \| x_0 - x_0' \| + \int_{0}^{t} L_x \| \xi_{x_0,u}(s) - \xi_{x_0',u'}(s)\|ds \nonumber\\
	&\hspace{0.5in}+\int_{0}^{t}\| u(s) - u'(s) \|ds \nonumber \\
	&\hspace{1in}\mbox{[by the Lipschitz continuity of $f$]} \nonumber\\
	&\leq \big( \| x_0 - x_0' \| + \int_{0}^{t} \|u(s) - u'(s)\| ds \big) e^{L_xt},
	\end{align}
	where the last inequality follows from the Bellman-Gronwall inequality (e.g. \cite{khalil-book-3ed}). Notice that we have a linear discrepancy function instead of the quadratic one
	we got in (\ref{eq:discrepFunction}). This means that the sensitivity of this system's trajectories with respect
	to changes in the input is smaller than that of nonlinear systems in general.
	
	
	
	\subsection{Approximating set construction}
	\label{sec:approxSetLinearConstruction}
	Let us fix $T$ and $\eps >0$ for this section.
	To construct an $(T,\eps)$-approximating function for a given trajectory, we use Algorithm~\ref{fig:approxFunct} again. The following lemma, as Lemma~\ref{lm:proveApproxFunc}, specifies the conditions that the values of $\delta_x, \delta_u$, and $\Talgo$ should satisfy for the output of Algorithm~\ref{fig:approxFunct}, $z$, to be an $(T,\eps)$-approximating for the system trajectory using the new discrepancy function.
	
	Before stating the lemma, let us define a new $g_c$ for this particular case:
	\begin{align}
	g_c^l(\delta_x, \delta_u, \Talgo) :=&g_{c,x}^l(\delta_x,\Talgo) + g_{c,u}^l(\delta_u,\Talgo),	\label{eq:feasible_set_linear}
	\end{align}
	where $g_{c,x}^l(\delta_x,\Talgo) := \delta_x e^{L_x\Talgo}$ and 
	$g_{c,u}^l( \delta_u, \Talgo) := \big(\Talgo \delta_u +\Talgo (\frac{\mu\Talgo}{2} + \jum) \big) e^{L_x\Talgo}$.
	

	
\begin{Lemma}
	\label{lm:approxFunctionLinear}
Given $\eps > 0$, fix $\delta_x$, $\delta_u$, and $\Talgo$ such that $g_c^l(\delta_x,\delta_u,\Talgo) \leq \eps$.
Then, for any $x_0 \in K$ and $u \in \mathcal{U}(\mu, \jum)$, for all $i \in [0;\floor{\frac{T}{\Talgo}}]$, and for all $t \in [i\Talgo,(i+1)\Talgo)$,
\begin{enumerate}[label=(\roman*)]
	\item $v_i \in S_{u,i}$,
	\item $\|u(t) - q_{u,i}\|\leq \mu (t - i\Talgo) + \jum + \delta_u$,
	\item $x_i \in S_{x,i}$, and
	\item $\| z_i(t - i\Talgo) - \xi_{x_i, u_i}(t - i\Talgo)\| \leq \eps$,
\end{enumerate}
where $u_i(t) := u(i\Talgo + t)$, the $i^{\mathit{th}}$ piece of the input signal of size $\Talgo$.
\end{Lemma}
	\begin{proof}
	The first two conclusions follow from the same proof of Lemma~\ref{lm:proveApproxFunc}. Now,
	fix $x_0\in K$ and $u\in \mathcal{U}(\mu, \jum )$ and let $t' = t - i\Talgo$. Then, by (\ref{eq:linearDiscrepFunc}),
	\begin{align}
	\label{eq:intermediatelinearapproxFuncbound}
	&\|z_i(t') - \xi_{x_i,u_i}(t')\|\nonumber \\
	&\leq \big(\| x_i - q_{x,i} \| + \int_{0}^{t'}\| u_i(s) -  q_{u,i} \| ds\big) e^{L_xt'} \nonumber\\
	&\hspace{0.8in}\mbox{[by (\ref{eq:linearDiscrepFunc})]}\nonumber\\
	&\leq  \big(\| x_i - q_{x,i} \|  + \int_{0}^{t'}\big(\| u_i(s) - u_i(0)\| \nonumber\\
	&\hspace{0.3in} + \|u_i(0) - q_{u,i} \|\big) ds\big) e^{L_xt'} \nonumber \\
	&\hspace{0.8in} \mbox{[by triangular inequality]} \nonumber \\
	&\leq \big(\delta_x  + \int_{0}^{t'} \| u_i(s) - u_i(0)\|ds + t' \delta_u \big) e^{L_xt'}, \nonumber \\
	&\hspace{0.8in} \mbox{[since $\| x_i - q_{x,i} \|\leq \delta_x$, $\|u_i(0) - q_{u,i} \| \leq \delta_u$]} \nonumber \\
	&\leq \big( \delta_x + \int_{0}^{t'} (\mu s + \jum) ds + t' \delta_u \big)e^{L_x t'} \nonumber\\
	&\hspace{0.8in}\mbox{[by (\ref{variationbound})]} \nonumber \\
	&\leq \big(\delta_x  + \Talgo \delta_u +\Talgo (\frac{\mu\Talgo}{2} + \jum)\big) e^{L_x\Talgo} \nonumber\\
	&\hspace{0.8in}\mbox{[since $t' \leq \Talgo$]} \nonumber \\
	&= g_c^l(\delta_x,\delta_u,\Talgo)\leq \eps,
	\end{align}

	 Hence, for all $i \in [0,\floor{\frac{T}{\Talgo}}]$ and $t \in [i\Talgo, (i+1)\Talgo]$, $x_i \in S_{x,i}$ and $\| \xi_{x_i,u_i}(t) - \xi_{q_{x,i}, q_{u,i}}(t)\| \leq \eps$. Thus, the lemma.
\end{proof}

	
\begin{corollary}
	\label{cor:linear_inter_bound}
Under the same conditions of Lemma \ref{lm:approxFunctionLinear}, the output $z$ of Algorithm~\ref{fig:approxFunct} is a $(T,\eps)$-approximating function of the corresponding trajectory of system (\ref{sys:linearInput}).
Moreover, since we are still using Algorithm~\ref{fig:approxFunct} to construct the approximating function, 
 we have the same upper bound on entropy of system (\ref{sys:linearInput}) as in Proposition \ref{prop:noise_entropy_upperbound_lip} in terms of the new values $\delta_x$, $\delta_u$ and $\Talgo$ that satisfy $g_c^l \leq \eps$ instead of $g_c \leq \eps^2$.
 \end{corollary} 

The following corollary relates the derived bound with that in  \cite{LM015:HSCC} when the input variation represented by $\mu$ and $\jum$ becomes negligible. It follows from a similar proof to that of Corollary~\ref{cor:relationToPrevBound}.
\begin{corollary}
	\label{cor:relationToPrevBound_linear}
	Given any $\eps > 0$, the entropy of system~(\ref{sys:linearInput}) $\underset{\mu,\jum \rightarrow 0}{\lim} h\est(\eps) \leq \frac{nL_x}{\ln 2}$. This bound exactly matches the bound on estimation entropy of closed systems in \cite{LM015:HSCC}.
\end{corollary}
\begin{proof}
	Follows from the same proof of Corollary~\ref{cor:relationToPrevBound} with replacing $\eps^2$ with $\eps$ and $g_c$ with $g_c^l$.
\end{proof}

%% file: PendulumExample.tex
\begin{example}[Pendulum]
\label{sec:pendulumExample}
\normalfont Consider a pendulum system:
\begin{align*}
	&\dot{x}_1 = x_2;\ \dot{x}_2 = -\frac{Mgl}{I} \sin x_1 + \frac{u}{I},
\end{align*}
where $I$ is the moment of inertia of the pendulum around the pivot point, $u$ is its input from a DC motor, $x_1$ is the angular position (with respect to  $y$-axis), $x_2$ is the angular speed, and $l$ is the length, and $M$ is the mass. 

Consider the case when $\frac{Mgl}{I} = 0.98$, $I = 1$, $u_{max} = 2$, $\mu = 0.1$ and $\jum = 1$. 
Jacobian matrices of $f$ are:
\[
J_x = 
\begin{bmatrix}
&0 &1 \\
&-\frac{Mgl}{I}\cos x_1 &0
\end{bmatrix}\ 
J_u = 
\begin{bmatrix}
0 \\
\nicefrac{1}{I}
\end{bmatrix}.
\]
Then, \hussein{$G_x = \nicefrac{n}{2}(2\lambda_{\max}(\frac{J_x + J_x^T}{2}) + 1) = 2 (|0.98 \cos(x_1) - 1| + \nicefrac{1}{2})$. Thus, $M_x = 4.96$.} \sayan{(was ($G_x$ equation was wrong): $\lambda_{\max}(\frac{J_x + J_x^T}{2}) = |0.98 \cos(x_1) - 1|$. Thus, $M_x = 1.98$)}
Moreover, \hussein{$G_u = \sqrt{m} \VERT J_u\VERT = \lambda_{max}(J_u^T J_u) = \frac{1}{I^2} = 1$, and $M_u = 1$.}  (\sayan{was $G_u = \VERT J_u\VERT = \lambda_{max}(J_u^T J_u) = \frac{1}{I^2} = 1$, and $M_u = 1$.})
%
We shall compute the entropy bounds for estimation accuracy of $\varepsilon = 0.01$.
Let us fix $\delta_u = 0.1$, $\Talgo = 2.5 \times 10^{-3}$, and $\delta_x = \frac{1}{\sqrt{2}}\eps e^{-M_x\Talgo}$
If we use the bound of Proposition~\ref{prop:noise_entropy_upperbound_lip}, then $g_c(\delta_x,\delta_u,\Talgo) = 0.003$ is not less than $\eps^2 = 10^{-4}$. However, $g_c^{l}(\delta_x,\delta_u,\Talgo) = 0.0098 \leq \eps = 0.01$. Thus, the bound per Corollary~\ref{cor:linear_inter_bound} would be 
  $h\est(0.01)\leq \hussein{673}$ \sayan{(was $1586$)}. If we choose $\Talgo = 4 \times 10^{-5}$ instead, we get $g_c(\delta_x,\delta_u,\Talgo) = 9.8 \times 10^{-5}$, which is less than $\eps^2$, and the bound we got is $h\est(0.01)\leq \hussein{42032}$ \sayan{(was $99114$)}. 
%
 Thus, Corollary~\ref{cor:linear_inter_bound} can give an entropy bound that's much tighter than that in Proposition~\ref{prop:noise_entropy_upperbound_lip}.
\end{example}


%

\section{Harrier Jet Jacobian matrices}
\label{sec:harrier_jacobian}
The Jacobian of the harrier jet dynamics in example~\ref{eg:harrier_jet} with respect to the state $x$ is: $J_x = $
\begin{align}
\begin{bmatrix}
	0 &0 &0 &1 &0 &0 \\
	0 &0 &0 &0 &1 &0 \\
	0 &0 &0 &0 &0 &1 \\
	0 &0 &-g\cos x_3 -\frac{u_1}{m'}\sin x_3 - \frac{u_2}{m'}\cos x_3 &-\frac{c}{m'} &0 &0 \\
	0 &0 &-g\sin x_3 +\frac{u_1}{m'}\cos x_3 - \frac{u_2}{m'}\sin x_3 &0             &-\frac{c}{m'} &0 \\
	0 &0 &0 &0 &0 &0
\end{bmatrix}. 
\end{align}
Hence, $(J_x + J_x^T)/2 = $
\begin{align}
	\begin{bmatrix}
		0 &0 &0 &1 &0 &0 \\
		0 &0 &0 &0 &1 &0 \\
		0 &0 &0 &\frac{y_1}{2} &\frac{y_2}{2} &1 \\
		1 &0 &\frac{y_1}{2} &-\frac{c}{m'} &0 &0 \\
		0 &1 &\frac{y_2}{2} &0 &-\frac{c}{m'} &0 \\
		0 &0 &1 &0 &0 &0
	\end{bmatrix},
\end{align}
where $y_1 = -g\cos x_3 -\frac{u_1}{m'}\sin x_3 - \frac{u_2}{m'}\cos x_3$ and $y_2 = -g\sin x_3 +\frac{u_1}{m'}\cos x_3 - \frac{u_2}{m'}\sin x_3$.
The unique eigenvalue of $(J_x + J_x^T)/2$ would be $-\frac{c}{m'}$.

The Jacobian with respect to the input $u$ is: $J_u = $
\begin{align}
\begin{bmatrix}
0 &0 \\
0 &0 \\
0 &0 \\
\frac{\cos x_3}{m'} & \frac{-\sin x_3}{m'} \\
\frac{\sin x_3}{m'} & \frac{\cos x_3}{m'} \\
\frac{r}{J} &0
\end{bmatrix}.
\end{align}

The singular values of $J_u$ can be obtained by computing the square root of the absolute values of the  eigenvalues of $J_u J_u^T$. They are: $0$, $\sqrt{\frac{r^2}{J^2}+ \frac{1}{m'^2}}$, and $\frac{1}{m'}$. Therefore, $\hussein{\VERT J_u \VERT_2} = \sqrt{\frac{r^2}{J^2}+ \frac{1}{m'^2}}$ \hussein{ and thus $\VERT J_u \VERT \leq \sqrt{\frac{r^2}{J^2}+ \frac{1}{m'^2}}$}.

\section{Comparing $d(t)$ and $d_{co}$}
\label{sec:dt_vs_du}
\begin{theorem}
\label{thm:dt_vs_dco}
For any $t > 0$, $d(t)$ of system~(\ref{sys:switched_system}) is upper bounded by $t d_{co} e^{\max_{p\in[N]}L_{p} t}$.
\end{theorem}
\begin{proof}
Recall that
\begin{align*}
\dist(t) := &\max_{p_1,p_2 \in [N]} \sup_{x \in \reach_\sw(K, T_d, \infty)} \nonumber \\ 
&\int_{0}^{t} \|f_{\sw}(\xi_{x,p_1}(s),p_1) - f_{\sw}(\xi_{x,p_2}(s),p_2) \| ds.
\end{align*}

Then, for any $x \in \mathbb{R}^n$, $p_1$ and $p_2 \in [N]$ and $t > 0$,
\begin{align}
&\int_{0}^{t} \|f_{\sw}(\xi_{x,p_1}(s),p_1) - f_{\sw}(\xi_{x,p_2}(s),p_2) \| ds \nonumber\\
&\leq \int_{0}^{t} \|f_{\sw}(\xi_{x,p_1}(s),p_1) - f_{\sw}(\xi_{x,p_1}(s),p_2) \| ds + \nonumber\\
&\hspace{0.5in} \int_{0}^{t} \|f_{\sw}(\xi_{x,p_1}(s),p_2) - f_{\sw}(\xi_{x,p_2}(s),p_2) \| ds \nonumber\\
&\hspace{0.5in}\mbox{[triangular inequality]} \nonumber \\
&\leq t d_{co} + \int_{0}^{t} \|f_{\sw}(\xi_{x,p_1}(s),p_2) - f_{\sw}(\xi_{x,p_2}(s),p_2) \| ds \nonumber\\
&\hspace{0.5in}\mbox{[$d_{co}$ upper bounds the first integral's argument]}\nonumber\\
&\leq t d_{co} + \int_{0}^{t} L_{p_2} \|\xi_{x,p_1}(s) - \xi_{x,p_2}(s) \| ds \nonumber\\
&\hspace{0.5in}\mbox{[$L_{p_2}$ is the Lipschitz constant for $f_\sw$ with mode $p_2$]} \nonumber\\
&\leq t d_{co} + \int_{0}^{t} L_{p_2} d(s) ds \nonumber\\
&\leq td_{co} e^{L_{p_2} t},
\end{align}
where the last inequality follows from Bellman-Grownall inequality.
\end{proof}

%% file: root.bbl
\begin{thebibliography}{10}
\providecommand{\url}[1]{#1}
\csname url@samestyle\endcsname
\providecommand{\newblock}{\relax}
\providecommand{\bibinfo}[2]{#2}
\providecommand{\BIBentrySTDinterwordspacing}{\spaceskip=0pt\relax}
\providecommand{\BIBentryALTinterwordstretchfactor}{4}
\providecommand{\BIBentryALTinterwordspacing}{\spaceskip=\fontdimen2\font plus
\BIBentryALTinterwordstretchfactor\fontdimen3\font minus
  \fontdimen4\font\relax}
\providecommand{\BIBforeignlanguage}[2]{{%
\expandafter\ifx\csname l@#1\endcsname\relax
\typeout{** WARNING: IEEEtran.bst: No hyphenation pattern has been}%
\typeout{** loaded for the language `#1'. Using the pattern for}%
\typeout{** the default language instead.}%
\else
\language=\csname l@#1\endcsname
\fi
#2}}
\providecommand{\BIBdecl}{\relax}
\BIBdecl

\bibitem{LM:TAC2018}
D.~{Liberzon} and S.~{Mitra}, ``Entropy and minimal bit rates for state
  estimation and model detection,'' \emph{IEEE Transactions on Automatic
  Control}, vol.~63, no.~10, pp. 3330--3344, 2018.

\bibitem{LM015:HSCC}
D.~Liberzon and S.~Mitra, ``Entropy and minimal data rates for state estimation
  and model detections,'' in \emph{HSCC 2016}.\hskip 1em plus 0.5em minus
  0.4em\relax Vienna: ACM, April 2016.

\bibitem{sibai-mitra-2017}
\BIBentryALTinterwordspacing
H.~Sibai and S.~Mitra, ``Optimal data rate for state estimation of switched
  nonlinear systems,'' in \emph{Proceedings of the 20th International
  Conference on Hybrid Systems: Computation and Control}, ser. HSCC '17.\hskip
  1em plus 0.5em minus 0.4em\relax New York, NY, USA: Association for Computing
  Machinery, 2017, p. 71–80. [Online]. Available:
  \url{https://doi.org/10.1145/3049797.3049799}
\BIBentrySTDinterwordspacing

\bibitem{Savkin-Petersen-2003}
A.~V. Savkin and I.~R. Petersen, ``Set-valued state estimation via a limited
  capacity communication channel,'' \emph{IEEE Transactions on Automatic
  Control}, vol.~48, no.~4, pp. 676--680, April 2003.

\bibitem{sibai-mitra-2018}
\BIBentryALTinterwordspacing
H.~Sibai and S.~Mitra, ``State estimation of dynamical systems with unknown
  inputs: Entropy and bit rates,'' in \emph{Proceedings of the 21st
  International Conference on Hybrid Systems: Computation and Control (Part of
  CPS Week)}, ser. HSCC '18.\hskip 1em plus 0.5em minus 0.4em\relax New York,
  NY, USA: Association for Computing Machinery, 2018, p. 217–226. [Online].
  Available: \url{https://doi.org/10.1145/3178126.3178150}
\BIBentrySTDinterwordspacing

\bibitem{nair-entropy-tac-ncs}
G.~N. Nair, R.~J. Evans, I.~M.~Y. Mareels, and W.~Moran, ``Topological feedback
  entropy and nonlinear stabilization,'' \emph{IEEE Trans.\ Automat.\ Control},
  vol.~49, pp. 1585--1597, 2004.

\bibitem{nair-...-survey}
G.~N. Nair, F.~Fagnani, S.~Zampieri, and R.~J. Evans, ``Feedback control under
  data rate constraints: An overview,'' vol.~95, pp. 108--137, 2007.

\bibitem{colonius-kawan-09}
F.~Colonius and C.~Kawan, ``Invariance entropy for control systems,''
  \emph{SIAM J.\ Control Optim.}, vol.~48, pp. 1701--1721, 2009.

\bibitem{colonius-kawan-nair-equivalence}
F.~Colonius, C.~Kawan, and G.~Nair, ``A note on topological feedback entropy
  and invariance entropy,'' \emph{Systems Control Lett.}, vol.~62, pp.
  377--381, 2013.

\bibitem{katok-haselblatt}
A.~Katok and B.~Hasselblatt, \emph{Introduction to the Modern Theory of
  Dynamical Systems}.\hskip 1em plus 0.5em minus 0.4em\relax Cambridge
  University Press, 1995.

\bibitem{yang2015stabilizing}
{G. Yang} and D.~{Liberzon}, ``Stabilizing a switched linear system with
  disturbance by sampled-data quantized feedback,'' in \emph{2015 American
  Control Conference (ACC)}, 2015, pp. 2193--2198.

\bibitem{Yang-Liberzon-2016}
\BIBentryALTinterwordspacing
G.~Yang and D.~Liberzon, ``Finite data-rate stabilization of a switched linear
  system with unknown disturbance**this work was supported by the nsf grants
  cns-1217811 and eccs-1231196.'' \emph{IFAC-PapersOnLine}, vol.~49, no.~18,
  pp. 1085 -- 1090, 2016, 10th IFAC Symposium on Nonlinear Control Systems
  NOLCOS 2016. [Online]. Available:
  \url{http://www.sciencedirect.com/science/article/pii/S2405896316318997}
\BIBentrySTDinterwordspacing

\bibitem{yang_liberzon_stabilization_switched_tac2018}
G.~{Yang} and D.~{Liberzon}, ``Feedback stabilization of switched linear
  systems with unknown disturbances under data-rate constraints,'' \emph{IEEE
  Transactions on Automatic Control}, vol.~63, no.~7, pp. 2107--2122, 2018.

\bibitem{Rungger-Zamani-2017}
\BIBentryALTinterwordspacing
M.~Rungger and M.~Zamani, ``Invariance feedback entropy of nondeterministic
  control systems,'' in \emph{Proceedings of the 20th International Conference
  on Hybrid Systems: Computation and Control}, ser. HSCC '17.\hskip 1em plus
  0.5em minus 0.4em\relax New York, NY, USA: ACM, 2017, pp. 91--100. [Online].
  Available: \url{http://doi.acm.org/10.1145/3049797.3049801}
\BIBentrySTDinterwordspacing

\bibitem{Tomar_Zamani_Networkinvarianceentropy_2020}
M.~S. {Tomar} and M.~{Zamani}, ``Compositional quantification of invariance
  feedback entropy for networks of uncertain control systems,'' \emph{IEEE
  Control Systems Letters}, vol.~4, no.~4, pp. 827--832, 2020.

\bibitem{Tomar_Zamani_numerical_2020}
M.~S. Tomar, C.~Kawan, P.~Jagtap, and M.~Zamani, ``Numerical estimation of
  invariance entropy for nonlinear control systems,'' 2020.

\bibitem{MSThesis:Schmidt}
J.~Schmidt, ``Topological entropy bounds for switched linear systems with lie
  structure,'' Master's thesis, Electrical and Computer Engineering from the
  University of Illinois at Urbana-Champaign, 2016.

\bibitem{Guosong_Schmidt_Liberzon_Hespanha_2020}
\BIBentryALTinterwordspacing
G.~Yang, A.~J. Schmidt, D.~Liberzon, and J.~P. Hespanha, ``Topological entropy
  of switched linear systems: general matrices and matrices with commutation
  relations,'' \emph{Mathematics of Control, Signals, and Systems}, vol.~32,
  no.~3, pp. 411--453, 2020. [Online]. Available:
  \url{https://doi.org/10.1007/s00498-020-00265-9}
\BIBentrySTDinterwordspacing

\bibitem{Yang_Schmidt_Liberzon_CDC2018}
G.~{Yang}, A.~{James Schmidt}, and D.~{Liberzon}, ``On topological entropy of
  switched linear systems with diagonal, triangular, and general matrices,'' in
  \emph{2018 IEEE Conference on Decision and Control (CDC)}, 2018, pp.
  5682--5687.

\bibitem{Yang_Hespanha_Liberzon_hscc2019}
\BIBentryALTinterwordspacing
G.~Yang, J.~a.~P. Hespanha, and D.~Liberzon, ``On topological entropy and
  stability of switched linear systems,'' in \emph{Proceedings of the 22nd ACM
  International Conference on Hybrid Systems: Computation and Control}, ser.
  HSCC '19.\hskip 1em plus 0.5em minus 0.4em\relax New York, NY, USA:
  Association for Computing Machinery, 2019, p. 119–127. [Online]. Available:
  \url{https://doi.org/10.1145/3302504.3311815}
\BIBentrySTDinterwordspacing

\bibitem{Vicinansa_Liberzon_regular_switched_entropy_2019}
G.~S. {Vicinansa} and D.~{Liberzon}, ``Estimation entropy for regular linear
  switched systems,'' in \emph{2019 IEEE 58th Conference on Decision and
  Control (CDC)}, 2019, pp. 5754--5759.

\bibitem{liberzon_interconnected_entropy_2021}
D.~Liberzon, ``On topological entropy of interconnected nonlinear systems,''
  \emph{IEEE Control Systems Letters}, vol.~5, no.~6, pp. 2210--2214, 2021.

\bibitem{Gao-Liberzon-Basar-2015}
X.~Gao, D.~Liberzon, J.~Liu, and T.~Basar, ``Connections between stability
  conditions for slowly time-varying and switched linear systems,'' in
  \emph{2015 54th IEEE Conference on Decision and Control (CDC)}, Dec 2015, pp.
  2329--2334.

\bibitem{Hespanha-Morse-1999}
J.~P. Hespanha and A.~S. Morse, ``Stability of switched systems with average
  dwell-time,'' in \emph{Proceedings of the 38th IEEE Conference on Decision
  and Control (Cat. No.99CH36304)}, vol.~3, 1999, pp. 2655--2660 vol.3.

\bibitem{khalil-book-3ed}
H.~K. Khalil, \emph{Nonlinear Systems}, 3rd~ed.\hskip 1em plus 0.5em minus
  0.4em\relax New Jersey: Prentice Hall, 2002.

\bibitem{colonius-siam-2012}
F.~Colonius, ``Minimal bit rates and entropy for exponential stabilization,''
  \emph{SIAM J.\ Control Optim.}, vol.~50, pp. 2988--3010, 2012.

\bibitem{Matveev2009}
\BIBentryALTinterwordspacing
A.~S. Matveev and A.~V. Savkin, \emph{Stabilization of Linear Multiple Sensor
  Systems via Limited Capacity Communication Channels}.\hskip 1em plus 0.5em
  minus 0.4em\relax Boston: Birkh{\"a}user Boston, 2009, pp. 1--64. [Online].
  Available: \url{https://doi.org/10.1007/978-0-8176-4607-3_3}
\BIBentrySTDinterwordspacing

\bibitem{Huang-Fan-Mitra-16}
\BIBentryALTinterwordspacing
Z.~Huang, C.~Fan, and S.~Mitra, ``Bounded invariant verification for
  time-delayed nonlinear networked dynamical systems,'' \emph{Nonlinear
  Analysis: Hybrid Systems}, vol.~23, pp. 211 -- 229, 2017. [Online].
  Available:
  \url{http://www.sciencedirect.com/science/article/pii/S1751570X16300279}
\BIBentrySTDinterwordspacing

\bibitem{Fan_Mitra_2016}
\BIBentryALTinterwordspacing
C.~Fan, J.~Kapinski, X.~Jin, and S.~Mitra, ``Locally optimal reach set
  over-approximation for nonlinear systems,'' in \emph{Proceedings of the 13th
  International Conference on Embedded Software}, ser. EMSOFT '16.\hskip 1em
  plus 0.5em minus 0.4em\relax New York, NY, USA: ACM, 2016, pp. 6:1--6:10.
  [Online]. Available: \url{http://doi.acm.org/10.1145/2968478.2968482}
\BIBentrySTDinterwordspacing

\bibitem{Fan_Mitra_2017}
\BIBentryALTinterwordspacing
{C. Fan}, {J. Kapinski}, {X. Jin}, and {S. Mitra}, ``Simulation-driven
  reachability using matrix measures,'' \emph{ACM Trans. Embed. Comput. Syst.},
  vol.~17, no.~1, Dec. 2017. [Online]. Available:
  \url{https://doi.org/10.1145/3126685}
\BIBentrySTDinterwordspacing

\bibitem{this_paper_arxiv}
\BIBentryALTinterwordspacing
H.~Sibai and S.~Mitra, ``State estimation of open dynamical systems with slow
  inputs: Entropy, bit rates, and relation with switched systems,'' 2020.
  [Online]. Available: \url{https://arxiv.org/abs/2011.10496}
\BIBentrySTDinterwordspacing

\bibitem{Astrom-2008}
K.~J. Astrom and R.~M. Murray, \emph{Feedback Systems: An Introduction for
  Scientists and Engineers}.\hskip 1em plus 0.5em minus 0.4em\relax Princeton,
  NJ, USA: Princeton University Press, 2008.

\bibitem{berger_jungers_cdc20_infinite_entropy}
G.~O. {Berger} and R.~M. {Jungers}, ``Finite data-rate feedback stabilization
  of continuous-time switched linear systems with unknown switching signal,''
  in \emph{2020 59th IEEE Conference on Decision and Control (CDC)}, 2020, pp.
  3823--3828.

\bibitem{MARGALIOT20068}
\BIBentryALTinterwordspacing
M.~Margaliot and D.~Liberzon, ``Lie-algebraic stability conditions for
  nonlinear switched systems and differential inclusions,'' \emph{Systems \&
  Control Letters}, vol.~55, no.~1, pp. 8--16, 2006. [Online]. Available:
  \url{https://www.sciencedirect.com/science/article/pii/S0167691105000824}
\BIBentrySTDinterwordspacing

\end{thebibliography}
